\documentclass{article}
\usepackage[utf8]{inputenc}

\pdfoutput=1 
\usepackage{jheppub}
\usepackage{graphicx,verbatim}
\usepackage{psfrag, times}
\usepackage[T1]{fontenc}
\usepackage{epsf}
\usepackage{mathtools}
\usepackage[vcentermath]{youngtab}

\usepackage{amsmath}
\usepackage{amsfonts}
\usepackage{amssymb}
\usepackage{mathrsfs}
\usepackage{slashed}
\usepackage{xcolor}
\def\pa{\partial}
\def\ii{\textrm i}
\def\spch{XXX_{\mathfrak{sl}_2}}
\def\half{\frac 12}
\def\CalS{\mathcal{S}}
\def\CalF{\mathcal{F}}

\def\bE{\mathbb{E}}
\def\xb{\bar{x}}
\def\kq{\mathfrak{q}}

\def\ba{\mathbf{a}}
\def\rx{\mathrm{x}}
\def\ri{\mathrm{i}}
\def\rj{\mathrm{j}}

\def\bm{\mathbf{m}}
\def\bz{\mathbf{z}}
\def\by{\mathbf{y}}
\def\bx{\mathbf{x}}
\def\BP{\mathbb{P}}
\def\BC{\mathbb{C}}
\def\BR{\mathbb{R}}
\def\BZ{\mathbb{Z}}
\def\BN{\mathbb{N}}
\def\CalN{\mathcal{N}}
\def\CalV{\mathcal{V}}
\def\CalP{\mathcal{P}}
\def\CalR{\mathcal{R}}
\def\CalZ{\mathcal{Z}}
\def\CalW{\mathcal{W}}
\def\CalU{\mathcal{U}}
\def\CalO{\mathcal{O}}
\def\CalC{\mathcal{C}}
\def\qe{\mathfrak{q}}
\def\CalB{\mathcal{B}}
\def\CalL{\mathcal{L}}
\def\CalX{\mathcal{X}}
\def\CalM{\mathcal{M}}

\def\Tr{{\rm Tr}}

\def\ta{\mathtt{a}}
\def\tb{\mathtt{b}}
\def\tc{\mathtt{c}}
\def\td{\mathtt{d}}
\def\tp{\mathtt{P}}
\def\rS{\mathrm{S}}

\def\beq{\begin{equation}}
\def\eeq{\end{equation}}

\title{Quantum Spin Systems and Supersymmetric Gauge Theories, I}
\author[]{Norton Lee${}^{\diamondsuit}$ and Nikita Nekrasov ${}^{\bullet}$}
\affiliation{$^{\diamondsuit}$\rm C. N. Yang Institute for Theoretical Physics, Stony Brook University, Stony Brook, NY, 11794-3636, USA}
\affiliation{$^{\bullet}$\rm Simons Center of Geometry and Physics, Stony Brook University, Stony Brook, NY, 11794-3636, USA\footnote{On leave of absence from:
Center for Advanced Studies, SkolTech, 1 Nobel Street, Moscow,  143026, Russia, and 
Kharkevich Institute for Information Transmission Problems RAS, Bolshoy Karetny 19, Moscow 127051 Russia}}
\emailAdd{norton.lee@stonybrook.edu}
\emailAdd{nnekrasov@scgp.stonybrook.edu}

\vspace{2cm}
\abstract{The relation between supersymmetric gauge theories in four dimensions and quantum spin systems is exploited to find an explicit formula for the Jost function of the $N$ site $\mathfrak{sl}_{2}$ $XXX$ spin chain (for infinite dimensional complex spin representations), as well as the $SL_N$ Gaudin system, which reduces, in a limiting case, to that of the $N$-particle periodic Toda chain. 
Using the non-perturbative Dyson-Schwinger equations of the supersymmetric gauge theory we establish relations between the spin chain commuting Hamiltonians with the twisted chiral ring of gauge theory.
Along the way we explore the chamber dependence of the supersymmetric partition function, also the expectation value of the surface defects, giving new evidence for the AGT conjecture.}

\begin{document}

\maketitle

\section{Introduction and Summary}

The BPS/CFT correspondence \cite{Nikita:I} relates the algebra and geometry of two dimensional conformal field theories, and their $q$-deformations, to the algebra and geometry of the moduli space of vacua of four dimensional ${\CalN}=2$ supersymmetric gauge theories, and their various deformations, such as $\Omega$-deformation, lift to higher dimensions, inclusion of extended objects and so on. In many respects the BPS/CFT correspondence is an analogue of the mirror symmetry, relating the count of (pseudo)holomorphic curves in symplectic manifolds to the periods of mirror complex manifolds. Here the analogue of curve counting is the enumeration of instantons in four dimensional gauge theory, while the role of the mirror complex geometry is played by the two dimensional conformal field theory. Indeed, thanks to the holomorphic factorization, the CFT calculations, especially in a semi-classical limit, quite often becomes a problem in complex geometry. 
\paragraph{}

Algebraic geometers consider the curve counting problems difficult, therefore the mirror map is a welcome simplification. 
With the higher genus corrections in place both sides become complicated. Sometimes additional dualities are available, mapping the problem of counting curves or quantizing the variations of Hodge structure to the problems of counting ideal sheaves or generalized gauge instantons. 
\paragraph{}

\subsection*{From gauge theory to a spin chain}

In this paper we explore a specific corner of the BPS/CFT correspondence, where the techniques developed in the four dimensional instanton counting are applied to a seemingly very distant problem: calculating a quantum mechanical wave-function of 
a many-body system, or a spin chain. 

\paragraph{}
The 
${\CalN}=2$ gauge theories in four dimensions have an intrinsic connection \cite{Gorsky:1994dj,DW1} to 
algebraic integrable systems \cite{hitchin1987}, usually called the Seiberg-Witten integrable systems after
\cite{Seiberg:1994aj,Seiberg:1994rs}. The gauge theory of our interest, the asymptotically $\CalN=2$ superconformal  SQCD in four dimension, reveals the structure which has dual (bi-spectral) descriptions. On the one hand, it is a complex generalization of the Heisenberg spin chain, based on the Lie algebra
$\mathfrak{sl}_{2}$, on the other hand it is a special case of Gaudin model (which is, in turn \cite{Nekrasov:1995nq}, a special case of the Hitchin system \cite{hitchin1987}), based on the $\mathfrak{sl}_{N}$ Lie algebra.  

The first hints of this correspondence, for the asymptotically free theories, were observed in ~\cite{Gorsky:1994dj,Martinec:1995by}, at the classical level. Much later, thanks to the development of localization methods in supersymmetric gauge theories \cite{Nekrasov:2002qd}, this correspondence was extended to the relation between the quantized integrable systems~\cite{Nikita-Shatashvili, Nekrasov:2011bc}, and $\Omega$-deformed supersymmetric gauge theories, including some of the asymptotically conformal theories.

For almost twenty years now the non-perturbative aspects of the $\CalN=2$ supersymmetric gauge theories can be extracted from the exact computations in the theory subject to the $\Omega$-background on $\BR^4$ (or a product of two cigars as in \cite{Nekrasov:2011bc}) with two parameters $\epsilon_1$ and $\epsilon_2$. The partition function $\CalZ$ and certain BPS observables can be computed exactly by localization for a large class of $\CalN=2$ supersymmetric gauge theories \cite{Nikita:I}. The limit ${\epsilon}_{1}, {\epsilon}_{2} \to 0$ reveals the classical integrable system whose phase space, incidentally, can be identified with the moduli space of solutions of some partial differential equations of gauge theoretic origin \cite{Nekrasov:2012xe}. In the NS limit ${\epsilon}_{2} \to 0$, ${\epsilon}_{1} = {\hbar}$ one expects to find the quantum version of that integrable system \cite{Nikita-Pestun-Shatashvili}. 
 
The quantization program has three aspects: the deformation of the commutative algebra of observables to the noncommutative associative algebra, with a big enough commutative subalgebra in the integrable case, the construction of the representation of the algebra of observables in the space of states, and, to make contact with the physical predictions of probabilities, endowing the space of states with the Hilbert space structure. 
The first step can be, in principle, analyzed with the help of two dimensional topological sigma model \cite{Kontsevich:1997vb} called the Poisson sigma model \cite{Cattaneo:1999fm}. However the second and the third steps do not seem natural in this approach. In the topological A model, using the so-called cc branes of \cite{Kapustin:2001ij} and the more familiar Lagrangian branes, one can, at least under some additional assumptions, produce both the algebra and its representation. 

One is naturally led to the question of computing the wavefunctions, in some specific representation, of quantum integrable systems, of the stationary eigenstates, i.e. the common eigenvectors of the quantum integrals of motion. This is where the four dimensional supersymmetric gauge theory, as opposed to the two dimensional sigma model, seems to give an advantage. First of all, the cc branes lift to the pure geometry (at the tip of the cigar). The Lagrangian branes can be interpreted as the boundary conditions at infinity on the first cigar. The stationary states of the quantum integrable system, under the Bethe/gauge correspondence \cite{Nekrasov:2009uh, Nekrasov:2009ui}, are the vacua of the effective two dimensional
${\CalN}=(2,2)$ gauge theory. In order to get the wavefunction of the stationary state we compute the expectation value
of the special local observable in this effective two dimensional theory -- the surface defect in the four dimensional theory. The parameters of the surface defect become the coordinates on which the wavefunction depends. As we will review in section \ref{sec:BPS/CFT}, introduction of a surface defect proves to be a very useful tool  when studying quantum version of the Bethe/gauge correspondence. The four dimensional theory with the co-dimension two surface defect can be viewed as a theory on an orbifold. The localization techniques generalize so as to compute the defect instanton partition function \cite{K-T} and also the expectation values of some observables. Our scope is the class of $qq$-character observables, which are fractionalized in the presence of the surface defect \cite{Nikita:V}. The main statement in \cite{Nikita:II} proves a certain vanishing theorem for the expectation values of the $qq$-characters, with or without surface defect inserted. These vanishing equations, called the \emph{non-perturbative Dyson-Schwinger equations}, can be used to derive the KZ-equations \cite{Knizhnik:1984nr} satisfied by the defect partition function \cite{NT}. Furthermore, in the NS-limit, these Dyson-Schwinger equations become the Schr{\"o}dinger-type equations  satisfied by the defect instanton partition function \cite{Nikita:V,Jeong:2017pai} (for pure ${\CalN}=2$ theory this has been observed to hold on purely algebraic grounds in \cite{Braverman:2004cr}).
The localization computation of the surface defect partition function therefore provides a systematic way of constructing both the spectrum and the eigenstate wavefunction for the corresponding quantum integrable model.

This story is a infinite-dimensional generalization of the correspondence between the strictly two dimensional ${\CalN}=2$ theories and finite dimensional quantum systems, where 
the Bethe Ansatz Equations of the quantum integrable system can be recovered from either the saddle point equation of the corresponding supersymmetric gauge theory instanton partition function \cite{Nikita-Shatashvili,Chen:2019vvt,HYC:2011}, or from the properties of the $qq$-characters \cite{Nikita:I,Nikita:V}. 
\paragraph{}

\subsection*{Classical and quantum integrability}

According to \cite{Nikita-Shatashvili} the algebraic integrable system governing the special K{\"a}hler geometry of the vectormultiplet moduli space of the four dimensional theory is deformation quantized, the Planck constant $\hbar$ being the $\Omega$-deformation parameter ${\epsilon}_{1}$. 
The quantum system remains integrable, with the spectrum of the commutative subalgebra of the algebra of observables being the twisted chiral ring of the effective two dimensional theory. 

Now, the subtle point, which is best understood by relating the four dimensional gauge theory to the two dimensional sigma model as in \cite{Nekrasov:2011bc}, 
is that the ``spectrum'' of the previous sentence, is understood in the algebraic geometry sense. It becomes the physical spectrum, typically isolated once the additional data such as the choice of supersymmetric boundary conditions at infinity, is made. 

In this paper we shall not pursue this line. Instead, we shall study the analogue of the continuous spectrum problem, 
the construction of the scattering states wavefunctions (sometimes called the Jost functions). 
The gauge theory analogue of this problem is the following. Suppose we fix the vacuum with the special coordinates ${\ba} = ( a_{1}, \ldots , a_{N}) $ on the Coulomb branch (these determine the masses of the $W$-bosons, say). In this vacuum we compute the expectation values of the gauge invariant observables built out of the vector
multiplet scalars
\beq
{\CalO}_{k} ({\ba}) = \langle {\Tr} {\sigma}^{k} \rangle_{\ba}
\label{eq:cok}
\eeq
Using the Bethe/gauge dictionary, this expectation value is identified, as in ~\cite{Nikita-Shatashvili,Dorey:2011pa,HYC:2011}, with the eigenvalues $H_{k}(a)$ of the commuting 
Hamiltonians ${\hat H}_{k}$, $k = 1, 2, \ldots $ :
\beq
{\hat H}_{k} {\Psi}_{\ba} = H_{k}({\ba}) {\Psi}_{\ba}
\eeq
where ${\Psi}_{a}$ is the wavefunction of the state characterized by the spectral parameters we identify with $\ba$. The expectation values \eqref{eq:cok} receive contributions from the all instanton sectors. If we ignore all the instanton contributions, then the expectation values \eqref{eq:cok} are given by the classical expressions
\beq
{\CalO}_{k} ({\ba})^{\rm pert} = \sum_{i=1}^{N} a_{i}^{k}
\eeq
which are the eigenvalues of the free Hamiltonians, e.g. $\sum_{i=1}^{N} ( - {\ii}{\pa}_{\rx_{i}} )^{k}$ acting on the plane wave function 
\beq
{\Psi}_{\ba}^{\rm free} \sim e^{{\ii} \sum a_{i}\rx_{i}}
\eeq
With the instanton contributions included this function is dressed up into the scattering state wavefunction we are after, while the eigenvalues 
\eqref{eq:cok} become the complicated functions of $\ba$, the masses, the $\Omega$-deformation parameter, and the gauge coupling $\kq$. 
These contributions can be studied using the nonperturbative Dyson-Schwinger equations, which can be conveniently organized with the help of the $qq$-character observables~\cite{Nikita:I}. 
\paragraph{}

\subsection*{Surface defect and the wavefunction}

The main question is the choice of the polarization in which one to represent the wavefunction in question. Fortunately, here as well 
the gauge theory provides a candidate. Generalizing the disorder operators of the Ising model and the 't Hooft and Wilson loops of the conventional gauge theory, one introduces a codimension two defect operators ${\CalS}_{\Sigma, c}$, which are the instruction to perform the path integral over the singular gauge fields, having the
nontrivial holonomy around the small loops linking a codimension two surface $\Sigma$ in spacetime. The conjugacy class $c$ of the holonomy is fixed
throughout $\Sigma$ while the representative varies. Let $G$ denote the gauge group and let $G_{c}$ be the stabilizer of the conjugacy class $c$. Then the singularity at the defect is classified
by the set of equivalence classes $\left[  {\rm Maps} \left( {\Sigma}, G/G_{c} \right) \right]$. We can therefore  generally write
\beq
\langle {\CalS}_{\Sigma, c} {\ldots} \rangle_{\ba}  = \sum_{{\bf d} \in \left[  {\rm Maps} \left( {\Sigma}, G/G_{c} \right) \right]} e^{{\bf d} \cdot {\bf x}} 
\langle {\ldots} \rangle_{{\rm singular}\ {\rm gauge}\ {\rm fields}\ {\rm in\ the\ homotopy\ class\ of}\ {\bf d}, \ {\rm in\ the\ vacuum}\ {\ba}}
\eeq
We identify the wavefunction ${\Psi}_{\ba}({\bf x})$ with the normalized vev of ${\CalS}_{\sigma, c}$. Our main method is the supersymmetric
localization allowing to compute the unnormalized surface defect partition function ${\CalZ}^{\rm defect}$ in the four dimensional $\Omega$-background, with two parameters
${\epsilon}_{1}, {\epsilon}_{2}$, from which we extract, in a nontrivial manner sketched below, the wavefunction in question:
\beq
{\Psi}_{\ba}({\bx}; {\bm}, {\hbar}) = \lim_{{\epsilon}_{2} \to 0}\quad \frac{{\CalZ}^{\rm defect} ({\ba}, {\epsilon}_{1}, {\epsilon}_{2}, {\bm}, {\bx}, {\kq})}{{\CalZ}^{\rm bulk}({\ba}, {\epsilon}_{1}, {\epsilon}_{2}, {\bm}, {\kq})}
\eeq
with $\hbar = {\epsilon}_{1}$ being the Planck constant, and $\bm$ entering the quantum integrable system in an interesting way we describe below. 

\subsection*{One flew over the limit shape}

The limits of vanishing $\Omega$-deformation parameters are the main applications of the localization techniques. In the limit ${\epsilon}_{1}, {\epsilon}_{2} \to 0$
the $\Omega$-background approaches the flat space limit, where the supersymmetric gauge theory regains the full ${\CalN}=2$ supersymmetry. The $F$-terms
of the low-energy effective theory are recovered from the small $\epsilon$-expansion. 
The finite $\epsilon$ computation is often doable, reducing the complicated gauge theory path integral to a sum over an infinite yet
finite at each instanton order set. The set is ${\CalP}^{N}$, with ${\CalP}$ being the set of all partitions, or Young diagrams. 

The limit ${\epsilon} \to 0$, with appropriate choices for the parameters, such as the Coulomb moduli $\ba$, the masses $\bm$
etc. can be analyzed, by observing that one term in this infinite sum dominates, the limit shape phenomenon of Vershik-Kerov-Logan-Schepp. 
In particular, in \cite{NO1, Nekrasov:2012xe} the limit shape determining the prepotential ${\CalF}$ of the low-energy effective theory for a large class of theories was found. It is found in the limit ${\epsilon}_{1}, {\epsilon}_{2} \to 0$ from the asymptotics of the gauge theory supersymmetric partition function, which in this paper
we call the bulk partition function:
\beq
{\CalZ}^{\rm bulk} ({\ba}, {\epsilon}_{1}, {\epsilon}_{2}, {\bm}, {\kq}) \sim e^{\frac{{\CalF}({\ba}, {\bm}, {\kq})}{{\epsilon}_{1}{\epsilon}_{2}} +\ldots } \, , \qquad
{\epsilon}_{1}, {\epsilon}_{2} \to 0.
\label{eq:prep}
\eeq
The bulk partition function is invariant under the exchange ${\epsilon}_{1} \leftrightarrow {\epsilon}_{2}$. 

The choices mentioned in the previous paragraph are then dealt with by the use of analyticity of ${\CalZ}^{\rm bulk}$ which is a consequence of 
supersymmetry. The asymptotics \eqref{eq:prep} assumes the $({\ba}, {\bm}, {\kq})$ parameters are generic. If, however, the parameters 
are fine tuned to some special values, the asymptotics \eqref{eq:prep} gets much more interesting and complicated, reflecting the subtleties of the low-energy effective theory. 

In \cite{Nikita-Shatashvili,Nikita-Pestun-Shatashvili} this analysis is extended to the more complicated limit ${\epsilon}_{2} \to 0$, with
${\epsilon}_{1} = {\hbar}$ kept finite. In this case one obtains the effective twisted superpotential ${\CalW}$ of the effectively two dimensional ${\CalN}=(2,2)$
theory corresponding to the four dimensional theory subject to the two dimensional $\Omega$-background:
\beq
{\CalZ}^{\rm bulk} ({\ba}, {\epsilon}_{1}= {\hbar}, {\epsilon}_{2}, {\bm}, {\kq}) \sim e^{\frac{{\CalW}({\ba}, {\hbar}, {\bm}, {\kq})}{{\epsilon}_{2}} +\cdots }
\, , \qquad
{\epsilon}_{2} \to 0
\label{eq:super}
\eeq
As explained in \cite{Nekrasov:2002qd, NO1} the exponential asymptotics \eqref{eq:prep}, \eqref{eq:super}
can be interpreted as the fact that the supersymmetric partition function has the extensive behavior of the typical thermodynamic partition function, with $\frac{1}{{\epsilon}_{1}{\epsilon}_{2}}$ playing the role of the four dimensional volume and $\frac{1}{{\epsilon}_{2}}$ playing the role
of the two dimensional area. The area and the volume entering here are the measures of the space occupied by the instantons. 

Now, in the presence of the surface defect, the supersymmetric partition function gets modified to
$$
{\CalZ}^{\rm defect}({\ba}, {\epsilon}_{1}, {\epsilon}_{2}, {\bm}, {\bx}, {\kq})\ . 
$$
Again, the localization makes it a sum over a countable set. Actually the set is the same ${\CalP}^{N}$, but the sum is different. 

Assuming the defect is localized in the plane affected by the $\epsilon_2$-part of the $\Omega$-deformation the small $\epsilon_2$ asymptotics is not, at the leading order, modified, as the bulk instantons don't feel much of the defect:
\beq
{\CalZ}^{\rm defect} ({\ba}, {\epsilon}_{1}= {\hbar}, {\epsilon}_{2}, {\bm}, {\bx}, {\kq}) \sim e^{\frac{{\CalW}({\ba}, {\hbar}, {\bm}, {\kq})}{{\epsilon}_{2}}} \left( {\Psi}_{\ba}( {\bx} ; {\bm}, {\hbar}) + \ldots \right)
\, , \qquad
{\epsilon}_{2} \to 0
\label{eq:superdefect}
\eeq
In \cite{Nikita:IV, Nikita:V} an ${\infty}:1$ map ${\pi}_{N}: {\CalP}^{N} \longrightarrow {\CalP}^{N}$ is constructed, which represents
the map between the moduli space of instantons in the presence of the surface defect to the moduli space of instantons in the bulk (the map is a finite ramified cover in a fixed instanton sector). The sum giving the left hand side of \eqref{eq:superdefect} can be reorganized as the sum over the image of ${\pi}_{N}$ of the sums over the fibers. The former, in the ${\epsilon}_{2} \to 0$ limit, is dominated by one term, the limit shape of the bulk theory. The latter remains to be evaluated. This is the main objective of this paper. 

\subsection*{The sum-cracking secret}

Here is the strategy we employ. We first recall, that the sum the localization reduces the supersymmetric partition function to can and originally was represented as a series of countour integrals. Remarkably, the remaining sum we are to evaluate can also be represented as a series of contour integrals, which can be further intepreted as the series of integrals of the cohomological field theory type
over a sequence of moduli spaces of solutions to matrix equations, defined in a way, similar to the folded instanton constructions of \cite{Nikita:III}. These equations depend on some real parameters, the Fayet-Illiopoulos terms ${\zeta}_{\BR}$. The integrals over the moduli spaces do not change under the small variations of $\zeta_{\BR}$'s, however they may and do jump, as $\zeta_{\BR}$'s cross the walls of stability where the corresponding moduli space becomes singular.

{}The simplest example of such crossing is the moduli space of solutions to the equation $\sum_{i=1}^{N} |z_{i} |^{2} = {\zeta}_{\BR}$, 
with complex numbers $(z_{1}, \ldots , z_{N})$. If one divides by the symmetry $(z_{i}) \mapsto (e^{{\ii}{\theta}} z_{i})$, then, for
${\zeta}_{\BR} >0$ one gets the complex projective space ${\BC\BP}^{N-1}$ as the moduli space, with interesting topology captured by the integrals akin to the ones we study in this paper. For ${\zeta}_{\BR} < 0$ the moduli space is empty so all the reasonable integrals vanish on the occasion. 

The significance of the wall-crossing becomes obvious at the second step of our approach.  We move the contour of integration, letting it circle around the infinity and wrap around the set of poles one is ignoring in taking the integral over the original contour by residues. Remarkably, the residues at infinity can be summed up. 
The sum of the residues at other poles can be interpreted as integrals over the moduli spaces of the same folded instanton
equations but with the different sign of the $\zeta_{\BR}$ parameters. The moduli spaces in that case are non-trivial yet simpler, at least
at the level of the fixed points of the global symmetry group, to which the integrals localize. Notice, that in variance with \cite{Gorsky:2017hro}, we do not modify the original theory. We merely compute the original path integral by the contour manipulation.

{}To be specific, we shall be working with the four dimensional ${\CalN}=2$ supersymmetric gauge theory with the $SU(N)$ gauge group, and $N_f = 2N$ hypermultiplets in the fundamental $N$-dimensional representation. The number of the matter multiplets is precisely such that the theory is superconformal at high energy, as such it is characterized by the ultraviolet gauge coupling $g$, and the theta angle $\theta$. The latter
is a parameter since the axial anomaly is cancelled for $N_f = 2N$ as well.  It is convenient to combine $g$ and $\theta$ into the complex parameters $\tau$ and $\kq$:
\begin{align}
    \tau=\frac{\theta}{2\pi}+\frac{4\pi \ii}{g^2}, \quad \kq = \exp 2\pi \ii \tau,
\end{align}
The masses ${\bf m} = \left( m_{f}^\pm \right)_{f=1}^{N}$ of the fundamental hypermultiplets (the splitting to $+$ and $-$ masses will be commented on in the main text) are complex as well. It is useful to think of the masses as of the scalars in the vector multiplet of the global symmetry $U(N_f)$.

In this way we arrive at the main result of this paper: the formula for the wavefunction. Specifically, in the section \ref{sec:surfaceIPF} we demonstrate that the normalized vev of the surface defect partition function of $\CalN=2$ SQCD can be written in terms of $\frac{N(N-1)}{2}$ Mellin-Barnes-type contour integrals.  In the limit to the asymptotically free pure ${\CalN}=2$ theory our formula  becomes that of the periodic Toda lattice wavefunction \cite{Kharchev:2000ug, Kharchev:2000yj}.

As a by-product, and also as a warm-up,  we discuss the similar contour manipulation applied to the bulk partition function. 
For the pure $\CalN=2$ SYM, or the $\CalN=2$ SQCD with $N_f<2N-1$ flavors, and $\CalN=2^{*}$ theories with gauge group $U(N)$, the instanton partition function does not depend on the sign of the FI-parameter ${\zeta}_{\BR}$ entering the deformed ADHM equations \cite{Nakajima:1994nid} (the $B$-field in string theory realization \cite{Nekrasov:1998ss} of noncommutative instantons used in the localization approach). However, the supersymmetric gauge system we study does exhibit the $\zeta_{\BR}$-dependence. 
As we will discuss in detail in \ref{sec:surfaceIPF}, the change in the integrals over the moduli spaces corresponding to different
$\zeta_{\BR}$'s can be organized into an elegant crossing formula, confirming the $U(1)$-factors in the AGT-conjecture \cite{Alday:2009aq}. 

Furthermore, we find that in the chain-saw and hand-saw quiver extensions \cite{Nakajima:2011yq}, the instanton counting parameters of each quiver node are related in a non-trivial manner with different stability conditions. 
This leads to the transformations of the coordinates $\bx$ of the integrable system, looking vaguely similar to the cluster structures in \cite{goncharov2011dimers, Fock:2015, Marshakov:2019vnz}.

\subsection*{More on $\spch$ spin chain/SQCD correspondence}

Bethe/gauge correspondence identifies the quantum integrals of motion of some quantum integrable system with the elements of the twisted chiral ring of some
gauge theory with the ${\CalN}=(2,2)$, $d=2$ supersymmetry. Among such theories we find the four dimensional ${\CalN}=2$ supersymmetric theories subject to the
two dimensional $\Omega$-deformation. The limit ${\epsilon}_{1} = {\hbar} \to 0$ restores the four dimensional super-Poincare invariance while being the classical limit of the quantum system. In section \ref{sec:XXX/SQCD c} we relate the Darboux coordinates, which are natural
in the spin chain realization of the Seiberg-Witten integrable system describing the $N_f = 2N$ SQCD, to the parameters of the surface defect and the bulk theory. 
{}
In this limit our Mellin-Barnes-type integrals can be evaluated by the saddle point approximation. The latter can also be used to classify the possible contours of integration. We find the
the saddle point equations of the surface defect partition function  look like the nested Bethe equations, which can be solved in terms of the holonomy matrix of the classical limit of the $\spch$-spin chain. In this way we 
recover Sklyanin's separated variables \cite{Sklyanin:1992eu,sklyanin1995separation}.

We then extend the $\spch$/SQCD correspondence to the quantum level in the section \ref{sec:XXX/SQCD q}. Using the nonperturbative Dyson-Schwinger equations we are able to generate infinitely many bulk gauge invariant chiral ring observables, whose vacuum expectation values are the eigenvalues of the mutually commuting differential operators (Hamiltonians) acting on the surface defect partition functions, which are the higher quantum integrals of motion of the $\spch$ spin chain. We present the explicit calculation of the first three Hamiltonians. 

Along the way, we find that the inclusion of all the $qq$-characters, 
not only the fundamental ones \cite{Nikita:I}, is needed to recover all the indepedent Hamiltonians.

We conclude in the Section~\ref{sec:discussion}.
Various definitions and some of the computational details are given a series of Appendices.

\subsection*{Duality of correspondences, and correspondences of dualities: $\mathfrak{sl}_{N}$ Gaudin vs $\spch$-chain}

{}Quite often the Poisson-commuting Hamiltonains of the classical integrable system can be organized into a algebraic equation $R(x,z) = 0$ describing an algebraic curve. The values of the Hamiltonians are the parameters of the curve. Sometimes this algebraic equation is the characteristic polynomial of an operator ${\Phi}(z)$
depending on the additional parameter $z$, 
\beq
R(x, z) = {\rm Det}\left(x - {\Phi}(z)\right)
\label{eq:charpol}
\eeq
The gauge theory counterpart of the values of the Poisson-commuting Hamiltonians, has been observed in several cases to be the spectrum of the chiral ring, e.g. in the  $\CalN=2$ SQCD ~\cite{Gorsky:1995zq,Martinec:1995by,Seiberg:1996nz,Gorsky:1996hs}, and shown more generally in \cite{DW1}. 
When the four dimensional ${\CalN}=2$ theory is $\Omega$-deformed in two dimensions, the theory retains ${\CalN}=(2,2)$ two dimensional super-Poincare invariance, with the translational symmetry in two dimensions unaffected by the $\Omega$-background. Remarkably, the equation $R(x,z) = 0$ may have several interpretations like \eqref{eq:charpol}. This is related to the phenomena of dualities in integrable systems \cite{Hans:1986, Bispec, Fock:1999ae}, and bi-spectrality. This includes the Nahm duality between the integrable system on the moduli space of periodic monopoles and Gaudin model \cite{Cherkis:2000cj}, whose relation to the four dimensional gauge theory is demonstrated in \cite{Nekrasov:2012xe}, for all $A$-type quiver gauge theories. 
The same duality, in the $A_1$-case, with an excursion into the quantum realm, is  discussed recently in \cite{mironov2013spectral}. 

Let us explain this duality in the classical case. 
Consider the following version of the Hitchin system. Let ${\Phi}_{\xi}$, ${\xi} \in \{ 0, {\kq}, 1, {\infty} \}$ be the $N\times N$ traceless complex
matrices, with fixed eigenvalues, which we assume to be distinct for ${\xi} = 0, \infty$, and maximally degenerate yet non-trivial (i.e. with multiplicity 
$(N-1, 1)$) for $\xi = {\kq}, 1$. Define:
\beq
{\Phi}(z) = \sum_{\xi} \frac{{\Phi}_{\xi}}{z-{\xi}}
\label{eq:laxope}
\eeq
Let us require ${\Phi}(z)$ to be holomorphic outside $\{ 0, {\kq}, 1, {\infty} \}$, which means 
\beq
\sum_{\xi} {\Phi}_{\xi} = 0
\label{eq:sumres}
\eeq
and divide by the group $G_{\BC} = SL (N, {\BC})$ acting by the simultaneous conjugation 
\beq
\left( {\Phi}_{\xi} \right)_{0,{\kq},1,{\infty}} \mapsto \left( g^{-1} {\Phi}_{\xi} g  \right)_{0,{\kq},1,{\infty}}\ \text{for } g \in G_{\BC} \ .
\label{eq:groupdiv}
\eeq
The space of solutions to \eqref{eq:sumres} modulo \eqref{eq:groupdiv} is the phase space $\CalM$ of Gaudin model, 
\beq
{\CalM} = \left( {\CalO}_{0} \times {\CalO}_{\kq} \times {\CalO}_{1} \times {\CalO}_{\infty} \right)//G_{\BC}
\eeq
the latter notation suggesting $\CalM$ is a symplectic manifold. The symplectic structure ${\omega}_{\CalM}$ can be described in terms of
the Poisson brackets of functions of the matrix elements of the residues ${\Phi}_{\xi}$, 
\beq
\{ \left({\Phi}_{\xi}\right)_{i}^{j}, \left({\Phi}_{\xi'}\right)_{i'}^{j'} \} = {\delta}_{\xi, \xi'} \left( 
{\delta}_{i'}^{j}  \left({\Phi}_{\xi}\right)_{i}^{j'} - {\delta}_{i}^{j'} \left({\Phi}_{\xi}\right)_{i'}^{j} \right)
\label{eq:poissbr}
\eeq
It follows from \eqref{eq:poissbr} that the coefficients of the characteristic polynomial
\beq
R(x,z) : = {\rm Det}_{N} \left( x - {\Phi}(z) \right) = \sum_{i=0}^{N} x^{N-i} u_{i}(z)\, , \qquad u_{1}(z) \equiv 0 
\eeq
Poisson-commute for any value of $z$. Furthermore, the functions $u_{i}(z)$ only have poles at $z = 0, {\kq}, 1, {\infty}$, with the leading asymptotics determined by the fixed eigenvalues of the residues. It can be shown by the straightforward algebraic analysis that the number of independent parameters in $u_i(z)$ is equal to $N^2 -N+2(N-1)-N^2+1= N-1$, which is half the expected dimension of $\CalM$, meaning we have a complex integrable system. Moreover, one can recover a point on $\CalM$ given the curve $R(x,z) = 0$ and a point on its Jacobian, i.e. a holomorphic line bundle. This bundle is identified with the eigenline of ${\Phi}(z)$ corresponding to the eigenvalue $x$. With proper adjustments, all of the Jacobian, i.e. the complete abelian variety, is the fiber of the projection ${\CalM} \to {\BC}^{N-1}$
given by fixing the spectral curve $R(x,z) =0$, belongs to $\CalM$. This makes $\CalM$ an algebraic integrable system in the sense of \cite{hitchin1987}. The periods of the differential $xdz$ provide the action variables (there are many choices for the $N-1$ cycles on the spectral curve, leading to the special geometry and the prepotential). 

{}Another representation of the same algebraic integrable system is obtained by the Nahm transform. Namely, consider the moduli space of
solutions to the complex part of the $SU(2)$ Bogomolny equations:
\beq
{\ii} \, D_{\xb} {\tilde\Phi} \, + \,  F_{\xb u} = 0 
\label{eq:bogeq}
\eeq
where $u \sim u +1$ is a coordinate on $S^{1}$, $x, {\xb}$ are the coordinates on ${\BR}^{2}$, $F_{\mu\nu} = {\pa}_{\mu} A_{\nu} - {\pa}_{\nu} A_{\mu} + [A_{\mu}, A_{\nu}]$, and ${\tilde\Phi}, A_{\mu}$
are the adjoint-valued Higgs field and gauge field on $S^{1} \times {\BR}^{2}$, respectively. The Eqs. \eqref{eq:bogeq} imply that the 
spectrum of the complexified $SL(2, {\BC})$-valued holonomy
\beq
g(x, {\xb}) : = P{\rm exp}\oint \left( A_{u} + {\ii} {\tilde\Phi} \right) du
\eeq
varies holomorphically with $x$:
\beq
{\bar\pa}_{\xb} \, {\tilde R}(x, {\tilde z}) = 0 \ , \qquad {\tilde R}(x, {\tilde z}) = {\rm Det}_{2} \left( {\tilde z} - g(x, {\xb}) \right) \ .
\eeq
If we impose, in addition, the condition that at $x \to \infty$ the conjugacy class of $g(x, {\xb})$ approaches that of ${\rm diag}({\kq}^{\half}, {\kq}^{-\half})$, while at $x = {\mu}_{1}, \ldots , {\mu}_{N}$ there are singularities which can be modelled on the $U(1)$ Dirac singular monopoles embedded into the $SU(2)$ gauge fields, then 
\beq
{\kq} P_{+}(x) {\tilde R}(x, {\tilde z}) = (z-1)(z-{\kq}) R(x, z)
\label{eq:moncurve}
\eeq
with 
\beq
z = {\tilde z} \sqrt{{\kq} \frac{P_{+}(x)}{P_{-}(x)}} \ .
\eeq
Thus, the monopole spectral curve and the Hitchin-Gaudin spectral curves essentially coincide. The precise map between ${\Phi}$ and ${\tilde \Phi}, A$ data is obtained analogously to the usual Nahm transform \cite{Corrigan:1983sv}. 
By writing
\beq
\frac{(z-1)(z-{\kq})}{z} R({\tilde x},z) = z P_{-}(x) - (1+{\kq}) T(x) + {\kq} z^{-1}P_{+}(x)  
\eeq
with 
\beq
{\tilde x} = \frac{x}{z} - \sum_{\xi = 0, {\kq}, 1, {\infty}} \frac{{\rm m}_{\xi}}{z-{\xi}}
\eeq
with ${\rm m}_{\kq}, {\rm m}_{1}$ equal to multiplicity $N-1$ eigenvalues of ${\Phi}_{\kq}, {\Phi}_{1}$, respectively, one deduces \cite{Nekrasov:2012xe} that the monopole spectral curve data becomes the data of the algebraic integrable system associated with the $\spch$ spin chain with the complex spins $s_{f} = \frac{m_{f}^{+} - m_{f}^{-}}{\hbar}$, and the inhomogeneities ${\mu}_{f} = {\half} \left( m_{f}^{+} + m_{f}^{-} \right)$. 

The $\spch$ spin chain side of the story is addressed in this paper. The Hitchin-Gaudin representation is obtained from the ${\epsilon}_{2} \to 0$ limit of the Knizhnik-Zamolodchikov equation derived in the companion paper \cite{NT}. In this way we obtain a generalization of the results of 
\cite{Feigin:1994in}, which can be recovered for special values of masses and Coulomb parameters.

\subsection*{Acknowledgements} 
Research of NL is supported by the Simons Center for Geometry and Physics at Stony Brook University. NN thanks S.~Jeong, A.~Okounkov and O.~Tsymbalyuk for discussions. We thank D.~Gaiotto and M.~Dedushenko for informing us about their upcoming work on the relation of the $XXX$ spin chains to defects in ${\CalN}=4$ theory. 
We also thank M.~Dedushenko and T.~Kimura for their critical reading of our draft and for their comments. 

\section{The surface defect} \label{sec:BPS/CFT}

In this section we briefly recall the construction of the surface defect and study its vacuum expectation value.

\subsection{From gauge theory to a statistical model}

Localization technique reduces generally complicated supersymmetric gauge theory path integral into computation of an effective  statistical model, capturing the correlation functions of the BPS protected operators.

Let us consider the $\CalN=2$ $A_1$-quiver gauge theory in 4 dimensions, with the gauge group $SU(N)$ and $2N$ fundamental hypermultiplets. The Lagrangian is parametrized by the complexified gauge coupling 
\begin{align}
    \tau=\frac{\theta}{2\pi}+\frac{4\pi \ri}{g^2}, \quad \kq = \exp 2\pi \ri \tau ,
\end{align}
and by the choice $\bm$ of $2N$ masses, which we split into $N$  \emph{fundamental} ${\bm}^{+}=(m_{1}^{+},\dots,m_{N}^{+})$ and $N$ \emph{anti-fundamental} ${\bm}^{-}=(m_{1}^{-},\dots,m_{N}^{-})$ ones. The choice of the vacuum is parametrized by the $N$ Coulomb moduli parameters ${\ba}=(a_1,\dots,a_N)$, obeying 
\beq
\sum\limits_{{\alpha}=1}^{N} a_{\alpha} = 0 \ .
\eeq

{}The localization of the $\Omega$-deformed theory \cite{Nekrasov:2003rj,Nikita:I} produces the the statistical model whose configurations space is ${\CalP}^{N}$,  the set of all $N$-tuples of Young diagrams $\vec{\lambda} = (\lambda^{(1)},\dots,\lambda^{(N)})$. In turn,  each individual Young diagram $\lambda^{(\alpha)}$, $\alpha = 1, \ldots, N$, is a collection  $\lambda^{(\alpha)}=(\lambda^{(\alpha)}_1,\lambda^{(\alpha)}_2,\dots)$ of nonnegative numbers obeying
\begin{equation}
	\lambda_i^{(\alpha)}\geq\lambda^{(\alpha)}_{i+1}, \quad i = 1, 2, \dots.
\end{equation}
which can be represented geometrically as Young diagram, where each number ${\lambda}^{({\alpha})}_{i}$ is the $i$-th
row of that many identical squares $\square$, as in the Fig. \ref{fig:Young}.

The pseudo-measure associated to the instanton configuration $\vec{\lambda}$ is defined using the \emph{plethystic exponent} $\bE$ operator, which converts the additive Chern characters to the multiplicative classes
\begin{align}
    \mathbb{E}\left[\sum_{a} {\mathtt{m}}_a e^{{\xi}_a}\right] = \prod_a {\xi}_a^{-{\mathtt{m}}_a}
\end{align}
where ${\mathtt{m}}_{a} \in {\BZ}$ is the multiplicity of the Chern root $\xi_a$.
For $\vec{\lambda}$ the associated pseudo-measure is computed by:
\begin{align}
    \CalZ({\ba},{\bm}^\pm,\vec{\epsilon})[\vec{\lambda}]= \mathbb{E}\left[-\frac{\hat{S}\hat{S}^*}{P_{12}^*} + \frac{\hat{M}\hat{S}^*}{P_{12}^*} \right]
    \label{def:instparti}
\end{align}
where
\begin{align}
    \hat{N} = \sum_{\alpha=1}^N e^{a_{\alpha}}, \quad
    \hat{K} = \sum_{\alpha=1}^N\sum_{(i,j)\in \lambda^{(\alpha)}} e^{a_{\alpha}}q_1^{i-1}q_2^{j-1}, \quad \hat{S} = \hat{N} - P_{12} \hat{K}, \quad 
    \hat{M} = \sum_{f=1}^N e^{m_f^+} + e^{m_f^-}. 
\end{align}
$q_i=e^{\epsilon_i}$ are the exponentiated complex $\Omega$-deformation parameters $\epsilon_1, \epsilon_2 \in \BC$ \cite{Nekrasov:2002qd,Nekrasov:2003rj,Pestun:2016zxk}, and
\begin{align}
    P_i = 1 - q_i, \quad P_{12} = (1-q_1)(1-q_2).
\end{align}
Given a virtual character $\hat{X} = \sum_{a} {\mathtt{m}}_{a} e^{{\xi}_{a}}$ we denote by  $\hat{X}^{*} = \sum_{a} {\mathtt{m}}_{a} e^{-{\xi}_{a}}$  the dual virtual character.

{}The localization equates the supersymmetric partition function of the $\Omega$-deformed $A_1$ $U(N)$ theory to the conventional partition function  of the grand canonical ensemble 
\begin{align}
    \CalZ( {\ba} , {\bm}^\pm ,\kq , \vec{\epsilon}) = \sum_{\vec{\lambda}} \kq^{|\vec{\lambda}|} \CalZ({\ba},{\bm}^\pm,\vec{\epsilon})[\vec{\lambda}].
\end{align}

\paragraph{}
A recent development in BPS/CFT correspondence notices differential equations of two dimensional conformal theories, such as KZ equations \cite{Knizhnik:1984nr} and KZB equations \cite{Bernard:1987df} can be verified by adding a regular co-dimension two surface defect in the supersymmetric gauge theory \cite{Nikita:V}. These conformal equations become eigenvalue equations of integrable models in Nekrasov-Shatashivilli limit (NS-limit for short) $\epsilon_2\to0$ \cite{Jeong:2018qpc,Chen:2019vvt}.   

The co-dimension two surface defect is introduce in the form of a $\mathbb{Z}_N$ type orbifolding  \cite{Nikita:IV} acting on $\mathbb{R}^4=\mathbb{C}_1\times\mathbb{C}_2$ by $(\bz_1,\bz_2)\to(\bz_1,\eta\bz_2)$ with $\eta^N=1$. 
The orbifolding generates chainsaw quiver structure \cite{Nakajima:2011yq,K-T}.
Such a surface defect is characterized by a coloring function $c:[N]\to\mathbb{Z}_N$ that assigns a representation $\CalR_{c(\alpha)}$ of $\BZ_N$ to each color $\alpha=1,\dots,N$. 

{}Here and below ${\CalR}_{\omega}$ denotes  the one-dimensional complex irreducible representation of ${\BZ}_{N}$, where the generator ${\eta}$ is represented by the multiplication by ${\exp}\, \frac{2\pi\ii \omega}{N}$. 

{}In the presence of fundamental matter, additional coloring functions $\sigma^\pm:[N]\to\mathbb{Z}_N$ assign a representation $\CalR_{\sigma^{\pm}(f)}$ to each fundamental flavor $m_f^{\pm}$, $f=1,\dots,N$.
In the simplest example, it is enough to assume that the coloring functions $c(\alpha)$ and $\sigma^\pm(f)$ take the form
$$
    c(\alpha)=\alpha-1, \ \alpha=1,\dots,N; \quad \sigma^\pm(f) = f-1, \ f=1,\dots,N.
$$
In principle, one may consider arbitrary degree orbifolding as the quotient by $\mathbb{Z}_{n}$ with any integer $n$. The defect corresponding to the $\mathbb{Z}_N$, represented in the color and in the both fundamental and anti-fundamental flavor spaces in a regular representation is called the \emph{full-type/regular surface defect}, which is relevant for our purpose. More detailed discussions can be found in~\cite{Feigin:2011SM,Finkelberg:2010JEMS,K-T,Nikita:IV}.
The complex instanton counting parameter $\mathfrak{q}$ fractionalizes to $N$ couplings $({\kq}_{\omega})_{\omega = 0}^{N-1}$:
\beq
\mathfrak{q}=\mathfrak{q}_0\mathfrak{q}_1\cdots\mathfrak{q}_{N-1};\quad \mathfrak{q}_{\omega+N}. 
\label{eq:fracq}
\eeq
The coupling $\mathfrak{q}_\omega$ is assigned to the representation $\CalR_\omega$ of $\mathbb{Z}_N$ as fugacity for the chainsaw quiver nodes.
The surface defect partition function is the path integral over the $\BZ_N$-invariant fields:
\begin{align}
    \CalZ^{\rm defect} ({\ba},{\bm}^\pm,\vec{\kq})  =\sum_{\vec{\lambda}}\prod_\omega \kq_\omega^{k_\omega}\mathbb{E}\left[-\left(\frac{\hat{S}\hat{S}^*-\hat{M}\hat{S}^*}{P_1^*(1-q_2^{-\frac{1}{N}}\CalR_{-1})}\right)^{\mathbb{Z}_N}\right]
\end{align}
with the power $k_{\omega}$ of fractional coupling $\kq_{\omega}$ defined in Eq.~\eqref{kv-data}.

The expectation value of the surface defect partition function $\CalZ$ in the Nekrasov-Shatashivilli limit (NS-limit for short) $\epsilon_2\to0$ has the asymptotics
\begin{align}\label{def:psi-asymptotic}
    \CalZ^{\rm defect}({\ba},{\bm}^\pm,\tau,\vec{\kq};\epsilon_1,\epsilon_2)=e^{\frac{1}{\epsilon_2}\CalW({\ba},{\bm}^\pm,\tau;\epsilon_1)} \cdot (\Psi_{\ba}(\bx,{\ba},{\bm}^\pm,\tau,\epsilon_1)+\CalO(\epsilon_2))
\end{align}
with the singular part being identical to that of the bulk partition function ${\CalZ}$,
\begin{align}
    \CalW({\ba},{\bm}^\pm,\tau;\epsilon_1)=\lim_{\epsilon_2\to0} \epsilon_2 \log \CalZ^{\rm bulk}\ .
\end{align}
We denote the \emph{normalized vev of the surface defect} by
\begin{align}
    \Psi_{\ba}(\bm^\pm,\vec{\kq},\epsilon = \hbar) = \lim_{\epsilon_2\to0} \frac{\CalZ^{\rm defect}(\ba,\epsilon_1,\epsilon_2,\bm^\pm,\vec{\kq},\kq)}{\CalZ^{\rm bulk}(\ba,\epsilon_1,\epsilon_2,\bm^\pm,\kq)}
\label{def:nexpectation valuesp}
\end{align}
Indeed, the exponential asymptotics is the thermodynamic large volume limit,  $1/{\epsilon}_{2}$ playing the role of the volume, the free energy being the effective twisted superpotential $\CalW({\ba},{\bm}^\pm,\tau;\epsilon_1)$ of that $\CalN=(2,2)$ two dimensional theory. The presence of the surface defect does not change the leading asymptotics, as it is an extensive quantity. 

The properties of partition function $\CalZ$ of $A_1$ quiver gauge theory along with the twisted superpotential $\CalW$ are well studied in various papers \cite{Nikita-Pestun-Shatashvili,Nikita-Shatashvili, Nekrasov:1995nq}, see also \cite{Nekrasov:1998ss,Nekrasov:2002qd,Nekrasov:2009uh,Nekrasov:2009ui}. 
In comparison the normalized vev of the surface defect partition function is much less explored and understood. However,as we will see in later chapters, the normalized vev of the surface defect will be identified as the wavefonction of the
scattering states in the dual quantum integrable model. 

\subsection{Shifted moduli}

For convenience of further computation, we scale $\epsilon_2\to\frac{\epsilon_2}{N}$ and define shifted moduli parameters by 
\begin{align}\label{shifted moduli}
{a_\alpha}-\frac{c(\alpha)}{N}\epsilon_2 = {\tilde{a}_\alpha};\quad {m_f^\pm}-{\frac{\sigma^\pm(f)}{N}}\epsilon_2 = {\tilde{m}_f^\pm}.
\end{align}
The shifted moduli parameters $\{\tilde{a}_\alpha\}$ and fundamental matter masses $\{\tilde{m}_f^\pm\}$ are charged neutral under the orbifolding. All the ADHM characters can be expressed in terms of the shifted moduli:
\begin{subequations}
\begin{align}
    &\hat{N}=\sum_{\omega=0}^{N-1} \tilde{N}_\omega q_2^\frac{\omega}{N}\CalR_\omega,\quad \tilde{N}_\omega=\sum_{c(\alpha)=\omega}e^{\tilde{a}_\omega},\quad\tilde{N}=\sum_{\omega=0}^{N-1}\tilde{N}_\omega; \\
    &\hat{M}=\sum_{\omega=0}^{N-1}\tilde{M}_{\omega}q_2^\frac{\omega}{N}\CalR_\omega,\quad \tilde{M}_\omega=\sum_{\sigma(f)=\omega}e^{\tilde{m}_{\omega}^{\pm}},\quad \tilde{M}=\sum_{\omega=0}^{N-1}M_\omega; \\
    &\hat{K}=\sum_{\omega=0}^{N-1}\tilde{K}_\omega q_2^\frac{\omega}{N}\CalR_\omega,\quad \tilde{K}_\omega=\sum_{\alpha}e^{\tilde{a}_\alpha}\sum_{J}\sum_{\underset{c(\alpha)+j-1= \omega+NJ}{(i,j)\in\lambda^{(\alpha)}}}q_1^iq_2^J,\quad \tilde{K}=\sum_{\omega=0}^{N-1}\tilde{K}_\omega; \\
    & \hat{S}=\hat{N}-P_1\left(1-q_2^\frac{1}{N}\CalR_1\right)\hat{K}=\sum_\omega\tilde{S}_\omega q_2^\frac{\omega}{N}\CalR_\omega,\quad \tilde{S}=\sum_{\omega=0}^{N-1}\tilde{S}_\omega, \label{S}
\end{align}
\end{subequations}
with
\begin{subequations}
\begin{align}
\tilde{S}_\omega
&=\tilde{N}_\omega-P_1\tilde{K}_\omega+P_1\tilde{K}_{\omega-1}, \, \omega=1,\dots,N-1; \\
\tilde{S}_0
&=\tilde{N}_0-P_1\tilde{K}_0+q_2P_1\tilde{K}_{N-1}.
\end{align}
\end{subequations}

The surface defect partition function is the $\BZ_N$-invariant fields, which can be easily obtained from the bulk partition function in Eq.~\eqref{def:instparti}:  
\begin{align}
\mathcal{Z}({\ba},{\bm}^\pm, \kq, \vec{z})
& =\sum_{\vec{\lambda}}\prod_\omega \kq_\omega^{k_\omega}\mathbb{E}\left[-\left(\frac{\hat{S}\hat{S}^*-\hat{M}\hat{S}^*}{P_1^*(1-q_2^{-\frac{1}{N}}\CalR_{-1})}\right)^{\mathbb{Z}_N}\right] \nonumber\\
& =\sum_{\vec{\lambda}}\prod_\omega \kq_\omega^{k_\omega}\mathbb{E}\left[-\frac{\tilde{S}\tilde{S}^*-\tilde{M}\tilde{S}^*}{P_{12}^*}+\frac{\sum_{\omega_1<\omega_2}\tilde{S}_{\omega_1}\tilde{S}_{\omega_2}^* - \tilde{M}_{\omega_1}\tilde{S}_{\omega_2}^*}{P_1^*}\right] \nonumber\\
& =\sum_{\vec{\lambda}}\prod_{\omega=0}^{N-1} \kq_\omega^{k_\omega}\mu_\text{bulk}({\ba},{\bm}^\pm)[\vec{\lambda}]\mu_\text{surface}({\ba},{\bm}^\pm)[\vec{\lambda}]. 
\label{eq:Z-defect}
\end{align}
The bulk contribution  
\begin{align}\label{def:bulk}
    \mu_\text{bulk}({\ba},{\bm}^\pm,\kq)[\vec{\lambda}]=\mathbb{E}\left[-\frac{\tilde{S}\tilde{S}^*-\tilde{M}\tilde{S}^*}{P_{12}^*}\right]
\end{align}
depends only on the bulk Young diagram $\tilde{S} = \tilde{N} - P_{12} \tilde{K}_{N-1}$.
Dependence on the fractional $\tilde{K}_{\omega}$ lies in the surface contribution
\begin{align}\label{def:surface}
    \mu_\text{surface}({\ba},{\bm}^\pm)[\vec{\lambda}]=\mathbb{E}\left[\frac{\sum_{\omega_1<\omega_2}\tilde{S}_{\omega_1}\tilde{S}_{\omega_2}^*-\tilde{M}_{\omega_1}\tilde{S}_{\omega_2}^*}{P_1^*}\right].
\end{align}

We define a new set of {\it virtual characters}:
\begin{align}
    \Gamma_\omega:=\tilde{S}_{0}+\cdots+\tilde{S}_{\omega-1},\quad \omega=1,\dots,N.
\end{align}
The surface contribution can be rewrite using $\{\Gamma_{\omega}\}$:
\begin{align}
    \mu_\text{surface}[\vec{\lambda}]=\mathbb{E}\left[\frac{\sum_{\omega=1}^{N-1}\Gamma_\omega(\Gamma_{\omega+1}-\Gamma_\omega)^*}{P_1^*}-\frac{\sum_{\omega=1}^{N-1}\tilde{M}_{\omega-1}(\Gamma_N-\Gamma_\omega)^*}{P_1^*}\right].
\end{align}

In the NS-limit $\epsilon_2\to0$ with $\epsilon_1\equiv \epsilon$ fixed. The bulk contribution is locked to the {\it limit shape instanton configuration} $\vec{\Lambda}_*$ (See appendix \ref{sec:limitshape} for detail about limit shape) which satisfies 
\begin{align}
     \mu_\text{bulk}({\ba},{\bm}^\pm,\kq)[\vec{\Lambda}_*]=\mathbb{E}\left[-\frac{\tilde{S}[\vec{\Lambda}_*]\tilde{S}^*[\vec{\Lambda}_*]-\tilde{M}\tilde{S}^*[\vec{\Lambda}_*]}{P_{12}^*}\right] = \exp \left( \frac{1}{\epsilon_2}\CalW({\ba},{\bm}^\pm,\tau,\epsilon_1)\right).
\end{align}

The character $\Gamma_N = \tilde{S}$ denotes the limit shape configuration in the bulk, while the remaining $\Gamma_{\omega}$, $\omega=1,\dots,N-1$ involves any surface structure on top of the bulk limit shape. In particular we find the virtual characters $\Gamma_{\omega}$'s of the from
\begin{subequations}
\begin{align}
    \Gamma_N&=\sum_{\alpha}e^{A_\alpha}+\sum_\alpha\sum_{\{J'\}}e^{\tilde{a}_\alpha}q_2^{J'}q_1^{\Lambda^{t,(\alpha)}_{*,J'+1}}\left(1-q_1\right)=F_N+P_1W, \\
    \Gamma_\omega
	&=\sum_{c(\alpha)<\omega} e^{A_\alpha}q_1^{\lambda^{t,(\alpha)}_{{\rm tail},\omega-c(\alpha)}}+\sum_{\alpha}\sum_{\{J'_\omega\}}e^{\tilde{a}_\alpha}q_2^{J'_\omega}q_1^{\Lambda^{t,(\alpha)}_{*,J'_\omega+1}}\left(1-q_1\right)=F_\omega+P_1U_\omega.
\end{align}
\end{subequations}
The $F_\omega$'s denote the $N-2$ Young diagrams $\vec{\lambda}_{\rm tail} = ({\lambda}^{(0)}_{\rm tail},{\lambda}^{(1)}_{\rm tail},\dots,{\lambda}^{(N-2)}_{\rm tail})$ attaching to first $J=0$ of limit shape $\vec{\Lambda}$, which we call \emph{tail}. Each tail Young diagram $\lambda^{(\omega)}$ for $\omega=0,1,\dots,N-2$ is the collection of row of boxes of non-negative length
$\lambda^{(\omega)}_{{\rm tail}} = (\lambda^{(\omega)}_{{\rm tail},1},\lambda^{(\omega)}_{{\rm tail},2},\dots)$ obeying
\begin{subequations}
\begin{align}
    & \lambda^{(\omega)}_{{\rm tail},1} \leq N-1-\omega; \nonumber \\ 
    & \lambda^{(\omega)}_{{\rm tail},i} \geq \lambda^{(\omega)}_{{\rm tail},i+1}, \quad i=1,2,\dots . \nonumber
\end{align}
\end{subequations}

The set of \emph{jumps in the bulk} $\{J'\}$ is defined by
\begin{align}
    & \{J'\} = \left\{ J\in\BN \ | \ \Lambda_{*,J}^{t,(\alpha)} - \Lambda_{*,J+1}^{t,(\alpha)} = 1 \right\}, \quad \{J'_1\} \subset \{J'_2\} \subset\cdots\subset \{J'_N\} = \{J'\}.
\end{align}

\begin{figure}
    \centering
    \includegraphics[width=0.85\linewidth]{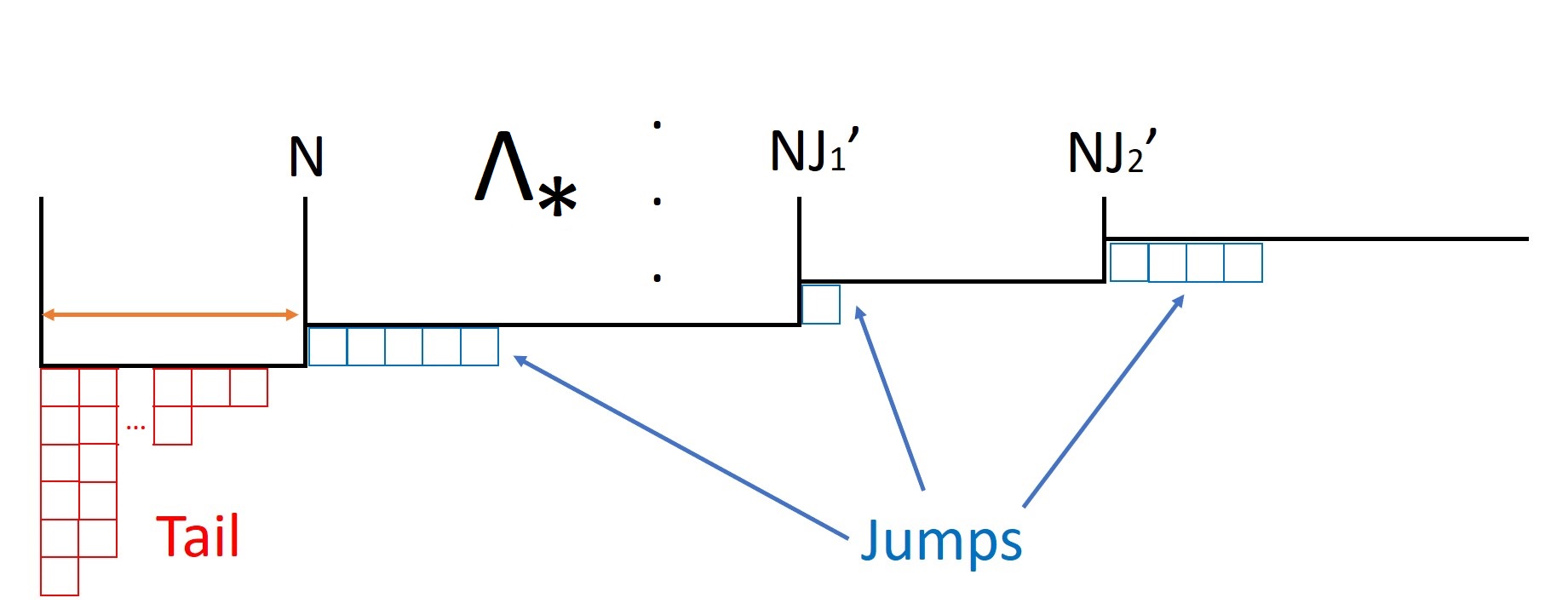}
    \caption{Tail and jumps that forms $\vec{\lambda}$ on top of limit shape $\vec{\Lambda}_*$. }
    \label{fig:jumps and tail}
\end{figure}

The normalized vev of the surface defect $\Psi_{\ba}$ is identified as an ensemble over all allowed surface configurations, namely the arrangements of jumps $\{U_{\omega}\}$ and tail Young diagrams $\lambda^{t,(\alpha)}_{{\rm tail},\omega-c(\alpha)}$ connected to the very bottom of limit shape $\vec{\Lambda}_{*}$. See Fig.~\ref{fig:jumps and tail} for illustration.
\begin{align}\label{chi}
	\Psi_{\ba}
	& = \sum_{\vec{\lambda}}\prod_{\omega=0}^{N-2} \kq_{\omega}^{k_{\omega} - k_{N-1}} \mu_\text{surface}[\vec{\lambda}] \nonumber\\
	&=\sum_{\rm \vec{\lambda}_{\rm tail}} \sum_{\{U_{\omega}\}}\prod_{\omega=0}^{N-2}\kq_\omega^{k_\omega-k_{N-1}}\mathbb{E}\left[\frac{\sum_{\omega=1}^{N-1}\Gamma_\omega(\Gamma_{\omega+1}-\Gamma_\omega)^*}{P_1^*}-\frac{\sum_{\omega=1}^{N-1}M_{\omega-1}(\Gamma_N-\Gamma_\omega)^*}{P_1^*}\right].
\end{align}

\section{The integral representation} \label{sec:surfaceIPF}

In this chapter, we demonstrate how we to simplify the expression for the normalized vev $\Psi_{\ba}$ of the surface defect in the vacuum characterized by the Coulomb moduli $\ba$ in \eqref{chi}. We shall cast it in the form of the $\frac{N(N-1)}{2}$-fold Mellin-Barnes contour integral. In the asymptotically free limit our integral approaches that of the eigenfunction of quantum periodic Toda chain \cite{Kharchev:2000ug, Kharchev:2000yj}, as it should. 

In this chapter and onward, $\{a_{\alpha}\}$ and $\{m_{f}^\pm\}$ always denote the shifted moduli parameters and fundamental/antifundamental multiplet masses.

\subsection{The emerging quiver structure} \label{sec:new quiver}
Define the \emph{dual character} $V_\omega$:
\begin{align}\label{Def: V_w}
    \Gamma_N-\Gamma_\omega=\sum_{l\geq\omega}e^{A_l}+P_1V_\omega,\quad V_\omega=(W-U_\omega)+\sum_{l<\omega}e^{A_l}\frac{1-q_1^{\lambda^{t,(l)}_{{\rm tail},\omega-l}-\lambda^{t,(l)}_{{\rm tail},N-l}}}{1-q_1}.
\end{align}
We see that $V_\omega$ is a pure character, i.e. it is a sum of monomials with positive coefficients.
The normalized vev of the surface defect $\Psi_{\ba}$ in \eqref{chi} can be rewritten in terms of the  $V_\omega$'s as follows:
\begin{align}\label{Wave function: V}
    \Psi_{\ba}=\sum_{\vec{\lambda}}&\prod_{\omega=0}^{N-2}\kq_\omega^{k_\omega-k_{N-1}}\ \mathbb{E}\left[\sum_{\omega=1}^{N-1}\left(\Gamma_N-M_{\omega-1}-\sum_{l\geq\omega}L_l\right)\frac{L_\omega^*}{P_1^*}\right] \nonumber\\
    &\times\mathbb{E}\left[P_1WV_1^*+\sum_{\omega=1}^{N-1}\left({L_{\omega-1}}V_{\omega}^*+q_1{L_\omega^*}V_\omega-P_1V_\omega(V_\omega-V_{\omega+1})^*-M_{\omega-1}V_{\omega}^*\right)\right], \quad L_{\omega} = e^{A_{\omega}}.
\end{align}
The new ADHM-like quiver system can be \emph{reconstructed} from the $\{V_\omega\}$ dependence in the Eq.~\eqref{Wave function: V}. With a little bit of a work one arrives at the data which consists of the following vector spaces:
\begin{itemize}
    \item $N-1$ complex vector spaces ${\bf V}_\omega \approx \BC^{v_{\omega}}$ whose character is denoted by $V_\omega$,
    \item $N$  complex vector spaces ${\bf L}_\omega \approx \BC^1$ whose character is denoted by $L_\omega=e^{A_\omega}$, 
    \item $N-1$ mass nodes ${\bf M}_\omega$, and
    \item one complex vector space ${\bf W} \approx \BC^{|W|}$ whose character is denoted by $W$, 
\end{itemize}
the maps between them (and their contributions to the character in \eqref{Wave function: V}): 
\begin{itemize}
    \item The map ${\bf I} : {\bf W} \to  {\bf V}_1$ \qquad ( $WV_1^*$ );
    \item The map ${\bf i}_\omega : {\bf L}_\omega \to {\bf V}_{\omega+1}$  \qquad ( $L_\omega V_{\omega+1}^*$ ); 
    \item The map ${\bf j}_\omega : {\bf V}_\omega \to {\bf L}_\omega$ \qquad ($q_1L_\omega^*V_\omega$ );
    \item The map $B^{(\omega)}_1 : {\bf V}_\omega \to {\bf V}_\omega$ \qquad ($q_1V_\omega V_\omega^*$ ); 
    \item The map $\beta_\omega : {\bf V}_\omega \to {\bf V}_{\omega+1}$ \qquad ($V_\omega V_{\omega+1}^*$) ,  
    \end{itemize}
a couple of the  ADHM-like equations (and their contributions to the character in \eqref{Wave function: V}):
\begin{itemize}
    \item The equation $B^{(\omega+1)}_1\beta_\omega-\beta_\omega B^{(\omega)}_1+{\bf i}_\omega {\bf j}_\omega=0$ \qquad ($-q_1V_\omega V_{\omega+1}^*$);
    \item The equation $B_1^{(1)}{\bf I}=0$ \qquad ($-q_1WV_1^*$)\ , 
    \end{itemize}
the gauge symmetry    $U({\bf V}_\omega)$, which contributes $( -V_\omega V_\omega^* )$ to the character, and, finally, one supplements the measure by the Euler class of the vector bundle of maps ${\bf M}_{\omega-1} \to {\bf V}_\omega$, effectively contributing $( -M_{\omega-1}V_\omega^* )$ to the character. 

\begin{figure}
    \centering
    \includegraphics[width=0.85\linewidth]{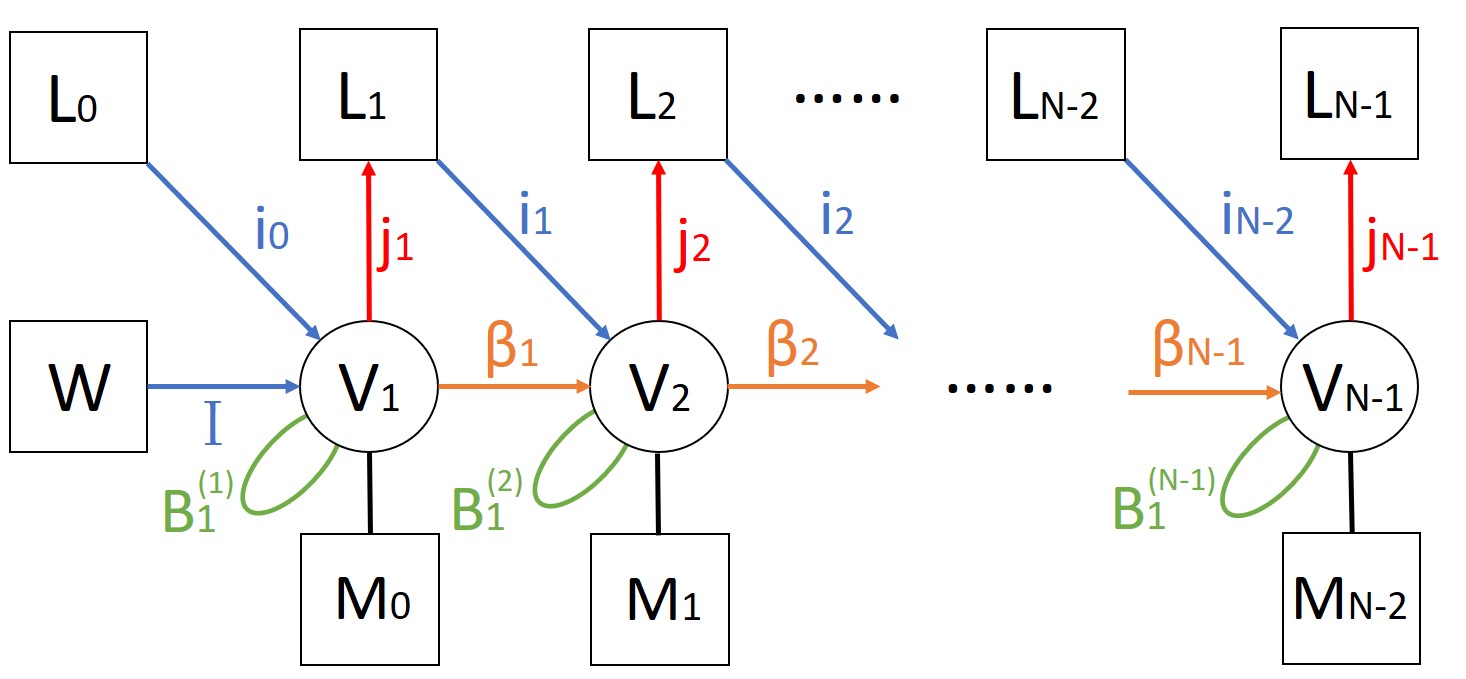}
    \caption{The quiver emerging from the $V_{\omega}$ dependence of the normalized vev ${\Psi}_{\ba}$.}
    \label{fig:newquiver}
\end{figure}

See Fig.~\ref{fig:newquiver} for the illustration of the new quiver, resembling, in particular, the handsaw quiver \cite{Nakajima:2011yq,K-T}.

The moduli space corresponding to this new quiver is given  by
\begin{align}
    \CalM=&\left\{\text{Hom}({\bf W} ,{\bf V}_1) \oplus\bigoplus_{\omega=1}^{N-2}\text{Hom}({\bf V}_\omega,{\bf V}_{\omega+1}) \oplus\bigoplus_{\omega=1}^{N-1}\text{Hom}({\bf V}_\omega,{\bf V}_\omega) \oplus\bigoplus_{\omega=0}^{N-1}\text{Hom}({\bf L}_\omega,{\bf V}_{\omega+1})\right. \nonumber \\
    &\left.\quad\oplus\bigoplus_{\omega=1}^{N-1}\text{Hom}({\bf V}_\omega, {\bf L}_\omega)+\text{eqs.}\right\}/[U({\bf V}_1)\times U({\bf V}_2)\times\cdots\times U({\bf V}_{N-1})]
\end{align}
with the global symmetry
\begin{align}
    U({\bf W})\times U({\bf L}_0)\times U({\bf L}_1)\times\cdots\times U({\bf L}_{N-1}) \times U({\bf M}_0)\times\cdots \times U({\bf M}_{N-2}) \times U(1)_{q_1}.
\end{align}
The $U(1)_{q_1}$ symmetry acts by 
$$
(B_1^{(\omega)},\beta_\omega,{\bf i}_\omega,{\bf j}_\omega,I)\to (q_1B_1^{(\omega)},\beta_\omega,{\bf i}_\omega,q_1{\bf j}_\omega,I)
$$ 
for all $\omega=1,\dots,N-1$. 
The quotient with respect to 
\beq
U({\bf V}_1)\times U({\bf V}_2)\times\cdots\times U({\bf V}_{N-1})
\label{eq:eqgg}
\eeq
is accompanied by the real moment map equations:
\begin{align}\label{real moment1}
    \mu_{\mathbb{R},\omega}=[B_1^{(\omega)},B_1^{(\omega)\dagger}]+\beta_{\omega-1}\beta_{\omega-1}^\dagger-\beta_\omega^\dagger\beta_\omega+{\bf i}_{\omega-1}{\bf i}_{\omega-1}^\dagger-{\bf j}_\omega^\dagger {\bf j}_\omega=\zeta_{\BR, \omega} {\rm Id}_{{\bf V}_\omega}, \, 
\end{align}
with $\beta_0={\bf I}$ and $\beta_{N-1}=0$, and
\beq
{\bf\zeta} = \{\zeta_{\BR, \omega}\}  
\eeq
being the set of Fayet–Iliopoulos parameters. We call the $\zeta_{\BR,\omega}>0$ choice \emph{the positive stability chamber}. A representative would be
$$
    \zeta_{\BR,\omega} \equiv \zeta_{\BR}> 0 \quad \forall \omega.
$$
This emerging quiver variety is the special case of the moduli space of folded instantons \cite{Nikita:II,Nikita:III} on the $\mathbb{Z}_N$ orbifold of ${\BC}^{4}$ acting via $(\bz_1,\bz_2,\bz_3,\bz_4)\to (\bz_1,\eta\bz_2,\bz_3,\eta^{-1}\bz_4)$ with $\eta^N=1$ \cite{Nikita:IV}, with the following 
gauge origami data: the Chan-Paton spaces, as the $\BZ_N$-modules, are:
\begin{subequations}
\begin{align}
    {\bf N}_{12}&=\bigoplus_{\omega=0}^{N-1} {\bf L}_\omega\otimes {\CalR}_\omega, \\
    {\bf N}_{24}&={\bf W}\otimes {\CalR}_0, \\
    {\bf K}&=\bigoplus_{\omega=1}^{N-1}{\bf V}_\omega\otimes {\CalR}_{\omega-1}.
\end{align}
\label{eq:emqdec}
\end{subequations}
The ADHM gauge origami matrices are decomposed as 
\begin{subequations}
\begin{align}
    B_1 & = (B_{1}^{(\omega)})_{\omega-1}; \\
    B_2 & = (\beta_{\omega})_{\omega-1}; \\
    I_{12} & = ({\bf i}_{\omega-1})_{\omega-1}; \\
    J_{12} & = ({\bf j}_{\omega})_{\omega-1}; \\
    I_{24} & = ({\bf I})_0
\end{align}
\label{eq:emqmat}
\end{subequations}
for $\omega = 1,\dots,N-1$,
and 
\beq
J_{24} = B_3 = B_4 = 0\ .
\label{eq:emqeq2}
\eeq
They satisfy:
\begin{subequations}
\begin{align}
    &[B_1,B_2]+I_{12}J_{12}=0, \\
    & B_{1}I_{24}=0 
\end{align}
\end{subequations}
In the positive stability chamber the vector space ${\bf K}$ decomposes as: 
\begin{align}\label{pstability}
    {\bf K}=\mathbb{C}[B_1,B_2]I_{12}({\bf N}_{12})+\mathbb{C}[B_2]I_{24}({\bf N}_{24}).
\end{align}
The Eqs. \eqref{eq:emqdec}, \eqref{eq:emqmat}, \eqref{eq:emqeq2}  follow from the folded instantons equations \cite{Nikita:III} subject to the decomposition \eqref{eq:emqdec} supplemented by the real moment map equation (which is equivalent to \eqref{real moment1}): 
\begin{align}\label{real moment2}
    \mu_{\mathbb{R}}=[B_1,B_1^\dagger]+[B_2,B_2^\dagger]+I_{12}I_{12}^\dagger-J_{12}^\dagger J_{12}+I_{24}I_{24}^\dagger={\zeta}_{\BR} \operatorname
    {Id}_{\bf K}.
\end{align}
As a result, ${\Psi}_{\ba} \equiv {\Psi}_{\ba, +}$ is the cohomological field theory partition function which is obtained by integrating the equivariant Euler class of the bundle of all the equations above over the moduli space of matrices obeying the stability condition \eqref{pstability} modulo the complexified symmetry group \eqref{eq:eqgg}.

\subsection{On the other side}

The calculation of the partition function ${\Psi}_{\ba, -}$ defined in the same way with the flip of the sign of the  FI-parameters, i.e. for ${\zeta}_{\BR}<0$ in the real moment map equation \eqref{real moment2} is much simpler. Indeed, 
in the \emph{negative stability chamber}, i.e. for ${\zeta}_{\BR}<0$, 
the Eq.~\eqref{real moment2} implies
$$
    ||I_{12}||^2 + ||I_{24}||^2 -||J_{12}||^2= k \zeta_{\BR}<0.
$$
It implies that both $I_{12} = I_{24} = 0$ at the fixed points of the global symmetry, leaving the $B_1^\dagger$ and $B_2^\dagger$ commuting. 
The vector space ${\bf K}$ is generated by the image of $J_{12}^\dagger$:
\begin{align}
    {\bf K}=\mathbb{C}[B_1^\dagger,B_2^\dagger]J^\dagger_{12}({\bf N}_{12}).
\end{align}
The fixed points on the moduli space are characterized by $N-1$ Young diagrams  $\vec{\lambda}_{\rm dual}=(\lambda_{\rm dual}^{(1)}, \dots , \lambda_{\rm dual}^{(N-1)} )$ with the restrictions on their maximal height in the $B_2^\dagger$ direction. Each Young diagram $\lambda_{\rm dual}^{(\alpha)}=(\lambda_{{\rm dual},1}^{(\alpha)},\lambda_{{\rm dual},2}^{(\alpha)},\cdots)$ is a collection of rows of squares of non-negative length obeying
\begin{subequations}
\begin{align}
    &\lambda^{(\alpha)}_{{\rm dual},i} \geq \lambda^{(\alpha)}_{{\rm dual},i+1}, \quad i=1,2,\dots \\
    & \lambda^{(\alpha)}_{{\rm dual},1}\leq \alpha, \quad \alpha = 1,\dots,N-1
\end{align}
\label{eq:dualdiag1}
\end{subequations}
The transposed Young diagrams can be expressed by the collection $(\lambda^{(\alpha)}_{\rm dual})^t=(\lambda^{(\alpha),t}_{1}, \lambda^{(\alpha),t}_{2},\dots,\lambda^{(\alpha),t}_{\alpha})$ with non-negative entries such that
\begin{align}
    \lambda^{(\alpha),t}_{i}\geq \lambda^{(\alpha),t}_{i+1} , \ i=1,\dots,\alpha.
    \label{eq:dualdiag2}
\end{align}
The dual characters $V_{\omega}$ in the negative stability chamber are
\begin{align}
    V_\omega
    &=\sum_{\alpha\geq\omega}e^{A_\alpha} \, q_{1}^{-1}
    \, \frac{1-q_1^{-\lambda^{(\alpha),t}_{{\rm dual},\alpha-\omega+1}}}{1-q_1^{-1}}, \nonumber\\
    \implies \Gamma_N-\Gamma_\omega
    & =\sum_{l\geq\omega}e^{A_l}+P_1V_\omega=\sum_{l\geq\omega}e^{A_l}q_1^{-\lambda^{(\alpha),t}_{{\rm dual},\alpha-\omega+1}}=F_{\geq\omega}
\end{align}

Set $\epsilon_{1} = {\hbar}$. 
The normalized vev of the surface defect in that chamber is equal to the sum over the Gelfand-Zeitlin-like table \eqref{eq:dualdiag1}, \eqref{eq:dualdiag2}, similar to the sum over fluxes in the gauged linear sigma model corresponding to the complete flag variety, or as in \cite{Kharchev:2000yj}:
\begin{align}
    \Psi_{\ba,-}\ =\ \sum_{\vec{\lambda}_{\rm dual}}\prod_{\omega=0}^{N-2}\ {\kq}_{\omega}^{- \frac{1}{\hbar}{\rm ch}_1(F_{\geq\omega}) } \ \mathbb{E}\left[ \, \frac{1}{P_{1}^{*}} \left( \Gamma_N F_{\geq1}^*-\sum_{\omega} F_{\geq\omega}(F_{\geq\omega}-F_{\geq\omega+1})^* -\sum_{\omega}M_{\omega-1}F_{\geq\omega}^* \right) \, \right].
    \label{eq:flagsum}
\end{align}

\subsubsection{Mutation of fractional couplings}

For future use let us define 
the effective fractional couplings. Let: 
\beq
c_{\omega} := 1-\kq_{\omega} + \kq_{\omega}\kq_{\omega+1} +\cdots+(-1)^{N-1-\omega}\kq_{\omega}\cdots\kq_{N-2}\ , 
\label{eq:convcrf}
\eeq
for ${\omega} = 0, 1, \ldots, N-2$, with $c_{-1} = c_{N-1} = 1$, 
and 
\begin{subequations}\label{effective kq}
\begin{align}
 \kq_{0,\text{eff}}
    &=\kq_0 \, c_1  \ , \\
    \kq_{\omega,\text{eff}}
    &=\kq_{\omega}\, {c_{\omega+1}}\, c_{\omega-1}^{-1}\, ,\qquad \omega=1,\dots,N-3 \, ,  \\
   \kq_{N-2,\text{eff}}
    &=\kq_{N-2}\, c_{N-3}^{-1}\ ,  \\
    \kq_{N-1,\text{eff}}  & = \ {\kq} \, \prod\limits_{\omega=0}^{N-2}{\kq}_{\omega,\text{eff}}^{-1}\ ,
\end{align}
\end{subequations}
so that the product of effective fractional couplings is equal to the  bulk coupling $\kq$. We leave the study of the properties of the map ${\vec\kq} \mapsto {\vec\kq}_{\rm eff}$ to future work.

\subsubsection{The crossing of the normalized vev}

See the appendix \eqref{sec: moreexamples} for the $N= 2,3$ examples and for the toy model illustrating the transformation of the fractional couplings. Here we present the case of  general $N$

The normalized vev of the surface defect of a general gauge group $U(N)$ is of the form
\begin{align}
    \Psi_{\ba, +} & = \sum_{\vec{\lambda}}\prod_{\omega=1}^{N-1}\kq_\omega^{v_\omega} \ \mathbb{E}\left[\left(\Gamma_N-\sum_{l\geq\omega}e^{A_l}-P_1V_\omega\right)\left(\frac{e^{A_\omega}}{P_1}+V_\omega-V_{\omega+1}\right)^*+\frac{M_{\omega-1}(\sum_{l\geq\omega}e^{A_l}+P_1V_\omega)^*}{P_1^*}\right]
\end{align} 
We again rewrite $\Psi_{\ba,+}$ using the dual characters $\{V_{\omega}\}$ defined by the Eq.~\eqref{Def: V_w}.
The sum over the entries in $V_\omega$ can be expressed as the sum over the residues in the contour integral
\begin{equation}
\begin{aligned}
    \sum_{v_1,\dots,v_{N-1}=0}^{\infty} &\prod_{\omega=1}^{N-1}\, \frac{\kq_{\omega-1}^{v_{\omega}}}{v_{\omega}!}\, \oint_{\CalC_\omega}\prod_{i=1}^{v_\omega}\frac{d\phi_i^{(\omega)}}{2\pi {\ii}}\frac{1}{\hbar}
    \prod_{i>j}\frac{(\phi_i^{(\omega)}-\phi_j^{(\omega)})^2}{(\phi_i^{(\omega)}-\phi_j^{(\omega)})^2-\hbar^2}\prod_{j=1}^{v_{\omega+1}}
    \frac{\phi_i^{(\omega)}-\phi_j^{(\omega+1)}+\hbar}{\phi_i^{(\omega)}-\phi_j^{(\omega+1)}} \\
    & \qquad\qquad\qquad\qquad \times \frac{(\phi_i^{(\omega)}-m_{\omega-1}^+)(\phi_i^{(\omega)}-m_{\omega-1}^-)}{(\phi_i^{(\omega)}-A_{\omega-1})(\phi_i^{(\omega)}-A_{\omega}+\hbar)}\times\prod_{s=1}^{|W|}\frac{\phi_i^{(1)}-b_s-\hbar}{\phi_i^{(1)}-b_s}.
    \label{eq:contourN}
\end{aligned}
\end{equation}
The integration is evaluated by deforming the contours $\CalC_{\omega}$ in steps:
\begin{enumerate}
    \item We start at $\omega=N-1$.
    \item We choose $v_{\omega}-l_{\omega}$ integration variables $\{\phi^{(\omega)}_{i}\}$, $i=1,\dots,v_{\omega}-l_{\omega}$, for some $l_{\omega}=0,\dots,v_{\omega}$, to pick up the residues at the pole at infinity. 
    \item The residue at infinity is computed using Eq.~\eqref{R_k,l}. 
    \item The integral over the remaining variables $\{\phi_{i}^{(\omega)}\}$, $i=v_{\omega}-l_{\omega}+1,\dots,v_{\omega}$ is performed by computing the residues at $A_{\alpha}-j\hbar$, $j=1,2,\dots$, for some $\alpha = \omega,\dots,N-1$. These poles generate the dual Young diagram corresponding to a fixed point of the quiver variety in the negative FI-parameter chamber. 
    \item Sum over $(v_\omega,l_\omega)$. 
    \item Repeat the steps 2 to 5 with ${\omega} \to \omega-1$.
\end{enumerate}
The residues at infinity for a single $\phi^{(\omega)}$ are 
\begin{align}
    &{r}_\infty^{(\omega)}=\frac{1}{\hbar}( m_{\omega-1}^++m_{\omega-1}^--A_{\omega-1}-A_{\omega}+\hbar), \quad \omega=1,\dots,N-1.
\end{align}
In terms of the quantity defined by the Eq. \eqref{eq:convcrf}
the total crossing factor is given by the formula
\begin{align}
    c_0^{|W|}\times\prod_{j=0}^{N-2}c_j^{r_\infty^{(j+1)}}. \nonumber
\end{align}

\subsection{Integral representation of the normalized vev of the surface defect}

The normalized vev of the surface defect $\Psi_{\ba,+}$ in the positive ${\zeta}_{\BR}$ chamber is related to its negative chamber counterpart by the crossing formula,
\begin{align}\label{eq: Relation of chi's}
    {\Psi}_{\ba, +}
    &=c_0^{\frac{1}{\hbar}\sum_{\alpha=0}^{N-1}A_\alpha-a_\alpha}\left[\prod_{j=0}^{N-2}c_j^{r^{(j+1)}_\infty}\right]\Psi_{\ba,-}.
\end{align}
The physical meaning of the parameters $A_\alpha$'s in the quantum integrable system are the asymptotic momenta at the spatial infinity. 

The normalized vev of the surface defect $\Psi_{\ba, -}$ in the negative FI-parameter chamber can be represented both as
the discrete sum \eqref{eq:flagsum}, and as the integral over the  set  $\by=(y_1,y_2,\dots,y_{N-1})$ of ${\BR}$-valued variables in the following form
\begin{align}\label{eq: chi<0}
    {\Psi}_{\ba, -} 
    &\ =\ \sum_{\vec{\lambda}}\prod_{\omega=0}^{N-1}\kq_{\omega,\text{eff}}^{k_\omega-k_{N-1}}\mathbb{E}\left[\frac{\Gamma_NF_{\geq1}^*}{P_1^*}-\frac{\sum_{\omega=1}^{N-1}F_{\geq\omega}(F_{\geq\omega}-F_{\geq\omega+1})^*}{P_1^*}-\frac{\sum_{\omega=1}^{N-1}M_{\omega-1}F_{\geq\omega}^*}{P_1^*}\right] \nonumber\\
    &\qquad\qquad =\ \int\mu(\by)C(\by)\mathbf{U}(\by)d\by
\end{align}
where
\begin{enumerate}
    \item The set $||y_{n,j}||_{1 \leq j \leq n  \leq N-1}$ is the lower left triangle of an $(N-1)\times(N-1)$ matrix. The function $\Psi(\by)$ is defined with the identification $\by=(y_1,\dots,y_{N-1})=(y_{N-1,1},\dots,y_{N-1,N-1})$ by the Mellin-Barnes integral   
    \begin{align}\label{eq: Mellin fund}
        {\bf U}(\by)=\int_\mathcal{C}
        \prod_{n=1}^{N-2} \frac{\kq_{N-n-1,\text{eff}}^{-\sum\limits_{j=1}^n\frac{y_{n,j}}{{\hbar}}}}{n!}
        &\frac{\prod\limits_{j=1}^n\prod\limits_{k=1}^{n+1}(-{\hbar})^{\frac{y_{n,j}-y_{n+1,k}}{{\hbar}}} \Gamma\left(\frac{y_{n,j}-y_{n+1,k}}{{\hbar}}\right)}{\prod\limits_{1\leq j< k\leq n}\Gamma\left(\frac{y_{n,j}-y_{n,k}}{{\hbar}}\right)\Gamma\left(\frac{y_{n,k}-y_{n,j}}{{\hbar}}\right)}
        \frac{1}{\prod\limits_{j=1}^n {\mathfrak{M}}_{N-n-1}(y_{n,j})}\prod_{n=1}^{N-2}\prod_{j=1}^n \frac{dy_{n,j}}{2\pi {\hbar}{\ii}}.
    \end{align}
    The integral is understood as follows: first integrate out the variable $y_{1,1}$ along a straight line 
    $$
        \mathcal{C}_{1,1}:=\left\{ y_{1,1} : \; \text{Re}\left(\frac{y_{1,1}}{{\hbar}}\right)=\text{constant}, \; \text{Re}\left({y_{1,1}}/{{\hbar}}\right)>\text{max}\{\text{Re}\left({y_{2,1}}/{{\hbar}}\right),\text{Re}\left({y_{2,2}}/{{\hbar}}\right)\} \right\}
    $$
    followed by integration with respect to the variables $(y_{2,1},y_{2,2})$ over the product of two straight lines $\{\CalC_{2,1},\CalC_{2,2}\}$ parallel to $\CalC_{1,1}$,
    $$
        \mathcal{C}_{2,i}:=\left\{ y_{2,1} : \; \text{Re}\left(\frac{y_{2,i}}{{\hbar}}\right)=\text{constant}, \;
        \text{Re}\left({y_{2,i}}/{{\hbar}}\right)>\text{max}_j\{\text{Re}\left({y_{3,j}}/{{\hbar}}\right)\}\right\}
    $$
    and so on. The last set of integrations with respect to the variables $(y_{N-2,1},\dots,y_{N-2,N-2})$ is performed along $N-2$ straight lines
    $$
        \mathcal{C}_{N-2,i}:=\left\{ y_{N-2,i} : \; \text{Re}\left(\frac{y_{N-2,i}}{{\hbar}}\right)=\text{constant}, \;
        \text{Re}\left({y_{N-2,i}}/{{\hbar}}\right)>\text{max}_j\{\text{Re}\left({y_{N-1,j}}/{{\hbar}}\right)\}\right\}.
    $$
The contribution to the integrand of the fundamental flavor multiplets is given by
    $$
        {\mathfrak{M}}_{n} (y)\ =\ \bE\left[\frac{M_ne^{-y}}{P_1^*}\right]=\prod_\pm{\Gamma\left(\frac{y-m_{n}^\pm}{{\hbar}}\right)}
        ({\hbar})^{\frac{y-m_{n}^\pm}{{\hbar}}}
    $$
    \item The function $C(\by)$ is given by
    \begin{align}
        C(\by)=\prod_{j=1}^{N-1}\frac{Q(y_j-{\hbar})}{{\mathfrak{M}}_{0}(y_j)}\frac{1}{\prod_{\alpha=1}^N\sin\pi\left(\frac{y_j-A_\alpha}{{\hbar}}\right)}
    \end{align}
    where $Q(y)$ is the solution of Baxter equation \eqref{Baxter Q Hyper} and $M_0(y)$ is given by 
    \begin{align}
        {\mathfrak{M}}_0(y)=\prod_\pm{\Gamma\left(\frac{y-m_{0}^\pm}{{\hbar}}\right)}
        ({\hbar})^{\frac{y-m_{0}^\pm}{{\hbar}}} \nonumber
    \end{align}
    Even though $C(\by)$ only has contributions from two fundamental masses, Eq.~\eqref{eq: chi<0} shall still have gamma function factor from $2N-2$ fundamental mass dependent on $\by$ by the pole structure of \eqref{eq: Mellin fund}. 
    \item The measure $\mu(\by)$, sometimes called the Sklyanin measure, is of the form
    \begin{align}
        \mu(\by)
        &=\kq_{0,\text{eff}}^{-\sum_{j=1}^{N-1}\frac{y_j}{{\hbar}}}\frac{1}{(2\pi {\ii}{\hbar})^{N-1}}\frac{1}{(N-1)!}\prod_{j\neq k}\frac{1}{\Gamma(\frac{y_j-y_k}{{\hbar}})} \nonumber\\
        &=\kq_{0,\text{eff}}^{-\sum_{j=1}^{N-1}\frac{y_j}{{\hbar}}}\frac{1}{(2\pi {\ii}{\hbar})^{N-1}}\frac{1}{(N-1)!}\prod_{1\leq j<k\leq N-1}\frac{(y_j-y_k)}{\pi{\hbar}}\sin\pi\left(\frac{y_j-y_k}{{\hbar}}\right)
    \end{align}
\end{enumerate}

We notice that the normalized vev of the surface defect \eqref{eq: chi<0} has the same structure as the $SL(2,\BR)$ spin wave function derived in \cite{Derkachov:2002tf}.   

\subsection{The limit to Toda}

Let us consider the limits $m^\pm_f\to \infty$, ${\kq} \to 0$, with $\kq \prod_{f=1}^{N}m_f^{+} m_f^{-} = \Lambda^{2N}$ kept finite. From the point of view of the gauge theory, we integrate out the fundamental hypermultiplets, arriving at
the pure super-Yang-Mills theory.  It is well-known to be dual to the $\hat{A}_{N-1}$ Toda lattice (periodic $N$-particle Toda chain) in the sense of  Bethe/gauge correspondence, 
\begin{equation}\label{H-Toda}
 \hat{H}_{\rm Toda}=  - \frac{{\hbar}^2}{2} \sum_{\alpha=1}^N \frac{\partial^2}{\partial \rx_\alpha^2} 
  +\Lambda^2\sum_{\alpha=1}^{N}e^{\rx_\alpha-\rx_{\alpha-1}},\quad \rx_{\alpha}\sim \rx_{\alpha+N}
\end{equation}

Integrating out the fundamental hypermultiplets results in several modification of the normalized vev of the surface defect \eqref{eq: Relation of chi's}. Since the integration over moduli space \eqref{eq:contourN} no longer has pole at the infinity, there will be no crossing factor. The surface partition functions in the positive and negative FI-parameter chambers are therefore identical
$$
    {\Psi}_{\ba,+} = {\Psi}_{\ba,-}.
$$
In addition, the fractional coupling $\kq_{\omega}$ will not be modified between the two chambers, namely 
$$
    {\kq}_{\omega,\text{eff}} = {\kq}_{\omega} = \Lambda^{2}e^{\rx_{\omega} - \rx_{\omega-1}}.
$$
In the mass decoupling limit, we recognize that the normalized vev of the surface partition fucntion coincides with the {integral formula for the eigenfunction of the $\hat{A}_{N-1}$ Toda lattice} \cite{Kharchev:2000ug,Kharchev:2000yj}. See appendix \ref{sec:Toda} for detail about reconstruct the Schr\"{o}dinger equation of $\hat{A}_{N-1}$ Toda lattice from $\CalN=2$ SYM with a co-dimension two surface defect.

\subsection{Crossing and AGT}

Here we work with both ${\epsilon}_{1}, {\epsilon}_{2}$ finite, so we restore the notation ${\epsilon}_{+} = {\epsilon}_{1} + {\epsilon}_{2}$. 
Let us consider the $SU(2)$ gauge theory. We can  start with the $U(2)$ gauge group with the  Coulomb moduli $a_1 = - a_2 = a$ having the zero trace. 
In the original AGT conjecture \cite{Alday:2009aq}, the $SU(2)$ vector multiplet is accompanied by two fundamental and two anti-fundamental hypermultiplets. The crossing formula in the Eq.~\eqref{eq:crossbulkN} considers all flavors in the fundamental representation of the gauge group. To match with the AGT convention, we change the hypermultiplets with masses $m_1^+$ and $m^-$ to the anti-fundamental representation of the gauge group, which modifies
$$
    m_1^\pm \to {\epsilon}_+ - m_1^\pm
$$
One special property of the $U(2)$ gauge theory with moduli parameter $a_1=-a_2=a$ is that the instanton partition functions in the positive and negative FI-parameter chambers are related by the additional symmetry $m \mapsto \epsilon_+ - m$. Such additional symmetry can be seen by identifying the instanton configuration $(\lambda_1,\lambda_2)$ in the ${\zeta}_{\BR}>0$ chamber with the instanton confugration $(\lambda_2,\lambda_1)$ in the $\zeta_{\BR} <0$ chamber, which results in
\begin{align}
    \CalZ_{U(2),+}(a,m_{i}^\pm;\kq) = \CalZ_{U(2),-}(a,\epsilon_+ - m_{i}^\pm;\kq).
\end{align}
At the same time, the crossing formula in Eq.~\eqref{eq:crossbulkN} predicts
\begin{align}
    \CalZ_{U(2),+}(a,m_{i}^\pm;\kq) = (1-\kq)^{-r_\infty} \CalZ_{U(2),-}(a,m_{i}^\pm;\kq). \nonumber
\end{align}
See the appendix \ref{sec:cross-bulk} for the derivation of the crossing formula in the bulk. 

The $r_\infty$ in the new convention of flavor becomes
$$
    r_\infty = \frac{\epsilon_+}{\epsilon_1\epsilon_2} ( m_0^+ + m_0^- - m_1^+ - m_1^- ).
$$
We denote the masses of the fundamental and anti-fundamental flavors of the effective $U(1)$ theory by
\begin{align}
    \mu_0 = \frac{m_0^+ + m_0^-}{2} , \, \mu_1 = \frac{m_1^+ + m_1^-}{2} \implies r_\infty = 2\epsilon_+(\mu_0 - \mu_1). \nonumber
\end{align}
A $U(1)$ instanton partition function with one fundamental flavor $\mu_{0}$ and one anti-fundamental flavor $\mu_1$ is equal to \cite{Nikita:I}: 
$$
    \CalZ_{U(1),+}(\mu_i;\kq) = (1-\kq)^{\frac{\mu_0(\epsilon_+-\mu_1)}{\epsilon_1\epsilon_2}}.
$$ 
The symmetry $m\mapsto\epsilon_+ - m$ can be restored by decoupling overall $U(1)$ instanton partition from the $U(2)$ instanton partition function,
\begin{align}
    \CalB(a,m_{i}^\pm;\kq) = \left(\CalZ_{U(1),+}(\mu_i,\kq)\right)^{-2}\CalZ_{U(2),+}(a,m_{i}^\pm;\kq) = (1-\kq)^{-\frac{2\mu_0(\epsilon_+-\mu_1)}{\epsilon_1\epsilon_2}}\CalZ_{U(2),{\zeta}_{\BR}>0}(a,m_{i}^\pm;\kq),
\end{align}
such that $\CalB(a,m_i^\pm;\kq)$ enjoys the $m\mapsto\epsilon_+ - m$ symmetry,
\begin{align}
    \CalB(a,\epsilon_+ - m_i^\pm;\kq)
    & = (1-\kq)^{-\frac{2(\epsilon_+ - \mu_0)\mu_1}{\epsilon_1\epsilon_2}}\CalZ_{U(2),+}(a,\epsilon_+-m_{i}^\pm;\kq) \nonumber\\
    & = (1-\kq)^{-\frac{2(\epsilon_+ - \mu_0) \mu_1}{\epsilon_1\epsilon_2}}\CalZ_{U(2),+}(a,m_{i}^\pm;\kq) \nonumber\\
    & = (1-\kq)^{-\frac{2(\epsilon_+ - \mu_0) \mu_1}{\epsilon_1\epsilon_2}}(1-\kq)^{r_\infty}\CalZ_{U(2),+}(a,m_{i}^\pm;\kq) \nonumber\\
    & = (1-\kq)^{-\frac{2\mu_0(\epsilon_+-\mu_1)}{\epsilon_1\epsilon_2}}\CalZ_{U(2),+}(a,m_{i}^\pm;\kq) \nonumber\\
    & = \CalB(a,m_{i}^\pm;\kq).
\end{align}

We can recover exactly the $U(1)$ factor given by the original AGT conjecture \cite{Alday:2009aq}. We also need to take special value of $\Omega$-parameters $\epsilon_1 = b$, $\epsilon_2 = b^{-1}$. As shown in \eqref{eq:liouville}, the $U(2)$ gauge theory is associated to the Liouville conformal theory. The momenta of vertex operators in Liouville theory are 
\begin{subequations}
\begin{align}
    & \mu_0 - \frac{\epsilon_+}{2} = \frac{m_0^+ + m_0^- - \epsilon_+}{2}, \quad \tilde{\mu}_0 - \frac{\epsilon_+}{2} = \frac{m_0^+ - m_0^-}{2}, \\
    & \mu_1 - \frac{\epsilon_+}{2} = \frac{m_1^+ + m_1^- - \epsilon_+}{2}, \quad \tilde{\mu}_1 - \frac{\epsilon_+}{2} = \frac{m_1^+ - m_1^-}{2}.
\end{align}
\end{subequations}
based on the identification in \eqref{eq:Deltas}, 

Instead of decoupling the $U(1)$ factor, an alternative choice to restore the $m\mapsto\epsilon_+ - m$ symmetry is by coupling the $U(1)$ instanton partition function in the opposite FI-parameter chamber
\begin{align}
    \tilde{\CalB}(a,m_{0,1}^\pm;\kq) := \CalZ_{U(1),-}(\mu_i,\kq)^{2}\CalZ_{U(2),+}(a,m_{i}^\pm;\kq)
\end{align}
with 
\begin{align}
    \CalZ_{U(1);-}(\mu_i;\kq) = (1-\kq)^\frac{(\epsilon_+ - \mu_0)\mu_1}{\epsilon_1\epsilon_2}.
\end{align}
Changing $m\mapsto\epsilon_+-m$ swaps the FI-parameter chambers the $U(2)$ and $U(1)$ instanton partition functions reside in.
The choice of fundamental masses of the $U(1)$ instanton partition function ensures that the crossing factor from $\CalZ_{U(1),{\zeta}_{\BR}<0}$ and $\CalZ_{U(2),{\zeta}_{\BR}>0}$ cancel each other, leaving $\tilde{\CalB}$ invariant.

The main statement of the AGT correspondence identifies $\CalB$ with the $4$-point conformal block of Liouville conformal theory on a sphere. See \cite{zamolodchikov2007lectures,Zamolodchikov:1995aa,Seiberg:1990eb} for the lectures on Liouville theory, and \cite{Alday:2009aq,Fateev:2009aw,Wyllard:2009hg} for details about the AGT correspondence.

\section{Classical $\spch$/SQCD correspondence} \label{sec:XXX/SQCD c}

A connection between the $\spch$ spin chain and the $\CalN=2$ SQCD in four dimensions has been anticipated a long time ago. Various hints were presented first in \cite{Gorsky:1996hs, Gorsky:1997mw}, then in \cite{Nekrasov:2009uh,Nekrasov:2009ui,Nikita-Shatashvili, Dorey:2011pa,HYC:2011}, for fine tuned parameters of the theory. In this paper we show in full generality that
the classical $\spch$ spin chain is the Seiberg-Witten geometry of the theory, in particular we establish relations between the $\mathfrak{sl}_2$ spin chain coordinate systems and the defect gauge theory parameters.


\paragraph{}
Let us briefly review the classical  $\spch$ spin chain Lax matrices and the monodromy matrix. Let $x$ be a local coordinate on the $\mathbb{CP}^1$. The Lax operators are defined as a set of $GL_2$-valued functions \cite{Faddeev:1996iy}
\begin{align}
    L_\omega^{\rm XXX}(x) = x - \mu_\omega + \CalL_\omega, 
    \quad \omega = 0,\dots,N-1
\end{align}
where $\CalL_\omega = \ell_\omega^{0}\sigma_0 + \ell_{\omega}^+\sigma_{+} + \ell_{\omega}^-\sigma_-$ are $\mathfrak{sl}_2$ matrices. The $\mu_i$'s are $N$ points on $\mathbb{CP}^1$ which are called the \emph{inhomogeneities}. The Lax matrix $L_\omega^{\rm XXX}(x)$ is assigned to the $(\omega+1)$-th site of $\spch$ spin chain lattice with a Poisson structure defined on each site
\begin{align}
    \{ \ell^0_\alpha , \ell^\pm_\beta \} = \pm \ell_{\alpha}^\pm \delta_{ \alpha \beta }, \, \{ \ell^+_{\alpha} , \ell^-_{\beta} \} = 2 \ell^0_{\alpha} \delta_{ \alpha \beta } . \label{eq:spin-poisson}
\end{align}
The monodromy matrix is defined as a product over Lax matrices
\begin{align}
    {\bf T}_{\rm SC}(x) = K(\kq) L_{N-1}^{\rm XXX}(x) \cdots L_{0}^{\rm XXX}(x).
    \label{eq:TSC}
\end{align}
The \emph{twist matrix} $K(\kq)$ is a constant $GL_2$-valued matrix. The spectral curve of spin chain is defined by introduction of spectral parameter $z$
\begin{align}
    \det (z - {\bf T}_{\rm SC}(x)) = 0.
\end{align}
Expanding the $2\times 2$ determinant explicitly gives
\begin{align}
    z^2 - z \Tr {\bf T}_{\rm SC}(x) + {\det K(\kq) P(x)} = 0, \quad P(x) = \prod_{\omega=1}^{N-1} ( x - \mu_\omega + s_\omega ) ( x - \mu_\omega - s_{\omega} ).
\end{align}

\subsection{Constructing the monodromy matrix} \label{chap:transfer matrix}

We now demonstrate how one can recognize the Eq. \eqref{eq:TSC} in the four dimensional supersymmetric gauge theory. We take both ${\epsilon}_{1}, {\epsilon}_{2} \to 0$ (classical, or flat space) limit of
the non-perturbative Dyson-Schwinger equations \eqref{eq:DS-frac} in the presence of the regular defect:
\begin{equation}
    Y_{{\omega}+1} + {\kq}_{\omega} \frac{P_{\omega}(x)}{Y_{\omega}} = (1+{\kq}_{\omega}) t_{\omega}(x)
    \label{eq:fracDS}
\end{equation}
where
$$
    P_{\omega}(x) = (x-m_{\omega}^{+})(x-m_{\omega}^{-}), \quad t_{\omega}(x) = x - {\rho}_{\omega}.
$$
Let us define 
\begin{equation}
    Y_{\omega}= (x-m_{\omega}^{+}) \frac{{\psi}_{\omega}}{{\psi}_{\omega -1}}
\end{equation}
such that Eq.~\eqref{eq:fracDS} becomes a second degree difference equation of the $\psi_{\omega}$'s.
\begin{equation}
   (x-m_{\omega + 1}^{+}) {\psi}_{\omega+1} + {\kq}_{\omega} (x-m_{\omega}^{-}){\psi}_{\omega -1} = (1+{\kq}_{\omega}) (x - {\rho}_{\omega}) {\psi}_{\omega},
   \label{eq:frbax}
\end{equation}
with twisted periodicity constraint imposed on ${\psi}_{\omega}$'s
\begin{equation}
    {\psi}_{\omega +N} = z {\psi}_{\omega}
    \label{eq:ztwist}
\end{equation}
for some complex $z$. 
Eq. \eqref{eq:frbax} can be rewrite as a first-order difference equation by defining a vector
\beq
{\chi}_{\omega} = \left( \begin{matrix} \psi_{\omega+1} - \psi_{\omega} \\ {\psi}_{\omega} \end{matrix} \right) \, , \qquad {\chi}_{\omega + N} = z {\chi}_{\omega}, \nonumber
\eeq
such that $\chi_{\omega}$ obeys
\begin{align}
    \chi_\omega
    &=\frac{1}{x-m_{\omega+1}^{+}}
    \begin{pmatrix}
    -\rho_\omega+m_{\omega+1}^++\kq_\omega(x-\rho_\omega) 
    & -\rho_\omega+m_{\omega+1}^++\kq_{\omega}(-\rho_{\omega}+m_{\omega}^{-}) \\
    x-m_{\omega+1}^{+} & x-m_{\omega+1}^{+}
    \end{pmatrix} 
    \chi_{\omega-1}.
    \label{eq:step1}
\end{align}
We consider a gauge transformation ${\Pi}_{\omega} = h_{\omega} {\chi}_{\omega}$ satisfying
\beq \label{def:gaugeh}
h_{\omega} \begin{pmatrix} {\kq}_{\omega} & 0 \\ 1 & 1 \end{pmatrix} h_{\omega -1}^{-1} = \begin{pmatrix} 1 & 0 \\ 0 & 1 \end{pmatrix}.
\eeq
In other words
\beq
h_{\omega}\begin{pmatrix} {\kq}_{\omega} & 0 \\ 1 & 1 \end{pmatrix}  \begin{pmatrix} {\kq}_{\omega-1} & 0 \\ 1 & 1 \end{pmatrix} \ldots
\begin{pmatrix} {\kq}_{0} & 0 \\ 1 & 1 \end{pmatrix}  = h_{\omega}\begin{pmatrix} u_{\omega+1}^{\vee} - u_{\omega}^{\vee} & 0 \\ u_{\omega}^{\vee} & 1 \end{pmatrix}  =  h_{-1} \nonumber
\eeq
where 
\beq
u_{\omega}^{\vee} = 1+{\kq}_{0} + {\kq}_{0}{\kq}_{1} + {\kq}_{0}{\kq}_{1}{\kq}_{2} + \ldots + {\kq}_{0}{\ldots}{\kq}_{\omega -1}.
\eeq
The twisted matrix is defined based on the gauge transformation
\begin{align}
    & h_{N-1}^{-1}  =  \prod_{\omega = 0}^{N-1} \begin{pmatrix} {\kq}_{\omega} & 0 \\ 1 & 1 \end{pmatrix} \ h_{-1}^{-1} := h_{-1}^{-1} K \implies K = h_{-1}\begin{pmatrix} \kq & 0 \\ u_{N-1}^\vee & 1 \end{pmatrix} (h_{-1})^{-1}.
\label{eq:twistK}
\end{align}
The first order difference equation Eq.~\eqref{eq:step1} can be written in terms of the vector $\Pi$,
\beq
{\Pi}_{\omega} = \left( 1  + \frac{1}{x - m_{\omega+1}^{+}} L_{\omega} \right) {\Pi}_{\omega -1}= \frac{L_{\omega}^\text{XXX}(x)}{x-m_{\omega+1}^+}\Pi_{\omega-1}
\eeq
with the Lax matrix of the form
\begin{align}\label{def:Lax}
    L_{\omega}
    &=h_{\omega}
    \begin{pmatrix}
    (1+\kq_{\omega})(-\rho_\omega+m_{\omega+1}^+) &\qquad (-\rho_\omega+m_{\omega+1}^+)+\kq_{\omega}(-\rho_{\omega}+m_{\omega}^-) \\
    0 & 0
    \end{pmatrix}
    \begin{pmatrix}
    \kq_\omega & 0 \\ 1 & 1
    \end{pmatrix}^{-1}h_{\omega}^{-1} \nonumber \\
    & = - s_{\omega} + {\mathcal{L}}_{\omega}\, , \qquad {\mathcal{L}}_{\omega} = \begin{pmatrix}
    {\ell}_{\omega}^{0} & {\ell}_{\omega}^{-}  \\
    {\ell}_{\omega}^{+} & - {\ell}_{\omega}^{0} 
    \end{pmatrix}. 
\end{align}
The spins $\{s_\omega\}$ and the inhomogeneities $\{\mu_\omega\}$ of the $\spch$ spin chain are expressed  through the masses of the fundamental and anti-fundamental flavors in the gauge theory via
$$
    s_{\omega} = \frac{-m_{\omega+1}^{+} + m_{\omega}^{-}}{2}, \quad \mu_{\omega}=\frac{m_{\omega+1}^++m_{\omega}^-}{2}.
$$
The quadratic Casimir operator of the $\mathfrak{sl}_2$ spin vector $\vec{\ell}_{\omega}$ is 
\beq \label{eq:spinnorm}
(\vec{\ell}_{\omega})^2 =  ({\ell}_{\omega}^{0})^{2} + {\ell}_{\omega}^{+} {\ell}_{\omega}^{-} = s_{\omega}^2.
\eeq
We now construct the spin chain monodromy matrix ${\bf T}(x)$ based on the Lax operators,
\begin{align}
    & {\bf T}(x) = K \prod_{\omega=0}^{N-1}\frac{L_{\omega}^{\rm XXX}(x)}{x-m_{\omega+1}^+} = \frac{1}{P_{+}(x)} \,  {\bf T}_{\rm SC}(x)
    \label{def:trans XXX}
\end{align}
with $P_\pm(x) = \prod_{\omega}(x-m_{\omega}^\pm)$. The twist matrix $K$ is defined in Eq.~\eqref{eq:twistK}. 
The spectral curve of the $\spch$ spin chain is 
\begin{align}
    0 &=
    \det(z-{\bf T}(x)) =z^2-z\text{Tr}{\bf T}(x)+\det{\bf T}(x) 
    =z^2- z (1+\kq)\frac{T(x)}{P_+(x)}+\kq \frac{P_-(x)}{P_+(x)}.
\end{align}
After the substitution $Y = z P_{+}(x)$, the spectral curve of the spin chain coincides with the bulk Seiberg-Witten curve when $Y\neq0$
\beq
{\det} ( Y - {\bf T}_{\rm SC}(x) ) = Y^2  - Y \text{Tr}\ {\bf T}_{\rm SC}(x) + {\kq} P(x)
\eeq
with
\beq
\text{Tr} {\bf T}_{\rm SC}(x) = (1+{\qe}) T(x)|_{\rm SW}. \nonumber
\eeq

\subsection{Canonical coordinates}
The components of spin vector $\vec{\ell}_{\omega}$ obeys Eq.\eqref{eq:spinnorm}. In what follows, we shall consider the representation for the spin components which is build upon two independent parameters $\beta_{\omega}$ and $\gamma_{\omega}$,
\beq
{\ell}_{\omega}^{0} =  {\beta}_{\omega}{\gamma}_{\omega}-s_{\omega}, \quad {\ell}_{\omega}^{+} = 2 s_{\omega} {\gamma}_{\omega} - {\beta}_{\omega} {\gamma}_{\omega}^{2}  , \quad {\ell}_{\omega}^{-} =   {\beta}_{\omega}.
\eeq
Given the spin component Poisson structure \eqref{eq:spin-poisson}, parameters $\{\beta_{\omega}, \gamma_{\omega}\}$ are canonical coordinate pairs subject to the Poisson relation
\begin{align}
    \{ \gamma_{\omega}, \beta_{\omega'}\} = \delta_{\omega\omega'}.
\end{align}

Our next objective is to identify canonical coordinates $\{\beta_{\omega},\gamma_{\omega}\}$ in terms of gauge theory parameters. The monodromy matrix \eqref{def:trans XXX} is gauge $h_{\omega}$ dependent. All the gauge matrices $h_{\omega}$ are generated by a single $h_{-1}$ based on their definition in Eq.~\eqref{def:gaugeh},
\begin{align}
    h_\omega=
    \begin{pmatrix}
    \ta_{\omega} & \tb_{\omega} \\
    \tc_{\omega} & \td_{\omega}
    \end{pmatrix}
    =h_{-1}\frac{1}{u_{\omega+1}^{\vee}-u_{\omega}^{\vee}} 
    \begin{pmatrix}
    1 & 0 \\ -u_{\omega}^{\vee} & u_{\omega+1}^{\vee}-u_{\omega}^{\vee}
    \end{pmatrix}. \nonumber
\end{align}
The Lax matrix $\CalL_{\omega}$ in Eq.~\eqref{def:Lax} is defined based on the gauge matrix $h_{\omega}$   
\begin{align}
    \CalL_\omega
    &= 
    \begin{pmatrix}
    \beta_{\omega}\gamma_{\omega}-s_{\omega} & \beta_{\omega} \\
    2s_{\omega}\gamma_{\omega}-\beta_{\omega}\gamma_{\omega}^2 & -\beta_{\omega}\gamma_{\omega}+s_{\omega}
    \end{pmatrix} \nonumber \\
    &=\frac{1}{\ta_{\omega}\td_{\omega}-\tc_{\omega}\tb_{\omega}}
    \begin{pmatrix}
    -\ta_{\omega}\tc_{\omega}\tp_{\omega}+(\ta_{\omega}\td_{\omega}+\tc_{\omega}\tb_{\omega})s_{\omega}
    & \ta_{\omega}^2\tp_{\omega}-(2\ta_{\omega}\tb_{\omega})s_{\omega} \\
    -\tc_{\omega}^2\tp_{\omega}+2\tc_{\omega}\td_{\omega}s_{\omega} &
    \ta_{\omega}\tc_{\omega}\tp_{\omega}-(\ta_{\omega}\td_{\omega}+\tc_{\omega}\tb_{\omega})s_{\omega}
    \end{pmatrix}
    \label{eq:XXXLmatrix}
\end{align}
where
\begin{align}
    \tp_{\omega}=\rho_{\omega}-m_{\omega+1}^+ + \kq_{\omega}(\rho_{\omega}-m_{\omega}^-)=(1+\kq_{\omega})(\rho_{\omega}+\mu_{\omega})+(\kq_{\omega}-1)s_{\omega}.
    \label{eq:P_w}
\end{align}
The spin chain canonical coordinates $\{\beta_{\omega},\gamma_{\omega}\}$ are identified in terms of the gauge transform matrix components
\begin{align} \label{eq:betagamma}
    \beta_{\omega}
    =\frac{\ta_{\omega}^2\tp_{\omega}- 2\ta_{\omega}\tb_{\omega}s_{\omega}}{\ta_{\omega}\td_{\omega}-\tc_{\omega}\tb_{\omega}}, \quad
    \gamma_\omega=-\frac{\tc_{\omega}}{\ta_{\omega}}.
\end{align}
The $\tp_{\omega}$ given in \eqref{eq:P_w} is related to the zeros of the fractional $t_{\omega}(x)$, which can be found by considering large $x$ expansion of Eq.~\eqref{eq:fracDS}:
\begin{align}
    \tp_{\omega}=-\left[\epsilon z_{\omega}\left(\frac{\partial}{\partial z_{\omega}}-\frac{\partial}{\partial z_{\omega-1}}\right)-a_{\omega+1}+\frac{z_{\omega}}{z_{\omega-1}}a_{\omega}-m_{\omega+1}^{+} + \frac{z_{\omega}}{z_{\omega-1}}m_{\omega}^{+} \right] \log \Psi_{\ba}. \nonumber
\end{align}
with the coordinates $\bz = (z_{0},\dots,z_{N-1})$ defined by
\begin{align}\label{def:z_w}
    \kq_{\omega} = \frac{z_{\omega}}{z_{\omega-1}}, \quad z_{\omega+N} = \kq z_{\omega}, \quad \nabla^z_{\omega} = z_{\omega}\frac{\partial}{\partial z_{\omega}}.
\end{align}
The normalized vev of the surface defect differs from $\Psi_{\ba}$ by the perturbative factor 
\begin{align}\label{def:Psi}
    \tilde{\Psi}=\left[\prod_{\omega=0}^{N-1}z_{\omega}^{-\frac{a_{\omega+1}-m_{\omega+1}^+}{\epsilon}}\right] \Psi_{\ba}
\end{align}
such that the $\tp_{\omega}$ becomes
\begin{align}
    \tp_{\omega} = \epsilon z_{\omega}\left(\frac{\partial}{\partial z_{\omega-1}}-\frac{\partial}{\partial z_{\omega}}\right)\log \tilde{\Psi} = z_{\omega} \left(\frac{\partial\rS}{\partial z_{\omega-1}}-\frac{\partial\rS}{\partial z_{\omega}}\right) .
\end{align}
The Hamilton-Jacobi function ${\rS}$ is defined by
\begin{align}
    \rS({\ba},{\bm}^\pm,\bz) = \lim_{\epsilon\to0} \epsilon\log \tilde{\Psi}({\ba},{\bm}^\pm,\bz) \nonumber
\end{align}
The twist matrix $K$ defined in Eq.~\eqref{eq:twistK} is also gauge dependent, which we shall use as the anchor for our gauge choice. In particular, we demand that the gauge transformation satisfies
\begin{align}
    K = h_{-1} \begin{pmatrix} \kq & 0 \\ u_{N-1}^\vee & 1 \end{pmatrix} (h_{-1})^{-1} = \begin{pmatrix} \kq & 0 \\ 0 & 1 \end{pmatrix},
    \label{def:K1}
\end{align}
which we solve out the gauge
\begin{align}\label{eq:h-1}
    h_{-1} = \begin{pmatrix} \frac{1}{z_{-1}} & 0 \\ \frac{u_{N-1}^\vee}{1-\kq} & 1 \end{pmatrix}.
\end{align}

The spin canonical coordinates $\{\gamma_{\omega},\beta_{\omega}\}$ with the gauge choice $h_{-1}$ in Eq.~\eqref{eq:h-1} can be identified using the gauge theory parameters. In particular, they satisfy the Hamiltonian-Jacobi equation
\begin{subequations}\label{eq:goodgamma}
\begin{align}
    & \gamma_{\omega} = \frac{z_{\omega} + z_{\omega+1} + \ldots + z_{\omega+N-1}}{{\kq}-1} = \frac{{\kq} ( z_{0} + \ldots + z_{\omega -1} ) + z_{\omega} + \ldots + z_{N-1}}{{\kq}-1}.  \\
    & \beta_{\omega} = \frac{\partial\rS}{\partial z_{\omega-1}} - \frac{\partial\rS}{\partial z_{\omega}} = \frac{\partial\rS}{\partial\gamma_{\omega}}
\end{align}
\end{subequations}

$\{\beta_{\omega},\gamma_{\omega}\}$ form Darboux coordinate pair are subject to the desired Poisson structure
\begin{align}
    \{\gamma_{\alpha},\beta_{\beta}\} = \delta_{\alpha\beta}.
\end{align}
Moreover, the coordinates $\{\gamma_{\omega},\beta_{\omega}\}$ obey the twisted periodicity condition
\beq
\gamma_{\omega + N} = {\kq} \gamma_{\omega}, \quad \beta_{\omega+N} = \frac{1}{\kq}\beta_{\omega}
\eeq

The only constrain on the twist matrix $K$ is that it is a $GL_2$ constant matrix with the determinant $\det(K) = \kq$. 
The choice of the twist matrix $K$ in \eqref{def:K1} is not unique. An alternative is  
\begin{align}
    K = \begin{pmatrix} \kq+1 & -\kq \\ 1 & 0 \end{pmatrix} = h_{-1}^{\rm new} \begin{pmatrix} \kq & 0 \\ u_{N-1}^{\vee} & 1 \end{pmatrix} (h_{-1}^{\rm new})^{-1}.
\end{align}
The gauge transformation that yields such twist matrix $K$ is
\begin{align} 
    h_{-1}^{\rm new} = \begin{pmatrix} u_{N-1}^\vee & 1 \\ 0 & 1 \end{pmatrix}.
\end{align}
The gauge matrix $h_{\omega}$ that defines the Lax matrices $\CalL_{\omega}$ now reads 
\begin{align}
    h_{\omega}^{\rm new} = \begin{pmatrix} \ta_{\omega} & \tb_{\omega} \\ \tc_{\omega} & \td_{\omega} \end{pmatrix}
    = \frac{1}{u_{\omega+1}^\vee - u_{\omega}^\vee} \begin{pmatrix} u_{N-1}^\vee-{u_{\omega}^\vee}{} & {u_{\omega+1}^\vee - u_{\omega}^\vee} \\ -{u_{\omega}^\vee} & {u_{\omega+1}^\vee - u_{\omega}^\vee} \end{pmatrix}.
\end{align}

A new set of coordinates $\{\beta_{\omega}^{\rm new}, \gamma_{\omega}^{\rm new}\}$ can be defined based on the new gauge, whose relation with the gauge theory parameters are less illuminating.

\subsection{An open spin chain inside the integral}

In this chapter, we study the normalized vev of the surface defect \eqref{eq: chi<0}  in the semi-classical limit $\hbar \to0$. To avoid the clutter, we denote
\beq
    \tilde{\kq}_{n}=-\kq_{N-n-1,\text{eff}}\, , \quad \tilde{m}_{n}^\pm = m_{N-n+1}^\pm\, , \qquad n=0,1,\dots,N-1.
\eeq
In the limit $\hbar\to0$, we use the Stirling approximation of the Gamma functions in the Mellin-Barnes integral representation \eqref{eq: Mellin fund} of the normalized vev of the surface defect \eqref{eq: chi<0},
\begin{align}
    {\bf U}({\bf y}) \approx & 
    \int_\mathcal{C} \, \prod_{n=1}^{N-2}\prod_{j=1}^n\frac{dy_{n,j}}{2\pi{\ii} j {\hbar}} \, e^{- \frac{1}{\hbar} W \left[ y_{n,j} \right]} \\
    & W \left[ y_{n,j} \right]  = \sum_{n=1}^{N-2}\log(-\tilde{\kq}_{n})\sum_{j=1}^n y_{n,j} -  \sum_{j=1}^n\sum_{k=1}^{n+1} \left( y_{n,j}-y_{n+1,k} \right)  \left( {\log}\left( y_{n,j}-y_{n+1,k}\right) - 1 \right) \\
    & \qquad + \sum_{1\leq j <  k\leq n} \, {\pm} {\pi}{\ii} \, \left( y_{n,j}-y_{n,k} \right) +   \sum_\pm\sum_{j=1}^n \left( y_{n,j}-m_{N-n-1}^\pm \right) \left( {\rm log} \left( y_{n,j}-m_{N-n-1}^\pm\right) - 1 \right).  \nonumber
\end{align}
The integration is dominated by the saddle point configurations with $y_{n,j}$ ($1\leq n<N-1$) satisfying ${\partial}W/{\partial} y_{n,j} = 0$. 
Exponentiating yields the system of nested Bethe equations
\beq
R_{n+1}(y_{n,j}) = -\tilde{\kq}_{n} R_{n-1}(y_{n,j}) \tilde{P}_{n}(y_{n,j})\, , \ n = 1, \ldots , N-2\, , j = 1, \ldots, n
\label{eq:minibethe}
\eeq
with
\beq
R_{n}(x) = \prod_{j=1}^{n} (x - y_{n,j})\, , \ n = 1, \ldots , N-1, \ R_{0}(x) = 1 \nonumber
\eeq
and $\tilde{P}_n(x)=(x-\tilde{m}_n^+)(x-\tilde{m}_n^-)$. These equations can be viewed as a discrete many-body system, admitting an interesting elliptic generalization \cite{Krichever:1996qd}, which arises in the six dimensional ${\CalN}=(1,0)$ analogues of the theory we studied. In this way one could rigorously justify some of the observations in  \cite{Chen:2020jla} based on M/string theory considerations. 

For the last integration variables $\{y_{N-1,j}\}$, we notice that in the $\hbar\to0$ limit, the functions $Q(x)$ and $Y(x)$ have the asymptotics
\begin{align}
    Q(x)\sim e^{\frac{1}{\hbar}\Sigma(x)}, \quad Y(x)\sim e^{\partial_x\Sigma(x)}.
\end{align}
The saddle point equation for $y_{N-1,j}$ is found by
\beq
Y( y_{N-1,j}) = - \tilde{\kq}_{N-1} R_{N-2}(y_{N-1,j}) \tilde{P}_{N-1}(y_{N-1,j})\, , \ j = 1, \ldots , N-1.
\label{eq:last}
\eeq
In the ${\epsilon}\to 0$ limit the function $Y(x)$ solves the algebraic equation,
\begin{align}
    Y(x)+\kq\frac{P(x)}{Y(x)}=(1+{\kq})T(x)|_{\rm SW}, \quad P(x)= \tilde{P}_{0}(x)\ldots \tilde{P}_{N-2}(x) \ ,  \tilde{P}_{N-1}(x).
    \label{eq: SW}
\end{align}
which defines the Seiberg-Witten curve of our theory. 

We can define a series of $T-Q$ Baxter equations based on the saddle point equations \eqref{eq:minibethe}
\beq
    R_{n+1}(x) = X_n(x)R_{n}(x) - \tilde{\kq}_{n} R_{n-1}(x) P_{N-n-1}(x)\, , \ n = 1, \ldots, N-2,
\label{eq:minibaxter}
\eeq
with $X_n(x) = (1 + \tilde{\kq}_{n})(x - w_{n})$ being a degree one polynomial. 
The $T-Q$ Baxter equation \eqref{eq:minibaxter} can be extended to the $n=0$ case by defining
\begin{align}
    X_0(x)=R_1(x)+\tilde{\kq}_0\tilde{P}_{0}^{-}(x)=(1+\tilde{\kq}_{0})(x-w_0)R_0(x)=R_1(x)+\tilde{\kq}_0\tilde{P}_{0}(x)R_{-1}(x)
\end{align}
with $\tilde{P}_{0}^+(x) R_{-1}(x) = R_{0}(x) = 1$.
Using Eq. \eqref{eq:minibaxter} we express the $X_n(x)$ function as
\begin{align}
    X_n(x)=\frac{R_{n+1}(x)}{R_n(x)}+\tilde{\kq}_{n}\frac{\tilde{P}_{n}(x)R_{n-1}(x)}{R_n(x)}
    \label{eq:character}
\end{align}
for $n=0,1,\dots,N-2$. We will see later that the $X_n(x)$ functions can be identified as $q$-characters.

\subsubsection{Construction of the holonomy matrix}

Let us rewrite the Eqs.~\eqref{eq:minibaxter} in terms of the first order difference equations by defining a vector
\beq
{\Xi}_{n} (x) =  \begin{pmatrix} R_{n+1}(x) \\ {\tilde P}_{n+1}^{+}(x) R_{n} (x) \end{pmatrix} = x^{n+1} \cdot 
\begin{pmatrix} 1 \\ 1 \end{pmatrix} + \ldots, \nonumber
\eeq
obeying the transport equation
\beq
{\Xi}_{n} (x) \ = \ \tilde{L}_{n}(x) \, {\Xi}_{n-1}(x) \ .
\label{eq:minitransfer}
\eeq
The local Lax operators $L_n$ are defined as gauge transforms of $\tilde{L}_{n}$:
\beq \label{eq:gaugeopen}
L_{n}(x) = h_{n+1} \tilde{L}_{n}(x) h_{n}^{-1} = x + h_{n+1}\begin{pmatrix} - w_{n} (1+{\tilde\kq}_{n}) & {\tilde\kq}_{n} {\tilde m}^{-}_{n} \\
- {\tilde m}^{+}_{n+1} & 0 \end{pmatrix} h_{n}^{-1}=x-\mu_n+\CalL_n.
\eeq
with
\begin{align}
    h_{n+1}\begin{pmatrix} 1 + {\tilde\kq}_{n} & - {\tilde\kq}_{n} \\ 1 & 0 \end{pmatrix} h_n^{-1} = \begin{pmatrix} 1 & 0 \\ 0 & 1 \end{pmatrix}.
\end{align}
The 
Eq.~\eqref{eq:last} does not fit for general pattern, of having a polynomial $X_{N-1}(x)$ since $Y(x)$ is multi-valued.  However, by observing the \eqref{eq: Mellin fund} is invariant under the shift of $y_{n,j}\to y_{n,j}+\hbar$ of the integration variables, which can be interpreted as one of the non-perturbative ``large'' contour modifications of \cite{Nikita:I},  subtracting one instanton in the corresponding quiver node ${\bf V}_{N-n-1}$ in the new quiver system. 
We recall the derivation of the corresponding nonperturbative Dyson-Schwinger equations in the form of the $qq$-character analyticity, in the appendix \ref{sec:basic}. A similar procedure can be applied to the integration representation of the normalized vev of the surface defect \eqref{eq: chi<0}. In this way 
we find $N$ $q$-characters\footnote{They are only $q$, not $qq$-characters, since we already have ${\epsilon}_{2} = 0$.} whose expectation values are degree one polynomials. In the classical limit $\hbar\to0$ these $q$-characters are
\begin{subequations}
\begin{align}
    X_{n}(x)
    &=\Upsilon_{n}(x)+\frac{\tilde{\kq}_n\tilde{P}_n(x)}{\Upsilon_{n-1}(x)}, \quad \Upsilon_n(x) = \frac{R_{n+1}(x)}{R_n(x)}, \ n=0,1,\dots,N-2, \label{eq:Xn}\\
    X_{N-1}(x)
    &=\frac{1}{R_{N-1}(x)}\left[(1+\kq)T(x)|_{\rm SW}+\tilde{\kq}_{N-1}\tilde{P}_{N-1}(x)R_{N-2}(x)+U(x)\right]. \label{eq:XN-1}
\end{align}
\end{subequations}
The function $U(x)$ is a degree $N$ polynomial
\begin{align}
    U(x)=R_{N-1}(x)\sum_{k=1}^{N-1}\frac{\prod_{j=1}^{k}\tilde{\kq}_{j-1}\tilde{P}_{j-1}(x)}{R_{k}(x)R_{k-1}(x)}
\end{align}
such that 
$$
    U(y_{N-1,j})=\frac{\kq P(y_{N-1,j})}{\tilde{\kq}_{N-1}\tilde{P}_{N-1}(y_{N-1,j})R_{N-2}(y_{N-1,j})}=-\frac{\kq P(y_{N-1,j})}{Y(y_{N-1,j})}.
$$
Inspired by the function $U(x)$, we define the dual $R$-functions by
\beq
{\tilde R}_{n}(x) = R_{n}(x)
\left[ \tilde{P}_{0}^+(x) + \sum_{k=1}^{{n}} \frac{\prod_{j=1}^{k} \tilde{\kq}_{j-1} \tilde{P}_{j-1}(x)}{R_{k}(x) R_{k-1}(x)}
\right]
\sim \tilde{u}_{n}^{\vee} x^{n+1}+\cdots, \quad  n=1,\dots,N-1
\label{eq:dualR}
\eeq
with
\beq
    \tilde{u}_{n}^\vee = 1 + \tilde{\kq}_0 + \tilde{\kq}_0\tilde{\kq}_1 + \dots + \tilde{\kq}_0\dots\tilde{\kq}_{n-1}.
\eeq

In addition, we extend the dual $R$-function to $n=0$ and $n=-1$ by 
$$
\tilde{R}_0(x)=\tilde{P}_{0}^+(x), \quad \tilde{R}_{-1}(x)=0.
$$

In particular, the $\tilde{R}_{n}(x)$ function defined in Eq.~\eqref{eq:dualR} is a degree $n+1$ polynomial
\begin{align}
    \tilde{R}_{n}(x)=\tilde{u}_{n}^{\vee}\tilde{P}_{0}^+(x) \times \prod_{j=1}^{n} (x-\tilde{y}_{n,j}).
    \label{def:dualR}
\end{align}
The dual $\Upsilon$-function is defined as ratio of the two dual $R$-functions
\begin{align}
    \tilde{\Upsilon}_n(x)=\frac{\tilde{R}_{n+1}(x)}{\tilde{R}_{n}(x)}.
\end{align}
We find that for $n=0,\dots,N-2$, the $\tilde{\Upsilon}_n(x)$ are the other solution to Eq.~\eqref{eq:character}
\begin{align}
    \tilde{\Upsilon}_{n}(x)+\frac{\tilde{\kq}_{n}\tilde{P}_{n}(x)}{\tilde{\Upsilon}_{n-1}(x)} = X_n(x).
\end{align}
Hence the dual $R$-functions $\{\tilde{R}_{n}(x)\}$ are the other linearly independent solution to the second order Baxter equations \eqref{eq:minibaxter}
\begin{align}
    X_n(x)\tilde{R}_{n}(x)=\tilde{R}_{n+1}(x)+\tilde{\kq}_n \tilde{P}_n(x) \tilde{R}_{n-1}(x).
    \label{eq:minibaxter-dual}
\end{align}
The zeros of dual $R$-functions satisfy the Bethe equations, 
\begin{align}
    \tilde{R}_{n+1}(\tilde{y}_{n,j})+\tilde{\kq}_{n} \tilde{P}_{n}(\tilde{y}_{n,j})\tilde{R}_{n-1}(\tilde{y}_{n,j})=0, \quad n=1,\dots,N-2.
    \label{eq:minibethe-dual}
\end{align} 

The Wronskians of the two linearly independent solutions $R_n(x),\tilde{R}_n(x)$ of the Baxter equations are
\begin{align}
    {\tilde{R}_{n+1}(x)}{R_{n}(x)}-{\tilde{R}_{n}(x)}{R_{n+1}(x)}={\prod_{j=0}^{n}\tilde{\kq}_{j}\tilde{P}_{j}(x)}
    \label{eq:rwronskian}
\end{align}

We define the vector $\tilde{\Xi}_n$ which obeys the same transfer relation in Eq.~\eqref{eq:minitransfer} as the vector $\Xi_n$,
\begin{align}
    \tilde{\Xi}_{n}(x)=
    \begin{pmatrix}
    \tilde{R}_{n+1}(x) \\ \tilde{P}_{n+1}^+(x) \tilde{R}_{n}(x)
    \end{pmatrix}=\tilde{L}_{n}(x)\tilde{\Xi}_{n-1}(x).
    \label{eq:minitransfer dual}
\end{align}

The first $N-2$ spin chain Lax matrices can be constructed based on $\tilde{L}_{n}$ by gauge transformation in Eq.~\eqref{eq:gaugeopen}.
One potential candidate for the last Lax matrix $\tilde{L}_{N-1}$ is by using the last $q$-character $X_{N-1}(x)$ in Eq.~\eqref{eq:XN-1}:
\begin{align}
    X_{N-1}(x)R_{N-1}(x)=Y(x)+\frac{\kq P(x)}{Y(x)}+\tilde{\kq}_{N-1}\tilde{P}_{N-1}(x)R_{N-2}(x)+\tilde{R}_{N-1}(x).
\end{align}
The $q$-character $X_{N-1}(x)$ is degree one polynomial with the single root at $w_{N-1}$,
\begin{align}
    X_{N-1}(x)=(1+\kq+\tilde{\kq}_{N-1}+\tilde{u}_{N-1}^{\vee})(x-w_{N-1}). \nonumber
\end{align}
However, it turns out that the correct way to construct the last Lax matrix of the chain involves not the $q$-character ${X}_{N-1}(x)$, but its dual. We explain in detail in the next section.

\subsubsection{The dual $Q$-function}

The second order equations such as $T-Q$ equation \eqref{Baxter Q Hyper} have two linearly independent solutions over the (quasi)constants, which in the present case stands for the $\epsilon$-periodic functions of $x$. The normalized vev of the  surface defect in Eq.~\eqref{eq: chi<0} involves one solution $Q(x)$ inherited from the gauge observable $Y(x)$ in the Seiberg-Witten equation. The other solution to the $T-Q$ equation, denoted as $\tilde{Q}(x)$, can be expressed in terms of $Q(x)$ via the series 
\begin{align}
    \tilde{Q}(x)=\kq^{\frac{x}{{\hbar}}}\sum_{k=0}^\infty\frac{\kq^k Q(x) M(x+k{\hbar}) }{Q(x+k{\hbar})Q(x+(k+1){\hbar})}
    \label{eq:dualQ}
\end{align}
where $M(x)=P(x)M(x-{\hbar})$ given by a product of $\Gamma$-functions. 
A straightforward computation verifies that $\tilde{Q}(x)$ is a solution to the Baxter equation \eqref{Baxter Q Hyper}
\begin{align}
    &(1+\kq)T(x)\tilde{Q}(x)
    =\tilde{Q}(x+{\hbar})+\kq P(x) \tilde{Q}(x-{\hbar}) \nonumber
\end{align}
with the Wronskian 
\begin{align}
    \tilde{Q}(x)Q(x+{\hbar})-\tilde{Q}(x+{\hbar})Q(x)=\kq^{\frac{x}{{\hbar}}}M(x).
\end{align}
The dual $\tilde{Y}(x)$ function is defined as a ratio of two $\tilde{Q}$ functions with a shifted argument:
$$
    \tilde{Y}(x)=\frac{\tilde{Q}(x)}{\tilde{Q}(x-{\hbar})}
$$
which in classical limit ${\hbar}\to0$ relates to the original $Y(x)$ by
\begin{align}
    \tilde{Y}(x)=\frac{\kq P(x)}{Y(x)}.
\end{align}

The normalized vev of the surface defect $\Psi_\ba$ in Eq.~\eqref{eq: chi<0} involves only one solution of the Baxter equation \eqref{Baxter Q Hyper}. A general solution of second order equations considers a linear combination of both the $Q(x)$ and the $\tilde{Q}(x)$,
\begin{align}
    \Psi_{\ba,-}=\int \mu(\by)C(\by){\bf U}(\by) d\by + \kappa \int\mu(\tilde{\by})\tilde{C}(\tilde{\by}){\bf U}(\tilde{\by}) d\tilde{\by}
\end{align}
where $\kappa$ is some constant to be determined by initial or boundary conditions of the specific system. The function $\tilde{C}(\tilde{\by})$ is the dual version of the function $C(\by)$ in Eq.~\eqref{eq: chi<0} 
$$
    \tilde{C}(\tilde{\by}) = \prod_{j=1}^{N-1}\frac{\tilde{Q}(\tilde{y_j}-{\hbar})}{M_0(\tilde{y}_j)}\frac{1}{\prod_{\alpha=1}^N \sin \pi \left( \frac{\tilde{y}-A_\alpha}{{\hbar}}\right)}.
$$

In the classical limit ${\hbar}\to0$, The saddle point equations of Mellin-Barnes integration in ${\bf U}(\tilde{\by})$ generate exactly the Bethe equations Eq.~\eqref{eq:minibethe-dual} for $\{\tilde{y}_{n,j}\}$, $n=1,\dots,N-2$, $j=1,\dots,n$. The saddle point equations of $\tilde{\by}$ variables generate the dual version of Eq.~\eqref{eq:last}:
\begin{align}
    \tilde{P}_{0}^{+}(\tilde{y}_{N-1,j})\tilde{Y}(\tilde{y}_{N-1,j}) = -\tilde{\kq}_{N-1}\tilde{P}_{N-1}(\tilde{y}_{N-1,j})\tilde{R}_{N-2}(\tilde{y}_{N-1,j}).
\end{align}

The dual of the last $q$-character $X_{N-1}(x)$ \eqref{eq:XN-1} is defined as a degree one polynomial
\begin{align}
    \tilde{X}_{N-1}(x)\tilde{R}_{N-1}(x)=\left(\tilde{Y}(x)+\frac{\kq P(x)}{\tilde{Y}(x)}\right)\tilde{P}_0^+(x)+\tilde{\kq}_{N-1}\tilde{P}_{N-1}(x)\tilde{R}_{N-2}(x) - \tilde{P}_0^+(x)^2R_{N-1}(x)
    \label{eq:last-dual}
\end{align}
where
\begin{align}
    \tilde{X}_{N-1}(x)=\frac{1+\kq+\tilde{\kq}_{N-1}\tilde{u}_{N-2}^{\vee}-1}{\tilde{u}_{N-1}^{\vee}}(x-\tilde{w}_{N-1})=\tilde{\kq}_{N-1}(x-\tilde{w}_{N-1}). \nonumber
\end{align}

We are now able to define the last Lax matrix $\tilde{L}_{N-1}$ base on dual $q$-character $\tilde{X}_{N-1}(x)$
\begin{align}
    \tilde{L}_{N-1}(x)=
    \begin{pmatrix}
    \tilde{X}_{N-1}(x)+\tilde{P}_0^+(x) & -\tilde{\kq}_{N-1}\tilde{P}_{N-1}^{-}(x) \\ \, \tilde{P}_0^+(x) & 0
    \end{pmatrix}.
\end{align}
The vectors $\Xi_{N-1}(x)$ and $\tilde{\Xi}_{N-1}(x)$ are defined based on the action of matrix $\tilde{L}_{N-1}$,
\begin{subequations}
\begin{align}
    &\Xi_{N-1}(x) := \tilde{L}_{N-1}(x)\Xi_{N-2}(x) = \begin{pmatrix} R_N(x) \\ \tilde{P}_{N-1}^+(x) R_{N-1}(x) \end{pmatrix}, \\
    & \tilde{\Xi}_{N-1}(x) := \tilde{L}_{N-1}(x) \tilde{\Xi}_{N-2}(x) = \begin{pmatrix} \tilde{R}_{N}(x) \\ \tilde{P}_{N-1}^+(x) \tilde{R}_{N-1}(x) \end{pmatrix}.
\end{align}
\end{subequations}
The polynomials $R_N(x)$ and $\tilde{R}_N(x)$ are defined by the action of the matrix $\tilde{L}_{N-1}$:
\begin{subequations}
\begin{align}
    & R_N(x) := \left(\tilde{X}_{N-1}(x) + \tilde{P}_0^+(x)\right) R_{N-1}(x) - \tilde{\kq}_{N-1}\tilde{P}_{N-1}(x) R_{N-2}(x), \\
    & \tilde{R}_{N}(x) : = \left(\tilde{X}_{N-1}(x) + \tilde{P}_0^+(x)\right) \tilde{R}_{N-1}(x) - \tilde{\kq}_{N-1} \tilde{P}_{N-1}(x) \tilde{R}_{N-2}(x).
\end{align}
\end{subequations}

With the last Lax matrix in place, the spin chain holonomy matrix ${\bf T}_{\rm SC}$(x) can be constructed by
\begin{align}
    \tilde{L}_{N-1}(x)\cdots \tilde{L}_{0}(x) =  h_0^{-1} KL_{N-1}(x)\cdots L_0(x) h_0 = h_0^{-1} {\bf T}_{\rm SC}(x) h_0
    \label{def:LSC}
\end{align}
The trace of the holonomy matrix ${\bf T}_{\rm SC}(x)=KL_{N-1}\cdots L_0$ is found by
\begin{align}
    \Tr{\bf T}_{\rm SC}(x)
    &=\Tr KL_{N-1}\cdots L_{0} \nonumber\\
    &=\Tr \tilde{L}_{N-1}(x)\cdots \tilde{L}_{0}(x) \nonumber\\
    &=\tilde{P}_0^+(x)R_{N-1}(x)+\frac{1}{\tilde{P}_0^+(x)}\left[\tilde{R}_{N}(x)-\tilde{P}_0^+(x)\tilde{R}_{N-1}(x)\right] \nonumber\\
    &= \tilde{P}_0^+(x)R_{N-1}(x)+\frac{1}{\tilde{P}_0^+(x)}\left[(1+\kq)T(x)|_{\rm SW} \tilde{P}_0^+(x) - \tilde{P}_0^+(x)^2R_{N-1}(x) \right] \nonumber\\
    &=(1+\kq)T(x)|_{\rm SW}.
    \label{eq:SC=SW}
\end{align}

Finally we will choose the gauge 
$$
    h_0 = \begin{pmatrix} u_{N-1}^\vee & 1 - u_{N-1}^\vee \\ 0 & 1 \end{pmatrix}
$$
such that the twist matrix $K(\kq)$ is of the form 
$$
    K = \begin{pmatrix} \kq+1 & -\kq \\ 1 & 0 \end{pmatrix}.
$$
We will see in the next chapter that the choice of the gauge $h_0$ allows us to identify the variables $\{\tilde{y}_{N-1,j}\}$ around the saddle points as E. Sklyanin's separated variables.

The saddle point can be now found by using the Eq.~\eqref{eq:SC=SW}.  
The LHS of Eq.~\eqref{eq:SC=SW} is a degree $N$ polynomial whose coefficients depend on the root of the $q$-characters $\{w_n\}$, the fractional couplings $\{\tilde{\kq}_n\}$, and the fundamental flavors' masses $\{\tilde{m}_n^{\pm}\}$. The RHS of Eq.~\eqref{eq:SC=SW} is a degree $N$ polynomial with coefficients $\{E_n\}$, which are conserved quantities of the $\spch$ spin chain. The coefficients of degree $N$ polynomial in $x$ in \eqref{eq:SC=SW} give rise to $N$ equations on $\{w_n,E_n,\tilde{\kq}_n,\tilde{m}_n^\pm\}$.
The $N$ unknown $\{w_n\}$ can be solved in terms of $\{E_n,\tilde{\kq}_n, m_n^\pm\}$ to obtain the saddle point configuration.

The spin chain holonomy matrix constructed from the classical limit of the wavefunction \eqref{eq: chi<0} is an open spin chain with the initital to be either
$$
\Xi_{-1}=\begin{pmatrix} 1 \\ 1 \end{pmatrix}, \quad \text{or } \quad \tilde{\Xi}_{-1}=\tilde{P}^+_{0}(x)\begin{pmatrix} 1 \\ 0 \end{pmatrix}.
$$
which is different from the periodic spin chain constructed from the defect Seiberg-Witten curve in Eq.~\eqref{def:trans XXX}. 

\subsubsection{Sklyanin's separation of variables}

The separation of variable (SoV) is a  technique in basic elementary physical/mathematical curriculum. Briefly speaking, SoV reduces multidimensional problem to a set of many one dimensional problems. 
SoV was identified to be potentially the most universal tool to solve integrable models of both classical and quantum mechanics. In particular E. Sklyanin identified the standard construction of action-angle variable from Baker-Akhiezer function as variant of SoV \cite{Sklyanin:1992eu, sklyanin1995separation} in classical integrable systems, and in many particular models can be extended to quantum counterpart. In many cases the SoV can be related to T-duality in string theory, mapping the moduli space of higher dimensional $D$-branes (which is identified with the phase space of some integrable system, e.g. Hitchin system) to the moduli space of $D0$-branes \cite{Bershadsky:1995qy,Gorsky:1999rb}. 

Classical (complexified) Hamiltonian mechanics with finite $N$ degrees of freedom is Liouville integrable (algebraically integrable) if its phase space is a $2N$-dimensional symplectic manifold equipped with $N$ independent Hamiltonians $\{H_j\}$ commuting with respect to Poisson bracket
$$
    \{ H_j, H_k \}=0, \quad j,k=1,\dots,N.
$$
In addition, one requires the level sets $J_{h} = \{ \, (x,p) \, | \, H_{j}(x,p) = h_{j} \, \}$ of $H_j$'s to be compact (algebraic varieties). 
The system of Darboux coordinates $\{x_j,p_j\}$ 
$$
    \{x_j,x_k\}=0, \, \{p_j,p_k\}=0, \, \{x_j,p_k\}=\delta_{jk}
$$
are called \emph{separated variables} if there exist $N$ relations of the form 
\begin{align}
    f_j(x_j,p_j,H_1,\dots,H_N)=0 \ , 
\end{align}
connecting $(x_j,p_j)$ on the level set $J_h$. Suppose the commutative Hamilonians $\{H_j\}$ can be obtained from some $D\times D$ \emph{Lax matrix} ${\bf T}({\xi})$ whose elements are functions on the phase space and one additional parameter $\xi$ called
the \emph{spectral parameter}. The \emph{characteristic polynomial} of the matrix ${\bf T}({\xi})$
\begin{align}
    \det (z-{\bf T}({\xi})) = \sum_{n=0}^{D} t_n({\xi}) z^{D-n}, \, t_0({\xi}) = 1, \, t_D({\xi}) = \det {\bf T}({\xi}).
\end{align}
The \emph{characteristic equation}
$$
    \det (z-{\bf T}({\xi})) = 0
$$
defines the eigenvalue $z=z({\xi})$ of the Lax matrix ${\bf T}({\xi})$. The \emph{Baker-Akhiezer function} ${\bf\psi}({\xi})$ \cite{Krichever:BA1, Krichever:BA2} is defined as the eigenvector of ${\bf T}({\xi})$
\begin{align}
    {\bf T}({\xi}){\bf\psi}({\xi}) = z({\xi}){\bf\psi}({\xi})
\end{align}
associated to the eigenvalue $z({\xi})$.
In the case of the $\spch$ spin chain,  $D=2$, and  \cite{sklyanin1995separation} shows the Lax matrix ${\bf T}(x_{j})$ takes the upper triangular form
\begin{align}
    {\bf T}(x_j) = 
    \begin{pmatrix}
    z_j & {\bf T}_{12}(x_j) \\ 0 & {\bf T}_{22}(x_j)
    \end{pmatrix}, \quad z_j = z(x_j).
    \label{eq:Sklyanin}
\end{align}
for the separated variables $(x_{j})$. 
We show now that the $\hbar \to 0$ saddle point value of the integration variables $\{\tilde{y}_{N-1,j}\}$ are the classical Sklyanin's separated variables. 
The spin chain holonomy matrix ${\bf T}_{\rm SC}$ in \eqref{def:LSC} is a $2\times 2$ matrix 
\begin{align}
    {\bf T}_{\rm SC}(x)=\begin{pmatrix} {\bf T}_{11}(x) & {\bf T}_{12}(x) \\ {\bf T}_{21}(x) & {\bf T}_{22}(x) \end{pmatrix}.
\end{align}
The diagonal elements ${\bf T}_{11}(x)$ and ${\bf T}_{22}(x)$ are degree $N$ polynomials, while the off-diagonal ${\bf T}_{12}(x)$ and ${\bf T}_{21}(x)$ are degree $N-1$ polynomials. We denote
$$
    \tilde{\xi}_n=h_0\tilde{\Xi}_{n}
$$
such that $\tilde{\xi}_{-1}$ obeys
\begin{align}
    &\tilde{\xi}_{N-1} = {\bf T}_{\rm SC}(x) \tilde{\xi}_{-1} \nonumber\\ 
    \implies & \frac{1}{{{u_{N-1}^\vee}\tilde{P}_0^{+}(x)}} \begin{pmatrix} u_{N-1}^\vee & 1-u_{N-1}^\vee \\ 0 & 1 \end{pmatrix}
    \begin{pmatrix}
    \tilde{R}_{N}(x) \\ \tilde{P}_0^+(x)\tilde{R}_{N-1}(x)
    \end{pmatrix} = 
    \begin{pmatrix} {\bf T}_{11}(x) & {\bf T}_{12}(x) \\ {\bf T}_{21}(x) & {\bf T}_{22}(x) \end{pmatrix} \begin{pmatrix} 1 \\ 0 \end{pmatrix}
\end{align}
The matrix equation above becomes an eigenvalue equation of ${\bf T}_{\rm SC}(x)$ with the eigenvector $\bigl(\begin{smallmatrix}
1 \\ 0 
\end{smallmatrix} \bigr)$ when $x=\tilde{y}_{N-1,j}$. 
Thus the variables $\{\tilde{y}_{N-1,j}\}$ are the Sklyanin's separated variables, cf. Eq.~\eqref{eq:Sklyanin}. Furthermore, at the saddle point, the set  $\{\tilde{y}_{N-1,j}\}$ is the set of $N-1$ roots of the lower-left component ${\bf T}_{21}(x)$ of the holonomy matrix 
\begin{align}
    & {\bf T}_{21}(\tilde{y}_{N-1,j}) = 0 
\end{align}
with the associated eigenvalue/conjugate momentum
\begin{align}
    {\bf T}_{11}\left( {\tilde{y}}_{N-1,j} \right) = \tilde{Y}\left( {\tilde{y}}_{N-1,j} \right) = \tilde{z}_{j} \ .
\end{align}
For the other variables $\{y_{N-1,j}\}$, we denote the dual of vector $\tilde{\xi}_{n}$ by
$$
    \xi_{n}(x) = h_0 \Xi_n(x) \ , 
$$
so that
\begin{align}
    & \xi_{N-1} = {\bf T}_{\rm SC}(x) \xi_{-1} 
    \implies \begin{pmatrix} u_{N-1}^\vee & 1-u_{N-1}^\vee \\ 0 & 1 \end{pmatrix}
    \begin{pmatrix}
    {R}_{N}(x) \\ \tilde{P}_{0}^{+}(x){R}_{N-1}(x)
    \end{pmatrix} \nonumber  = \begin{pmatrix} {\bf T}_{11}(x) & {\bf T}_{12}(x) \\ {\bf T}_{21}(x) & {\bf T}_{22}(x) \end{pmatrix} \begin{pmatrix} 1 \\ 1 \end{pmatrix}.
\end{align}
We identify  $\{y_{N-1,j}\}$, $j=1,\dots,N-1$, and $\tilde{m}_{0}^+$ are the root of 
\begin{align}
    {\bf T}_{21}(y_{N-1,j}) + {\bf T}_{22}(y_{N-1,j}) = {\bf T}_{21}(\tilde{m}_0^+) + {\bf T}_{22}(\tilde{m}_0^+) = 0.
\end{align}
The sum ${\bf T}_{22}(x)+{\bf T}_{21}(x)$ is a degree $N$ polynomial $x$. The number of its roots is the number of unknowns $\{y_{N-1,j}\}$ plus one more, which fixes $\tilde{m}_{0}^+$. The variables $y_{N-1,j}$ are not the separated variables, they do not carry the associated conjugate momenta.

To go lower in the table of integration variables , i.e. for $\{y_{n,j},\tilde{y}_{n,j}\}$, with $n=1,\dots,N-2$, we consider the truncated holonomy matrix
\begin{align}
    {\bf T}^{(n)}(x) = K_n L_n(x)\cdots L_0(x) = 
    \begin{pmatrix}
    {\bf T}_{11}^{(n)}(x) & {\bf T}_{12}^{(n)}(x) \\
    {\bf T}_{21}^{(n)}(x) & {\bf T}_{22}^{(n)}(x)
    \end{pmatrix},
\end{align}
with 
\begin{align}
    K_n = h_{0} \prod_{n=0}^n \begin{pmatrix} 1+\tilde{\kq}_n & -\tilde{\kq}_n \\ 1 & 0 \end{pmatrix} h_{0}^{-1} = h_{0} \begin{pmatrix} \tilde{u}_{n+1}^\vee  & 1-\tilde{u}_{n+1}^\vee \\ \tilde{u}_{n}^\vee & 1-\tilde{u}_{n}^\vee \end{pmatrix} h_{0}^{-1},
\end{align}
so that
\begin{subequations}
\begin{align}
    & \xi_{n} = \begin{pmatrix} u_{N-1}^\vee & 1-u_{N-1}^\vee \\ 0 & 1 \end{pmatrix}\begin{pmatrix} R_{n+1}(x) \\ \tilde{P}_{n+1}^+(x)R_n(x) \end{pmatrix} = {\bf T}^{(n)}(x)\,  \xi_{-1}, \label{xi_n} \\
    & \tilde{\xi}_{n} = \begin{pmatrix} u_{N-1}^\vee & 1-u_{N-1}^\vee \\ 0 & 1 \end{pmatrix}
    \begin{pmatrix} \tilde{R}_{n+1}(x) \\ \tilde{P}_{n+1}^+(x)\tilde{R}_n(x) \end{pmatrix} = {\bf T}^{(n)}(x)\, \tilde{\xi}_{-1}. \label{xi_n2}
\end{align}
\end{subequations}
Eq.~\eqref{xi_n2} becomes eigenvalue equation when $x=\tilde{y}_{n,j}$, which is equivalent to the condition
\begin{align}
    {\bf T}^{(n)}_{21}(\tilde{y}_{n,j}) = 0.
\end{align}
In the case of $y_{n,j}$, we again identify 
\begin{align}
    {\bf T}^{(n)}_{21}(y_{n,j}) + {\bf T}_{22}^{(n)}(y_{n,j}) = 0.
\end{align}



\section{Quantum $\spch$/SQCD correspondence} \label{sec:XXX/SQCD q}

The $\spch$/SQCD duality is known to extend to the quantum level
\cite{Nikita-Pestun-Shatashvili,Nikita-Shatashvili}, in the sense that the $T-Q$ equation underlying the functional Bethe ansatz of the spin chain can be recovered from the NS-limit of the $A_1$ gauge theory. However, the conventional use of the $T-Q$ equation is mostly for the finite dimensional spin representations at the sites of the spin chain. In this paper we make the most general claim covering all possible $\spch$ spin chains, and their wavefunctions.  We match several quantum Hamiltonians with the commuting operators, for which the surface defect expectation value is a common eigenvector, and find the formula for its wavefunction. 

Since there are different claims in the literature concerning this duality, let us briefly recall, that the Lie algebra $\mathfrak{sl}_{2}$ has infinite dimensional representations of several types: there are Verma modules ${\CalV}^{\pm}_{h}$ of the lowest or highest weights, in which the spectrum of the operator $L_0$ (in the usual basis of $L_{-}, L_{0}, L_{+}$ generators) belongs to the set $h + n$, with $n \in {\BZ}_{\geq 0}$ or $n \in {\BZ}_{\leq 0}$, respectively, with $h\in {\BC}$ being the $L_0$ eigenvalue of the vacuum vector, annihilated by $L_{-}$, or $L_{+}$, respectively. For this modules the spin $s$ of the representation, defined through the value $s(s+1)$ of the Casimir operator $L_{-}L_{+} + L_{+}L_{-} + 2 L_{0}^{2}$, is determined by $h$. 
However, there are the modules ${\CalV}_{s,a}$, which are neither of the lowest nor of the highest weight, for which $L_{0} - a \in {\BZ}$. Such a module can be represented by the densities $f(z) z^{a} dz^{-s}$, via differential operators 
\beq
L_{0} = z{\partial}_{z} - s+a, \ L_{-} = {\partial}_{z} +a/z, \ L_{+} = (2 s-a) z - z^{2} {\partial}_{z} \ .
\label{eq:sasl2}
\eeq

For generic values of $a,s$ these modules are irreducible. However, for special, quantized values of $a$ and $s$ these modules contain $\mathfrak{sl}_{2}$-invariant submodules, allowing to take the quotients. For example, ${\CalV}_{s}^{+} \subset {\CalV}_{s,0}$, ${\CalV}_{s}^{-} \subset {\CalV}_{s,2s}$, and, for integer $2s \in {\BZ}_{+}$, ${\CalV}_{s, 2s} \approx {\CalV}_{s,0}$ allowing to take quotients leading to the familiar finite dimensional representations. 

The spin chains with the spin representations of finite dimensional or Verma module type were observed to be Bethe/gauge dual to some truncated versions of the $A_1$ theory long time ago in \cite{Nikita-Shatashvili, Dorey:2011pa,HYC:2011}. These identifications require fine tuning of the masses and Coulomb parameters. 

In the present work we don't impose any relations between the masses and Coulomb moduli. 


In several cases of Bethe/gauge correspondence,  reconstruction of Hamiltonians of quantum integrable system from their corresponding gauge theory with regular surface defect is within reach. This includes the Toda lattice/$\CalN=2$ SYM correspondence, and the Calogero-Moser system/$\CalN=2^*$ theory \cite{Nikita:V,Chen:2019vvt} duality. 
 
\paragraph{}
The quantum Hamiltonians of the $\spch$ spin chain $\{\hat{h}_i\}$, $i=1,\dots,N$ are computed, in the algebraic Bethe ansatz approach, from the monodromy matrix
${\bf T}_{\rm SC}(u)$ constructed in \eqref{def:trans XXX} by promoting the $\mathfrak{sl}_2$ spins
$\{\ell_j^0,\ell_j^+,\ell_j^-\}$ to operators and replacing the classical Poisson brackets \eqref{eq:spin-poisson} by the commutators
\begin{align}
    \left[\ell^0_j,\ell_k^\pm \right] = \pm \hbar \ell_j^\pm \delta_{jk}, \quad \left[\ell_j^+,\ell_k^-\right] = 2\hbar \ell_j^0 \delta_{jk}. 
\end{align}
The spin operators $\vec{\ell}_j$ can be realized as differential operators 
\begin{align}
    \ell_j^0 = \gamma_j\beta_j - s_j\hbar, \, \ell_j^- = \beta_j, \, \ell_j^+ = 2s_j\hbar \gamma_j - \gamma_j^2\beta_j
\end{align}
with canonical coordinates $(\gamma_n,\beta_n)$ obeying the commutation relation
$$
    [\gamma_k,\beta_j]= \hbar \delta_{jk} \implies \beta_j = -\hbar \frac{\partial}{\partial\gamma_j} \ , 
$$
up to a shift of ${\beta}_{j}$ by a $\gamma$-dependent term of the form ${\partial}_{\gamma_{j}} {\sigma}$, with ${\sigma}$ some function of $\gamma$. The $a$-dependence of \eqref{eq:sasl2} is an example of such shift.

The conserved Hamiltonians of $\spch$ spin chain are computed from the trace of the monodromy matrix
\begin{align}
    & \Tr {\bf T}_{\rm SC}(u) = (1+\kq) u^{N} + \hat{h}_1(\vec{\beta},\vec{\gamma}) u^{N-1} + \hat{h}_2(\vec{\beta},\vec{\gamma}) u^{N-2} + \cdots + \hat{h}_N(\vec{\beta},\vec{\gamma}).
\end{align}
Let $h_i$ denote the eigenvalue of $\hat{h}_i$, characteristic polynomial of monodromy matrix is
\begin{align}
    T(u)|_{\rm SC} = (1+\kq) u^{N} + h_1 u^{N-1} + h_2 u^{N-2} + \cdots + h_N.
\end{align}
The Casimir operator (quantum determinant) of the quantum $\spch$ spin chain is defined by
\begin{align}
    \Tr
    \left[{\bf T}_{\rm SC}(u) \wedge {\bf T}_{\rm SC}(u+{\hbar}) \right]= \prod_{\omega}(x-\mu_{\omega}+s_{\omega}{\hbar} + {\hbar}) (x-\mu_{\omega} -s_{\omega}{\hbar}) = P(x).
\end{align}
We identify the fundamental flavor masses $m_\omega^\pm$ with the spins $s_\omega$ and the inhomogeneities $\mu_{\omega}$ by
\begin{align}
    m_{\omega+1}^{+} + {{\hbar}} = \mu_{\omega} - s_{\omega}{\hbar} , \quad m_{\omega}^- = \mu_{\omega} + s_{\omega}{\hbar}
\end{align}

\subsection{Hamiltonians from the nonperturbative Dyson-Schwinger equation: from bulk to surface}

The $\Omega$-background supersymmetry protected gauge theory observables are also evaluated by the effective statistical mechanical system.  
We denote by $\CalO$ both the observable in supersymmetric gauge theory, and the statistical model observable to which it reduces thanks to the localization. 
Let $\CalO[\vec{\lambda}]$ be the corresponding evaluation at the state $\vec\lambda$. The expectation value of the observable $\CalO$ is computed  by
the average
\begin{align}
    \langle \CalO \rangle = \frac{1}{\CalZ({\ba},{\bm}^\pm,\kq,\vec{\epsilon})} \sum_{\vec{\lambda}}\kq^{|\vec{\lambda}|}\CalO[\vec{\lambda}]\CalZ({\ba},{\bm}^\pm,\vec{\epsilon})[\vec{\lambda}].
\end{align}
In particular, the $Y(x)$-observable, which is a local observable defined as the regularized characteristic polynomial of the adjoint scalar $\Phi$ in the vector multiplet evaluated at the origin $0\in\BC^2$
\begin{align}
    Y(x) = x^N \exp\left[\sum_{l=1}^\infty -\frac{1}{lx^l}\Tr \, \Phi^l \right].
    \label{def:Y}
\end{align}
reduces to the statistical mechanical observable, whose evaluation $Y(x)[{\vec\lambda}]$ computes as:
\begin{align}
    Y(x)[\vec{\lambda}] 
    & = \mathbb{E}\left[-e^x\tilde{S}^*\right]
\end{align}
The Ref. \cite{Nikita:I} introduced the fundamental $qq$-character observable
\begin{align}
    \CalX(x)[\vec{\lambda}] = Y(x+\epsilon_+)[\vec{\lambda}] + \kq \frac{P(x)}{Y(x)[\vec{\lambda}]}, \quad P(x) = \prod_{f=1}^N (x-m_f^+)(x-m_f^-)
    \label{eq:fundqq}
\end{align}
whose expectation value is shown to be a degree $N$ polynomial in $x$:
\begin{align}
    \langle \CalX(x) \rangle  = (1+\kq)T(x) = h_0x^N + h_1x^{N-1} + \cdots + h_N. \label{maintheorem}
\end{align}
The nonperturbative Dyson-Schwinger equations are the vanishing of the coefficients of the negative powers of $x$ in the Laurent expansion  $\CalX(x)$:
\begin{align}
    \left[x^{-I}\right]\langle \CalX(x) \rangle = 0 , \quad I=1,2,\dots.
\end{align}
See appendix \ref{sec:basic} for the derivation of the $qq$-character $\CalX(x)$ \eqref{eq:fundqq}.

In this paper the co-dimension two surface defect is introduced using the orbifold construction, as in \cite{Nikita:III, Nikita:IV}. Details can be found in appendix \ref{sec:defect}. The fundamental $qq$-character \eqref{eq:fundqq} splits into $N$ orbifolded fundamental $qq$-characters:
\begin{align}
    \mathcal{X}_\omega(x)[\vec{\lambda}]=Y_{\omega+1}(x+\epsilon_+)[\vec{\lambda}]+\kq_\omega\frac{P_\omega(x)}{Y_\omega(x)[\vec{\lambda}]}, \quad \omega=0,\dots,N-1. 
    \label{eq:DS-frac}
\end{align}
with $P_{\omega}(x) = (x-m_{\omega}^+)(x-m_{\omega}^-)$. The orbifolded $qq$-character $\CalX_{\omega}(x)$ obeys the same non-perturbative Dyson-Schwinger equation
$$
    \langle \CalX_{\omega}(x) \rangle = (1+\kq_{\omega}) t_{\omega}(x) = (1+\kq_{\omega}) (x - \rho_{\omega}).
$$
in other words, 
\begin{align}
    \langle [x^{-I}] \CalX_\omega(x) \rangle = 0\ , \qquad I > 0
    \label{eq:X_w=0}
\end{align}
The expectation value in the presence of a co-dimension two surface defect is defined as an average over the orbifolded pseudo-measure
\begin{align}
    \langle \CalO \rangle = \frac{1}{\CalZ^{\rm defect}({\ba},{\bm}^\pm,\vec{\kq},\vec{\epsilon})} \sum_{\vec{\lambda}}\prod_{\omega=0}^{N-1}\kq_\omega^{k_{\omega}}\CalO[\vec{\lambda}]\CalZ^{\rm defect}({\ba},{\bm}^\pm,\vec{\epsilon})[\vec{\lambda}].
\end{align}

To evaluate \eqref{eq:DS-frac}, we consider the expansion of the fractional $Y_{\omega}(x)$ function \eqref{eq:Y large x} in $x$:
\begin{align}
    Y_{\omega}(x)=(x-a_{\omega})\exp\left[\frac{\epsilon_1}{x}\nu_{\omega-1}+\sum_{I=1}^\infty \frac{\epsilon_1 D_{\omega-1}^{(I)}}{x^{I+1}}\right]
    \label{eq:Y large x}
\end{align}
where
\begin{align}
    D_{\omega}^{(I)} = \sigma_{\omega}^{(I)} - \sigma_{\omega+1}^{(I)} + \sum_{i=1}^{I}
    \begin{pmatrix} I \\ i \end{pmatrix} \left(\frac{\epsilon_2}{N}\right)^i \sigma_{\omega}^{(I-i)}, \quad \sigma_{\omega}^{(I)} 
    & = \sum_{(\alpha,\Box)\in K_{\omega-1}} \sum_{j=0}^{I} \begin{pmatrix} I \\ j \end{pmatrix} \frac{\epsilon_1^{I-j}}{I+1-j}(c_{\Box})^j
\end{align}
and $c_\Box = a_\alpha + ({\rm i}-1)\epsilon_1 + ({\rm j}-1) \epsilon_2$. The bulk $Y(x)$ function give rise to infinitely many bulk gauge invariant chiral ring observables $D^{(I)}$'s
\begin{align}
    Y(x) = \prod_{\omega}Y_{\omega}(x) = \left[\prod_{\omega}(x-a_\omega)\right] \times \exp 
    \left[ \sum_{I=1}^\infty \epsilon_1\frac{D^{(I)}}{x^{I+1}} \right], \quad D^{(I)} = \sum_{\omega} D^{(I)}_{\omega}
\end{align}

Let us define the observable $\CalU(x)$ as a linear combination of the fractional $qq$-characters $\CalX_{\omega}(x)$,
\begin{align}
    \label{def:U}
    \mathcal{U}(x)= \sum_{\omega \in {\BZ}_{N}} u_\omega\mathcal{X}_\omega(x)
\end{align}
with the linear combination coefficients $\{u_{\omega}\}$ given by
\begin{align}
    & u_\omega=1+\kq_{\omega+1}+\kq_{\omega+1}\kq_{\omega+2}+\cdots+\kq_{\omega+1}\cdots\kq_{\omega+N-1} \nonumber\\
    \implies & u_{\omega}-\kq_{\omega+1}u_{\omega+1}=1-\kq \quad \forall \, \omega=0,\dots,N-1.
    \label{eq:c_w}
\end{align} 
As a linear combination of the fractional $qq$-characters, the observable $\CalU(x)$ also satisfies the non-perturbative Dyson-Schwinger equation
\begin{align}
    \langle [x^{-I}]\CalU(x) \rangle = 0, \quad I=1,2,\dots. \nonumber
\end{align}

The choice of the coefficients $\{u_\omega\}$ ensures that $[x^{-I}]\CalU(x)$ always consists one bulk gauge invariant chiral ring observable $(1-\kq)\epsilon_1D^{(I)}$. 
The $I$-th Hamiltonian is defined as differential operator w.r.t variables $\{z_{\omega}\}$ acting on the surface defect partition function,
\begin{align}
    \frac{1}{\CalZ^{\rm defect} (\vec{z},{\ba},{\bm}^\pm,\vec{\kq},\vec{\epsilon})} \left[ \hat{\rm H}_I\CalZ^{\rm defect}(\vec{z},{\ba},{\bm}^\pm,\vec{\kq},\vec{\epsilon}) \right] := \langle - \left[ x^{-I+1} \right] \CalU(x) + (1-\kq)  \epsilon_1 D^{(I-1)} \rangle .
    \label{def:Hi}
\end{align}
which translates to an Schr\"{o}dinger-type equation of surface defect partition function $\CalZ^{\rm defect}$
\begin{align}
    \hat{\rm H}_I \CalZ^{\rm defect} (\vec{z},{\ba},{\bm}^\pm,\vec{\kq},\vec{\epsilon}) = (1-\kq) \langle \epsilon_1 D^{(I-1)} \rangle \CalZ^{\rm defect}(\vec{z},{\ba},{\bm}^\pm,\vec{\kq},\vec{\epsilon}).
    \label{eq:H-eigen}
\end{align}
The fact that all the Hamiltonians defined this way share the surface defect partition function as their common eigenfunction, all the Hamiltonians are mutually commuting.
\paragraph{}

We now evaluate the vacuum expectation value of the chiral ring observable $\langle D^{(I)} \rangle$  in the limit $\epsilon_2 \to 0$ with $\epsilon_1\equiv \epsilon=\hbar$ fixed (the so-called NS limit of \cite{Nikita-Shatashvili}). In the NS-limit, the 
four dimensional $\CalN=2$ theory effectively becomes $\CalN=(2,2)$ two dimensional theory, with the worldsheet $\mathbb{C}_2^1$. 
Such theory are known to be in correspondence with quantum integrable systems \cite{Nikita-Shatashvili,Nekrasov:2009uh,Nekrasov:2009ui}. 
\begin{align}
    D^{(I)} = \sum_{\omega=0}^{N-1} D_{\omega}^{(I)} = \sum_{\omega=0}^{N-1} \left[ \sum_{i=1}^{I}
    \begin{pmatrix} I \\ i \end{pmatrix} \left(\frac{\epsilon_2}{N}\right)^i \sigma_{\omega}^{(I-i)} \right] = \sum_{i=1}^I \begin{pmatrix} I \\ i \end{pmatrix} \left(\frac{\epsilon_2}{N}\right)^i \left[ \sum_{\omega=0}^{N-1}\sigma_{\omega}^{(I-i)} \right].
\end{align}
The summation over $\omega$ can be rearranged by
\begin{align}
    \sum_{\omega=0}^{N-1}\sigma_{\omega}^{(I-i)} = D_{0}^{(I-i)} + 2D_{1}^{(I-i)} + \cdots + (N-1)D_{N-2}^{(I-i)} + N \sigma_{N-1}^{(I-i)} + \CalO(\epsilon_2).
\end{align}
Contributions from the $D^{(I-i)}$'s are killed by the $\epsilon_2^i$ factor in the NS-limit $\epsilon_2\to0$, along with $\CalO(\epsilon_2)$ terms. The remaining $\{\sigma_{N-1}^{(i)}\}_{i=1}^I$ comes from the bulk
\begin{align}
    \sigma_{N-1}^{(I)} 
    & = \sum_{(\alpha,\Box)\in K_{N-1}} \sum_{j=0}^{I} \begin{pmatrix} I \\ j \end{pmatrix} \frac{\epsilon_1^{I-j}}{I+1-j}(c_{\Box})^j = \sum_{(\alpha,i,j)\in \vec{\Lambda}}\frac{1}{\epsilon_1} \left[ \frac{(c_\Box+\epsilon_1)^{I+1}}{I+1} -   \frac{c_{\Box}^{I+1}}{I+1} \right].
\end{align}

$\vec{\Lambda}$ is the limit-shape bulk Young diagrams defined in Eq.~\eqref{def:bulkyoung} in the NS-limit. Summation over $j$ can be approximated by integration in the NS-limit $\epsilon_2\to0$:
\begin{align}
    \sigma_{N-1}^{(I)}
    & = \sum_{\alpha, i} \frac{1}{\epsilon_1\epsilon_2} \int_0^{\xi_{\alpha i}} \, dw \left[\frac{(x_{\alpha i}^{(0)} + \epsilon_1 + w)^{I+1}}{I+1} - \frac{(x_{\alpha i}^{(0)}+w)^{I+1}}{I+1}\right], \quad x_{\alpha i}^{(0)} \equiv a_{\alpha} + (i-1)\epsilon_1 \\
    & = \sum_{\alpha,i} \frac{1}{\epsilon_1\epsilon_2}\frac{1}{(I+1)(I+2)} \left[ (x_{\alpha i}^{(0)} + \epsilon_1 + \xi_{\alpha_i})^{I+2} - (x_{\alpha i}^{(0)} + \epsilon_1 )^{I+2} - (x_{\alpha i}^{(0)} + \xi_{\alpha_i})^{I+2} + (x_{\alpha i}^{(0)})^{I+2}\right] \nonumber\\
    & = \sum_{\alpha, i} \frac{1}{\epsilon_1\epsilon_2}\frac{1}{(I+1)(I+2)} \left[ (x_{\alpha i}^{(0)} + \epsilon_1 + \xi_{\alpha_i})^{I+2} - (x_{\alpha i}^{(0)} + \xi_{\alpha_i})^{I+2} \right] + \sum_{\alpha} \frac{1}{\epsilon_1\epsilon_2}\frac{a_{\alpha}^{I+2}-A_\alpha^{I+2}}{(I+1)(I+2)}. \nonumber
\end{align}

In the NS-limit, the surface defect partition function has the asymptotics ~\eqref{def:psi-asymptotic}. Eq.~\eqref{eq:H-eigen} becomes an eigenvalue equations of the normalized vev of the surface defect $\Psi+{\ba,+}$
\begin{align}
    \hat{\rm H}_I \Psi_{\ba,+}({\ba},{\bm}^\pm,\bz;\kq) = (1-\kq) E_I({\ba},{\bm}^\pm;\kq) \Psi_{\ba,+}({\ba},{\bm}^\pm,\bz;\kq)
    \label{eq:chi-eigen}
\end{align}
with the eigenvalues coincide with the expectation value of the bulk gauge invariant chiral ring observables
\begin{align}
    E_I({\ba},{\bm}^\pm;\kq) 
    & =  \langle \epsilon_1 D^{(I-1)} \rangle  =  \epsilon_1 \frac{\epsilon_2}{N} (I-1) \cdot N\sigma_{N-1}^{(I-2)} \nonumber\\
    & = \sum_{\alpha, i}\frac{1}{I} \left[ (x_{\alpha i}^{(0)} + \epsilon_1 + \xi_{\alpha_i})^{I} - (x_{\alpha i}^{(0)} + \xi_{\alpha_i})^{I}\right] + \sum_{\alpha}\frac{a_{\alpha}^I-A_{\alpha}^I}{I} \\
    & = \sum_{\alpha, i}\frac{1}{I} \left[ (x_{\alpha i}^{(0)} + \epsilon_1 + \xi_{\alpha_i})^{I} - (x_{\alpha i}^{(0)} + \xi_{\alpha_i})^{I}\right] + \sum_{\alpha}\frac{(m_{\alpha}^+)^I-A_{\alpha}^I}{I} + E_{I}^{(0)}. \nonumber
\end{align}
By resetting the ground state energy, we may set $E_I^{(0)}=0$.
\begin{align}
    E_I = \sum_{\alpha, i}\frac{1}{I} \left[ (x_{\alpha i}^{(0)} + \epsilon_1 + \xi_{\alpha i})^{I} - (x_{\alpha i}^{(0)} + \xi_{\alpha i})^{I}\right] + \sum_{\alpha}\frac{(m_{\alpha}^+)^I-A_{\alpha}^I}{I}.
    \label{eq:E_Izero}
\end{align}
The generating function of the eigenvalues reads
\begin{align}
    \sum_{I=1}^{\infty} u^{-I} E_I 
    & = \sum_{I=1}^\infty \sum_{\alpha, i} \frac{u^{-I}}{I} \left[ (x_{\alpha i}^{(0)} + \epsilon_1 + \xi_{\alpha_i})^{I} - (x_{\alpha i}^{(0)} + \xi_{\alpha_i})^{I+1}\right] + \sum_{\alpha}\frac{u^{-I}}{I}\left[ (m_{\alpha}^+)^I-A_{\alpha}^I \right] \nonumber\\
    & = \sum_{\alpha, i} \log \left( 1 - \frac{x_{\alpha i}^{(0)}+\xi_{\alpha i}}{u} \right) - \log \left( 1 - \frac{x_{\alpha,i}^{(0)} + \epsilon_1 +\xi_{\alpha i}}{u} \right) + \sum_{\alpha} \log \left( 1 - \frac{A_{\alpha}}{u} \right) - \log \left( 1- \frac{m_\alpha^+}{u}\right) \nonumber\\
    & = \log \frac{Y(u)}{P_+(u)}
\end{align}
with  function $Y(u)$ defined based on the limit-shape Young diagram $\vec{\Lambda}$ in Eq.~\eqref{eq:Y-limit}. 
\paragraph{}
The Eq.~\eqref{eq:chi-eigen} is our main application of the power of exact calculations in gauge theory: The normalized vev of the surface defect
in the NS-limit is the eigenfunction of corresponding quantum integrable model. More precisely, it is the Jost function, namely, 
it is a suitably dressed scattering state, approaching the plane wave in one of the weak coupling corners of the parameter space. 

We denote the exponentiated generating function of the expectation value of chiral ring operators by
$$
    G(u) = \exp \left[ \sum_{I=1}^{\infty} u^{-I} E_I \right].
$$
The relation between conserving charges of gauge theory and the spin chain counter part is established using bulk T-Q equation \eqref{Baxter Q Hyper},
\begin{align}
    P_+(u+\hbar)G(u+\hbar) + \kq P_-(u) G(u)^{-1} = T(u)|_{\rm SC}=(1+\kq)x^N + h_1x^{N-1} + \cdots + h_N. 
    \label{eq:qq=SC}
\end{align}

The operator versions of generating function $G(u)$
\begin{align}
    &{\bf G}_+(u) := \exp \left[\sum_{I=1}^{\infty}u^{-I} \frac{\hat{\rm H}_I}{1-\kq}\right], \quad {\bf G}_-(u) := \exp \left[\sum_{I=1}^{\infty}u^{-I} \frac{\hat{\rm H}_I}{\kq-1}\right].
\end{align}
give the operator version of Eq.~\eqref{eq:qq=SC}
\begin{align}
    P_+(u+\hbar) {\bf G}_+(u+\hbar) + \kq P_-(u) {\bf G}_-(u) = \Tr {\bf T}_{\rm SC}(u).
    \label{eq:qq=SC2}
\end{align}

In the next couple of sections, we will verify the validity of Eq.~\eqref{eq:qq=SC2} in the first three quantum Hamiltonians $\hat{\rm H}_2$, $\hat{\rm H}_3$, and $\hat{\rm H}_4$. Along the way, $\hat{\rm H}_1$ comes as a welcome bonus.

\subsection{Second Hamiltonian}

The second Hamiltonian of the $\spch$ spin chain $\hat{h}_2$ can be expressed in terms of the coordinate systems $\gamma_{\omega}$ and $\beta_{\omega}$ established in Eq.\eqref{eq:goodgamma},
\begin{align}
    \hat{h}_2
    =&\Tr \begin{pmatrix} \kq & 0 \\ 0 & 1 \end{pmatrix} \sum_{\omega>\omega'} (-\mu_{\omega}+\CalL_{\omega})(-\mu_{\omega'}+\CalL_{\omega'}) \\
    =&\sum_{\omega>\omega'}(\kq\gamma_{\omega'}-\gamma_{\omega})(\gamma_{\omega}-\gamma_{\omega'})\beta_{\omega}\beta_{\omega'} + \sum_{\omega>\omega'} -\kq (m_{\omega}^-\gamma_{\omega'}\beta_{\omega'}+m_{\omega'}^-\gamma_{\omega}\beta_{\omega})+ \kq (m_{\omega'}^- - m_{\omega'}^+ + \epsilon )\gamma_{\omega'}\beta_{\omega} \nonumber\\
    & + \sum_{\omega>\omega'}  (m_{\omega+1}^{+}-\epsilon)\gamma_{\omega'}\beta_{\omega'}+(m_{\omega'+1}^+-\epsilon)\gamma_{\omega}\beta_{\omega}+(m_{\omega}^--m_{\omega+1}^+ + \epsilon)\gamma_{\omega}\beta_{\omega'} + \kq m_{\omega}^- m_{\omega'}^- + (m_{\omega+1}^+-\epsilon)(m_{\omega'+1}^+-\epsilon) \nonumber
\end{align}
such that
\begin{align}
    \hat{h}_2 \tilde{\Psi} = h_2 \tilde{\Psi}
\end{align}
where $\tilde{\Psi}$ is the properly normalized vev of the surface defect multiplying with a perturbative factor: 
\begin{align}
    \tilde{\Psi}=\left(\prod_{\omega=0}^{N-1}z_{\omega}^{-\frac{a_{\omega+1}-m_{\omega+1}^+}{\hbar}}\right) \Psi_{\ba,+}.
\end{align}
This factor is responsible for the appearence of the $a$-dependence in addition to the spins and inhomogeneities. 

The identification between the spin chain canonical coordinates $\{\gamma_{\omega},\beta_{\omega}\}$ and the surface defect gauge theory parameters $\{z_{\omega}\}$ are
\begin{align}
    u_{\omega}z_{\omega} = (\kq-1) \gamma_{\omega}, \quad \beta_{\omega} = -\hbar \frac{\partial}{\partial \gamma_{\omega}} = -\hbar \left( \frac{\partial}{\partial z_{\omega-1}} - \frac{\partial}{\partial z_{\omega}} \right).
\end{align}
where the coefficients $u_{\omega}$'s are defined in Eq.~\eqref{eq:c_w}.

To extract the Hamiltonians from non-perturbative Dyson-Schwinger equations  we consider the observable $\CalU(x)$ defined as a linear combination of fractional $qq$-character $\CalX_{\omega}(x)$ \eqref{eq:DS-frac}. The second Hamiltonian defined through \eqref{def:Hi} after resetting the zero point energy \eqref{eq:E_Izero} is 
\begin{align}
    \hat{{\rm H}}_2
    &=\sum_{\omega}-\frac{(1-\kq)}{2}\left[(\hbar\nabla^z_{\omega})^2 - 2m_{\omega+1}^{+}\hbar\nabla^z_{\omega} \right]-(\kq_{\omega+1} u_{\omega+1})\left(\hbar\nabla^z_\omega\right)\left(\hbar\nabla^z_\omega-m_{\omega+1}^{+}+m_{\omega+1}^{-} \right)
\end{align}
such that $\Psi$ is eigenfunction of $\hat{\rm H}_2$ with eigenvalue to expectation value of chiral ring operator
\begin{align}
    \hat{\rm H}_2\tilde{\Psi} = (1-\kq)E_2 \tilde{\Psi} = (1-\kq) \langle \hbar D^{(1)} \rangle \tilde{\Psi}.
\end{align}
See appendix \ref{sec:defect} for derivation detail of $\hat{\rm H}_2$.

The non-perturbative Dyson-Schwinger equation does not give a definition of the first Hamiltonian. Instead we simply define
\begin{align} \label{def:H1}
    \hat{\rm H}_1 = (\kq-1)\hbar\nabla_{c}^{z} = (\kq-1) \sum_{\omega=0}^{N-1} \hbar\nabla_{\omega}^z.
\end{align}
In particular, this definition agrees with the Eq.~\eqref{eq:qq=SC2}, with the first $\spch$ spin chain Hamiltonian given by
\begin{align}
    \hat{h}_1 & = \Tr \begin{pmatrix} \kq & 0 \\ 0 & 1 \end{pmatrix} \sum_{\omega} (-\mu_{\omega} + \CalL_{\omega}) \nonumber \\
    & = \sum_{\omega} (\kq-1) \gamma_{\omega}\beta_{\omega} - (\kq-1) s_{\omega}{\hbar}  - (\kq+1)\mu_{\omega} \nonumber\\
    & = \hat{\rm H}_1  - \kq m_{c}^- - (m_{c}^+ - N{\hbar}). \label{eq:h1=H1}
\end{align}
where $m_c^\pm = \sum_{\omega} m_{\omega}^\pm$.
The relation between the second Hamiltonian of the $\spch$ spin chain $\hat{h}_2$ and the gauge theory $\hat{H}_2$ is found by 
\begin{align}
    \hat{h}_2 = & \hat{\rm H}_2 + \frac{1+\kq}{2}\left(\frac{\hat{\rm H}_1}{\kq-1}\right)^2 + (N-1)\hbar\left(\frac{\hat{\rm H}_1}{1-\kq}\right) \nonumber\\
    & +(\kq m_c^- - m_c^+)\left(\frac{\hat{\rm H}_1}{1-\kq}\right)+ \sum_{\omega>\omega'} \kq m_{\omega}^-m_{\omega'}^- + (m_{\omega}^+-\hbar) (m_{\omega'}^+-\hbar)
    \label{eq:h2=H2}
\end{align}
Eq.~\eqref{eq:h2=H2} agrees with Eq.~\eqref{eq:qq=SC2}. The details can be found in appendix \ref{sec:h2=H2}. 


\subsection{Third Hamiltonian}

The third Hamiltonian $\hat{h}_3$ of the $\spch$ spin chain is
\begin{align}
    \label{eq:h3}
    \hat{h}_3 
    =& \Tr \begin{pmatrix} \kq & 0 \\ 0 & 1 \end{pmatrix} \sum_{\omega_1>\omega_2>\omega_3} (-\mu_{\omega_1}+\CalL_{\omega_1})(-\mu_{\omega_2}+\CalL_{\omega_2})(-\mu_{\omega_3}+\CalL_{\omega_3}) 
\end{align}

On gauge theory counter part, the third Hamiltonain is defined through Eq.~\eqref{def:Hi} with $I=2$. After resetting the zero point energy \eqref{eq:E_Izero}, the third Hamiltonian $\hat{\rm H}_3$ reads 
\begin{align}\label{eq:H3}
    \hat{\rm H}_3 = 
    & (\kq-1)\left[\sum_{\omega}\frac{1}{6}(\hbar\nabla^z_{\omega}-m_{\omega+1}^+)^3 - \frac{1}{3}(m_{\omega+1}^+)^3 \right]  \\
    & + \sum_{\omega} \hbar u_{\omega} \left[ \frac{(\hbar\nabla^z_{\omega}-m_{\omega+1}^+)^2}{2} - \frac{a_{\omega+1}^2}{2} + \hbar \langle D^{(1)}_{\omega} \rangle \right] \nonumber\\
    & -(u_{\omega}+\kq_{\omega+1}u_{\omega+1})\left[ \langle (\hbar\nabla^z_{\omega}-m_{\omega+1}^+)\hbar D^{(1)}_{\omega} \rangle - \frac{a_{\omega+1}^2}{2}(\hbar\nabla^z_{\omega}-m_{\omega+1}^+)  \right]   \nonumber
    \\
    &+\kq_{\omega+1}u_{\omega+1}\left[ (m_{\omega+1}^+ + m_{\omega+1}^-)\left(\frac{(\hbar\nabla^z_{\omega})^2}{2}+\frac{a_{\omega+1}^2}{2} - \hbar \langle D^{(1)}_{\omega} \rangle\right) -(m_{\omega+1}^+)^2\hbar\nabla^z_{\omega} + \frac{(m_{\omega+1}^+)^2}{2}(m_{\omega+1}^+-m_{\omega+1}^-) \right], \nonumber
\end{align}
such that
\begin{align}
    \hat{\rm H}_3 \tilde{\Psi} = (1-\kq) \langle \sum_{\omega} \hbar D_{\omega}^{(2)} \rangle \tilde{\Psi} = (1-\kq)E_3({\ba},{\bm}^\pm;\kq) \tilde{\Psi}.
\end{align}
The details of the construction of $\hat{\rm H}_3$ can be found in appendix \ref{sec:h3=H3}.

The third Hamiltonian $\hat{\rm H}_3$ consists $\langle D^{(1)}_{\omega} \rangle$ terms, which can be rewrite as a proper differential operator 
using the Dyson-Schwinger equations from $[x^{-1}]$ in Eq.~\eqref{eq:x-1coeff},
\begin{align}
    \hbar\langle D^{(1)}_{\omega} \rangle
    = & \frac{1}{\kq-1}\left[\sum_{n=0}^{N-1}\kq_{\omega}\cdots\kq_{\omega-n+1}\hbar\nabla^z_{\omega-n}(\hbar\nabla^z_{\omega-n}-m_{\omega-n+1}^++m_{\omega-n+1}^-)\right] \nonumber\\
    & + \frac{(\hbar\nabla^z_{\omega})^2}{2} + m_{\omega+1}^-\hbar\nabla^z_{\omega} + \frac{a_{\omega+1}^2}{2} - \frac{(m_{\omega+1}^+)^2}{2},
    \label{eq:D_w}
\end{align}
and
\begin{align}
    \hbar\langle \nabla^z_{\omega} D^{(1)}_{\omega} \rangle = 
    & \hbar\langle D^{(1)}_{\omega} \rangle \nabla^z_{\omega} + \hbar( \nabla^z_{\omega}\langle D^{(1)}_{\omega} \rangle). \nonumber
\end{align}
Eq.~\eqref{eq:H3} can now be defined properly as a third order differential operator in $z_{\omega}$ acting on the normalized vev of the surface defect $\Psi_{\ba,+}$.
After walking through the tedious calculation, we find a non-trivial relation between the $\spch$ spin chain Hamiltonain $\hat{h}_3$ and its gauge counter part $\hat{\rm H}_3$:
\begin{align}
    \hat{h}_3 =
    & \, \hat{\rm H}_3 + (1+\kq)\frac{\hat{\rm H}_2}{1-\kq} \frac{\hat{\rm H}_1}{1-\kq} + \frac{1-\kq}{6} \left(\frac{\hat{\rm H}_1}{1-\kq}\right)^3 - 2 \hbar \left(\frac{1}{2}\left(\frac{\hat{\rm H}_1}{1-\kq} \right)^2  +  \frac{\hat{\rm H}_2}{1-\kq}  \right) \nonumber\\
    & + (N\hbar - m_c^+) \left(\frac{1}{2}\left(\frac{\hat{\rm H}_1}{1-\kq} \right)^2 +  \frac{\hat{\rm H}_2}{1-\kq}  \right) - (\kq m_c^-) \left(\frac{1}{2}\left(\frac{\hat{\rm H}_1}{1-\kq} \right)^2 -  \frac{\hat{\rm H}_2}{1-\kq}  \right) \nonumber \\
    & + \left[\sum_{n>n'} (\hbar - m_{n}^+)(\hbar - m_{n'}^+) - \kq m_{n}^-m_{n'}^-\right]\frac{\hat{\rm H}_1}{1-\kq}  - \hbar(N\hbar - m_c^+)\frac{\hat{\rm H}_1}{1-\kq} + \hbar^2 \frac{\hat{\rm H}_1}{1-\kq} \nonumber\\
    & + \sum_{\omega_1>\omega_2>\omega_3} (\hbar-m_{\omega_1}^+)(\hbar-m_{\omega_2}^+)(\hbar-m_{\omega_2}^+) - \kq m_{\omega_1}^-m_{\omega_2}^-m_{\omega_3}^-.
    \label{eq:h3=H3}
\end{align}
which again agrees with Eq.~\eqref{eq:qq=SC2}. Details can be found in appendix \ref{sec:h3=H3}.

\subsection{Fourth Hamiltonian and second order $qq$-character}
Finally we will briefly demonstrate the relation between the fourth Hamiltonian of the $\spch$ spin chain $\hat{h}_4$ and the gauge theory counter part $\hat{\rm H}_4$. In particular the necessity of considering higher rank $qq$-characters for a proper definition of $\hat{\rm H}_4$ as a degree four differential operator. This also extends to potentially any $\hat{\rm H}_I$ with $I>4$.
The fourth Hamiltonian of spin chain $\hat{h}_4$ is
\begin{align}\label{eq:h4}
    \hat{h}_4 
    = & \Tr K \sum_{\omega_1>\omega_2>\omega_3>\omega_4} (-\mu_{\omega_1}+\CalL_{\omega_1}) (-\mu_{\omega_2}+\CalL_{\omega_2}) (-\mu_{\omega_3}+\CalL_{\omega_3}) (-\mu_{\omega_4}+\CalL_{\omega_4}) 
\end{align}

The fourth Hamiltonian $\hat{\rm H}_4$ is defined by Eq.~\eqref{def:Hi}:
\begin{align}
    \hat{\rm H}_4 = (\kq-1)\left[\sum_{\omega}\frac{\hbar^3}{2} \langle (\nabla^z_{\omega})^2 D^{(1)}_{\omega} \rangle \right] - \sum_{\omega} (u_{\omega}+\kq_{\omega+1}u_{\omega+1})\left[\frac{(\hbar\nabla^z_{\omega})^4}{4!} + \frac{\hbar^2}{2} \langle (D^{(1)}_{\omega})^2 \rangle + \hbar^2 \langle \nabla^z_{\omega} D_{\omega}^{(2)} \rangle \right] + \cdots ,
    \label{eq:H4}
\end{align}
so that the normalized vev of the surface defect is also an eigenfunction of $\hat{\rm H}_4$: 
\begin{align}
    \hat{\rm H}_4 \tilde{\Psi} = (1-\kq)\langle \hbar D^{(3)} \rangle \tilde{\Psi} = (1-\kq) E_4 \tilde{\Psi}.
\end{align}
To have $\hat{\rm H}_4$ as a properly defined differential operator acting on the $\Psi_{\ba,+}$, we need to rewrite the $\langle (D^{(1)}_{\omega})^2 \rangle$ and $\langle \nabla^z_{\omega} D_{\omega}^{(2)} \rangle$ similarly to what was done for the $\langle D^{(1)}_{\omega} \rangle$ in Eq.~\eqref{eq:D_w}. The expectation value of $\langle (\nabla^z_{\omega})^2D_{\omega}^{(1)}\rangle$ follows a similar procedure as for the $\langle \nabla^z_{\omega}D_{\omega}^{(1)} \rangle$. The expectation value of $\langle D^{(2)}_{\omega} \rangle$ can be derived using the Dyson-Schwinger equation in Eq.~\eqref{eq:[x-2]X} (see appendix \ref{sec:h4=H4} for detail). 
It is much complicated in the case of $\langle (D^{(1)}_{\omega})^2 \rangle$. It turns out that we need to consider the second order $qq$-character $\CalX^{(2)}(x)$:
\begin{align}
    \CalX^{(2)}(x;\nu)[\vec{\lambda}] =
     & Y(x+\epsilon_+)[\vec{\lambda}]Y(x+\nu+\epsilon_+)[\vec{\lambda}] + \kq R(\nu)Y(x+\nu+\epsilon_+)[\vec{\lambda}]\frac{P(x)}{Y(x)[\vec{\lambda}]} \nonumber\\
     & + \kq R(-\nu) Y(x+\epsilon_+)[\vec{\lambda}] \frac{P(x+\nu)}{Y(x+\nu)[\vec{\lambda}]} + \kq^2 \frac{P(x)P(x+\nu)}{Y(x)[\vec{\lambda}]Y(x+\nu)[\vec{\lambda}]}
     \label{eq:2ndqq1}
\end{align}
with one additional parameter $\nu\in\BC$ and
$$
    R(\nu) = \frac{(\nu+\epsilon_1)(\nu+\epsilon_2)}{\nu(\nu+\epsilon_+)}
$$
The main statement of \cite{Nikita:I} claims that the expectation value of $\CalX^{(2)}(x)$ is a polynomial of degree $2N$
\begin{align}
    \langle \CalX^{(2)}(x) \rangle = \sum_{n=0}^{2N} f_n x^{2N-n}.
\end{align}
The introduction of a regular co-dimension two surface defect splits the second order $qq$-character \eqref{eq:2ndqq1} into $N^2$ fractional $qq$-characters 
\begin{align}
    \CalX_{\omega_1,\omega_2}^{(2)}(x;\nu)=
    & \, Y_{{\omega}_{1}+1} (x+{\epsilon}_+)Y_{{\omega}_{2}+1}(x+{\nu} + {\epsilon}_+) +R_{\omega_1, \omega_2} ({\nu}) Y_{{\omega}_{2}+1}(x+{\nu} + {\epsilon}_+) {\kq}_{\omega_{1}} \frac{P_{\omega_{1}}(x)}{Y_{\omega_{1}}(x)} \nonumber \\
    &+ R_{\omega_2, \omega_1} (-{\nu}) Y_{{\omega}_{1}+1}(x+ {\epsilon}_+) {\kq}_{\omega_{2}} \frac{P_{\omega_{2}}(x+{\nu})}{Y_{\omega_{2}}(x+{\epsilon})}  +  {\kq}_{\omega_{1}} {\kq}_{\omega_2} \frac{P_{\omega_{1}}(x)P_{\omega_2}(x+{\nu})}{Y_{\omega_{1}}(x) Y_{{\omega}_{2}}(x+{\nu})}
\label{eq:2ndqq}
\end{align}
with $\omega_1,\omega_2=0,\dots,N-1$, and
\begin{align}
    R_{\omega_1,\omega_2}(\nu) = 
    \begin{cases}
    \frac{\nu+\epsilon_1}{\nu}, & \omega_2-\omega_1=0; \\
    \frac{\nu+\epsilon_2}{\nu+\epsilon_+}, & \omega_2-\omega_1=-1; \\
    1, & \text{otherwise}.
    \end{cases}
    \label{R}
\end{align}

We consider the $[x^{-2}]$ coefficient of the fractional $qq$-character $\CalX_{\omega_1,\omega_2}^{(2)}(x)$ in Eq.~\eqref{eq:[x-2]X2}.
In particular, we are interested in the case of $\omega_1=\omega_2=\omega$. After working through tedious calculation, we match the highest derivative terms between $\hat{h}_4$ and $\hat{\rm H}_4$ 
\begin{align}
    \hat{h}_4 = \hat{\rm H}_4 + (1+\kq)\frac{\hat{\rm H}_3}{1-\kq}\frac{\hat{\rm H}_1}{1-\kq} + \frac{1-\kq}{2}\frac{\hat{\rm H}_2}{1-\kq}\left(\frac{\hat{\rm H}_1}{1-\kq}\right)^2 + \frac{1+
    \kq}{4!}\left(\frac{\hat{\rm H}_1}{1-\kq}\right)^4 +\frac{1+\kq}{2}\left(\frac{\hat{\rm H}_2}{1-\kq}\right)^2 + \cdots
    \label{eq:h4=H4}
\end{align}
where $\cdots$ denotes any lower derivative terms. We again notice that Eq.~\eqref{eq:h4=H4} agrees with Eq.~\eqref{eq:qq=SC2}. Details can be found in the appendix \ref{sec:h4=H4}.

\section{Discussion} \label{sec:discussion}

In this paper we computed the wavefunctions ${\bf\Psi}_{\ba}({\bx})$ of scattering states of the $\spch$ chain corresponding to the infinite-dimensional spin sites. Our main tool was the application of the BPS/CFT correspondence. We identified the wavefunction with the normalized expectation value of the surface defect in the supersymmetric gauge theory in four dimensions with the gauge group $SU(N)$, $2N$ fundamental hypermultiplets, and $\Omega$-deformation in two dimensions along the surface defect. The masses and the Coulomb moduli, divided by the $\Omega$-deformation (equivariant) parameter $\hbar$ determine the spin  and inhomogeneity content of the $\spch$ chain. The four dimensional gauge coupling $\kq$ translates to the twist. 

We used the wall-crossing technique to express the normalized expectation value, given, a priori, by a very complicated sum involving the fine structure of the limit shape of the bulk theory (the limit shape in question was studied in \cite{Fucito:2011pn, Nikita-Pestun-Shatashvili, Poghossian:2016rzb, Fioravanti:2019awr}). Unlike the majority of the literature on the wall-crossing, including the seminal works \cite{Gaiotto:2009hg, Gaiotto:2011tf, Kontsevich:2013rda}, which focuses on the nonabelian structures emerging from the wall-crossing transformations, our formula is relatively simple, amounting to the simple multiplicative factor and a coordinate change. Our method, which consists of first finding an emerging quiver variety whose (rational limit of the ) ${\chi}_{y}$-genus gives the asymptotics ${\epsilon}_{2} \to 0$ of the normalized expectation value, then replacing the latter (viewed as a quotient of the set of stable points by the complexified gauge group) by the integral over the quotient of the unstable set. This is analogous to the computation in \cite{Witten:1992xu}. Another useful analogy is the computation of the equivariant integral over, e.g. ${\BC\BP}^{N-1}$, 
of, say, $c_{1}({\CalL})^{N-1}$, for ${\CalL} = {\CalO}(1)$. In the standard cohomological field theory calculation, as in \cite{Moore:1997dj}, one arrives at the contour integral:
\beq
\int_{{\BC\BP}^{N-1}} c_{1}({\CalL})^{N-1} = \frac{1}{2\pi\ii} \oint \frac{p^{N-1} dp}{(p - m_{1}) \ldots ( p -  m_{N})}
\eeq
where the contour is encircling the equivariant parameters (twisted masses in the ${\CalN}=(2,2), d=2$ language). Taking the sum over the residues is equivalent to using the Atiyah-Bott fixed point formula. If, instead, we pull the contour in the other direction, we get to pick a single pole at infinity, which corresponds, in the picture ${\BC\BP}^{N-1} = \left( {\BC}^{N} \right)^{\rm stable} // {\BC}^{*}$, to localizing at the unstable fixed point $0 \in {\BC}^{N}$. This is what we did in our paper. The same formalism (although it is not clear whether the same geometry is at play) is employed in \cite{Hori:2014tda}. 

Let us end this work by discussing a few loose ends in this note and commenting on future directions.

\begin{itemize}
    \item We would like to construct the $\spch$ spin chain monodromy matrix from the supersymmetry gauge theory at the quantum level. 
    \item The contour integration formula for the the instanton partition function does not forbid different integration variables to pick up poles in ${\zeta}_{\BR}>0$ chamber and ${\zeta}_{\BR}<0$ chamber simultaneously from different moduli parameters. This is equivalent to modifying the real moment map to
    \begin{align}
        \mu_{\BR} = {\zeta}_{\BR}\begin{pmatrix} 1_{k_+} & 0 \\ 0 & -1_{k_-} \end{pmatrix}, \quad k=k_++k_-,
    \end{align} 
    such that $k_+$ instantons are generated by $I({\bf N})$ and $k_-$ instantons are generated by $J^\dagger({\bf N})$. This situation is similar to the instanton partition function of the supergroup gauge theory \cite{Kimura:2019msw, Chen:2020rxu}. Such a modification breaks the $U(k)$ symmetry of the real moment map down to $U(k_+)\times U(k_-)$. The lost symmetry can be compensated by imposing additional $2k_+k_-$ complex equations. If one could come up with the natural set
    of such equations, an analogous trick would be of great help in trying to understand theories with the $SO$ and $Sp$ gauge groups.
    \item $\spch$ spin chain constructed from the non-perturbative Dyson-Schwinger equation is periodic, while the semiclassical limit of the normalized vev of the surface defect seems to be governed by the classical open spin chain. Are these two spin chain systems related? If so, how?
    \item The physical wave function of the quantum Toda lattice  \cite{Kharchev:2000ug,Kharchev:2000yj} is $L^2$ normalizable (once the center-of-mass motion is isolated), so is the wave function of the $SL(2,\BR)$ spin chain \cite{Derkachov:2002tf}. It will be nice to classify the convergence and normalizability constraints on  $\Psi_{\ba}$, perhaps using the integral representation we constructed. 
    
    \item
    It should be straightforward to generalize our work to the case of the $SL(2, {\BC})$ spin chains, in particular, to compare to the recent work \cite{Etingof:2019pni}. The complex spin group, further complexified, 
would correspond to the $XXX_{\mathfrak{sl}_2 \times \mathfrak{sl}_2}$ spin chain in our language, which is in some sense an (entangled?) product of two copies of the quantum field theories we just analyzed.     
    
    \item The quantum Hamiltonians $\hat{\rm H}_I$ are defined based on the non-perturbative Dyson-Schwinger equations of the orbifolded fundamental $qq$-character with a special linear combination defined in Eq.~\eqref{def:U}. In principle, one can extend such formulation to the higher rank $qq$-character. For instance, can one find for the orbifolded second order $qq$-character a set of coefficients $G_{\omega_1,\omega_2}$ (in analogue of $u_{\omega}$ in \eqref{eq:c_w})
    \begin{align}
        \CalU^{(2)}(x) = \sum_{\omega_1,\omega_2,\nu} G_{\omega_1,\omega_2}(\nu) \times \CalX_{\omega_1,\omega_2}^{(2)}(x;\nu)
    \end{align}
    so that the expectation value of $\CalU^{(2)}(x)$ only depends on $k_{\omega}$'s? And should such $\{G_{\omega_1,\omega_2}\}$ exist, what do the Dyson-Schwinger equations tell us for second and higher order $qq$-characters? 
    \item The new quiver system constructed in the section \ref{sec:new quiver} is helped us to simplify the expression for the normalized vev of the surface defect in the $\CalN=2$ gauge theory with fundamental flavors by noticing a much simpler pole structure on the other side of the contour integration. It is well-known that the 2d integrable system dual to the 4d $\CalN=2^*$ gauge theory is the Calogero-Moser system \cite{Nikita-Shatashvili,Nikita:V,Chen:2019vvt,DP1,DP2}. It is natural to ask if a similar procedure can be employed to solve the wave function of the quantum elliptic Calogero-Moser system. Unfortunately, unlike the $A_1$-type theory, the $\CalN=2^*$ (e.g. the ${\hat A}_{0}$-theory in the classification of \cite{Nekrasov:2012xe}) has almost equivalent pole structures on both sides of the contour of the integral representation, leading to no visible advantage in deforming the contour to see the new quiver structure.
    \item We would like to prove the validity of \eqref{eq:qq=SC2} to all orders. To do so, a systematic way of writing the gauge Hamiltonians $\hat{\rm H}_I$ as differential operators is necessary. We have demonstrated that any Hamiltonian $\hat{H}_I$ with $I\geq4$ requires taking higher order $qq$-character into consideration.
\end{itemize}

\newpage
\appendix

\section{Non-perturbative Dyson-Schwinger equation and fundamental $qq$-character} \label{sec:basic}

\begin{figure}
    \centering
    \includegraphics[width=0.4\linewidth]{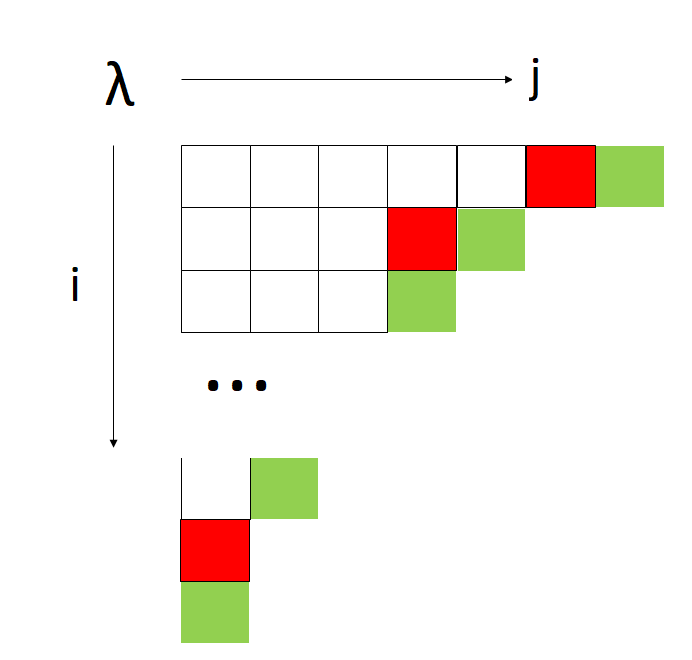}
    \caption{Outer boundary (colored in green) and inner boundary (colored in red) of a Young diagram $\lambda$}
    \label{fig:Young}
\end{figure}

The $Y(x)$-observable, which is a local observable defined as the regularized characteristic polynomial of the adjoint scalar $\Phi$ in the vector multiplet evaluated at the origin $0\in\BC^2$
\begin{align}
    Y(x) = x^N \exp\left[\sum_{l=1}^\infty -\frac{1}{lx^l}\Tr \, \Phi^l \right].
\end{align}
reduces to the statistical mechanical observable $Y(x)[{\vec\lambda}]$, whose evaluation computes as:
\begin{align}
    Y(x)[\vec{\lambda}] 
    & = \mathbb{E}\left[-e^x\hat{S}^*\right] \nonumber\\
    & = \prod_{\alpha=1}^{N}(x-a_{\alpha}) \prod_{(\ri,\rj)\in\lambda^{(\alpha)}}\frac{(x-a_\alpha - (\ri-1)\epsilon_1 - (\rj-1)\epsilon_2 -\epsilon_1)(x-a_\alpha - (\ri-1)\epsilon_1 - (\rj-1)\epsilon_2-\epsilon_2)}{(x-a_\alpha - (\ri-1)\epsilon_1 - (\rj-1)\epsilon_2)(x-a_\alpha - (\ri-1)\epsilon_1 - (\rj-1)\epsilon_2-\epsilon_+)} \nonumber\\
    & = \prod_{\alpha=1}^N \frac{\prod_{{\color{green} \blacksquare}\in \partial_+\lambda^{(\alpha)}}(x-c_{\color{green}\blacksquare} )}{\prod_{{\color{red}\blacksquare}\in \partial_-\lambda^{(\alpha)}} ( x - c_{\color{red} \blacksquare} - {\epsilon}_{+}) }
\end{align}
where $\epsilon_{+} = \epsilon_{1} + \epsilon_{2}$, $c_{\Box} = (\ri-1)\epsilon_1 + (\rj-1) \epsilon_2$, $\ri,\rj=1,2,\dots$. The outer boundary $\partial_+\lambda$ represents the position where potential new boxes can be added, and the inner boundary $\partial_-\lambda$ denotes boxes that can be removed. See Fig.~\ref{fig:Young} for illustration. 

The main statement of \cite{Nikita:I} is that there exist the $qq$-character observables $\CalX(x)$ as the Laurent polynomial in $Y(x)$ with possible shifted arguments:
\begin{align}
    \CalX(x)[\vec{\lambda}] = Y(x+\epsilon_+)[\vec{\lambda}] + \CalO(\kq).
\end{align}
The expectation value $\langle \CalX(x) \rangle$ is a degree $N$ polynomial in $x$: 
\begin{align}
    \langle \CalX (x) \rangle = T(x) = h_0 x^{N} + h_1 x^{N-1} + \cdots + h_N.
\end{align}

To construct a fundamental $qq$-character of the $A_1$ quiver gauge theory, we employ variation on the instanton configuration by adding a point like instanton, or conversely removing a point-like instanton. Inspection of the Fig.~\ref{fig:Young} shows that such a modification can only be achieved by either adding a box in the outer boundary ${\color{green} \blacksquare}\in \partial_+\lambda^{(\alpha)}$, or removing a box in the inner boundary ${\color{red}\blacksquare}\in \partial_-\lambda^{(\alpha)}$ for some $\alpha$. By adding a box $\xi = e^{c_{\color{green}\blacksquare}}$ in the outer boundary ${\color{green}\blacksquare}\in \partial_+\lambda^{(\alpha)}$, the pseudo-measure associated to the original Young diagram $\vec{\lambda}$ and the new one $\vec{\lambda}'$ differs by
\begin{align}
    \frac{\kq\CalZ[\vec{\lambda}']}{\CalZ[\vec{\lambda}]} 
    & = \kq \bE \left[-\frac{(\hat{S}-P_{12}\xi)(\hat{S}-P_{12}\xi)^*}{P_{12}^*} + \frac{\hat{M}(\hat{S}-P_{12}\xi)^*}{P_{12}^*} + \frac{\hat{S}\hat{S}^*}{P_{12}^*} - \frac{\hat{M}\hat{S}^*}{P_{12}^*}  \right] \nonumber\\
    & = \kq \bE \left[ \hat{S} [\vec{\lambda}]\xi^* + q_{12}\xi \hat{S}^*[\vec{\lambda}'] - \hat{M}\xi^* \right] \nonumber\\
    & = (-1)^{N-1} \frac{\kq P(c_{\color{green}\blacksquare})}{\left( {\rm Res}_{x=c_{\color{green}\blacksquare}}Y(x+{\epsilon}_{+})[\vec{\lambda}'] \right) Y^{\prime}(c_{\color{green}\blacksquare})[\vec{\lambda}]}
\end{align}
with $P(x) = \prod_{f=1}^{N} (x - m_f^+) (x-m_f^-)$. Additional $(-1)^{N}$ can be generated by changing $N$ fundamental hypermultiplets $m_f^-$, $f=1,\dots,N$, to anti-fundamental representation of gauge group $U(N)$. In contrast to fundamental matter, anti-fundamental matter contributes to the moduli space vector bundle with Eular class 
\begin{align}
    M^*K. \nonumber
\end{align}
The contribution of fundamental and anti-fundamental hypermultiplets are related by
\begin{align}
    \bE[ M^*K ] = (-1)^{N|\vec{\lambda}|} \bE [ MK^*].
\end{align}
In particular at the level of instanton partition function, the effect of changing fundamental matter to anti-fundamental is equivalent to choose the instanton counting parameter as $(-1)^N\kq$.  

We define the fundamental $qq$-character of $A_1$ theory by:
\begin{align}
    \mathcal{X}(x)[\vec{\lambda}]=Y(x+\epsilon_+)[\vec{\lambda}]+\frac{\kq P(x)}{Y(x)[\vec{\lambda}]}, \quad P(x) = \prod_{f=1}^{N}(x-m_{f}^+)(x-m_f^-).
\end{align}
The expectation value of $\CalX(x)$ now obeys Eq.~\eqref{maintheorem}. See \cite{Nikita:I, Haouzi:2020yxy} for the recent developments and generalizations to other root systems.

\section{Surface defect} \label{sec:defect}

The surface defect partition function is the $\BZ_N$-invariant contribution:
\begin{align}
    \CalZ({\ba},{\bm}^\pm,\vec{\kq})  =\sum_{\vec{\lambda}}\prod_\omega \kq_\omega^{k_\omega}\mathbb{E}\left[-\left(\frac{\hat{S}\hat{S}^*-\hat{M}\hat{S}^*}{P_1^*(1-q_2^{-\frac{1}{N}}\CalR_{-1})}\right)^{\mathbb{Z}_N}\right]
\end{align}
with the definition
\begin{subequations}
\begin{align}
    & K_\omega:=\{(\alpha,(i,j)) \mid \alpha=1,\dots,N;\quad(i,j)\in\lambda^{(\alpha)};\quad\alpha+j-1\equiv\omega \ \text{mod} \ N \}, \label{K} \\ 
    & k_\omega=|K_\omega|,\qquad\nu_\omega=k_\omega-k_{\omega+1}. \label{kv-data}
\end{align}
\end{subequations}

The $Y(x)$-observable \eqref{def:Y} splits into $N$-tuples of $Y_{\omega}(x)[\vec{\lambda}]$ under the orbifolding
$$
    Y(x)[\vec{\lambda}]=\prod_{\omega=0}^{N-1} Y_{\omega}(x)[\vec{\lambda}], \quad Y_{\omega+N}(x)[\vec{\lambda}]=Y_{\omega}(x)[\vec{\lambda}]
$$
Each orbifolded copy is given by
\begin{align}\label{Y_w}
Y_\omega(x)[\vec{\lambda}]
=(x-a_{\omega})&\prod_{(\alpha,(\ri,\rj))\in K_{\omega}}
\left[\frac{(x-c_\Box-\epsilon_1)}{(x-c_\Box)}\right] \prod_{(\alpha,(\ri,\rj))\in K_{\omega-1}} 
\left[\frac{(x-c_\Box-\epsilon_2)}{(x-c_\Box-\epsilon_2-\epsilon_1)}\right].
\end{align}
with $c_\Box = a_\alpha + ({\rm i}-1)\epsilon_1 + ({\rm j}-1) \epsilon_2$. 

To calculate \eqref{eq:X_w=0}, we explore the large $x$ behavior of the fractional $Y_{\omega}(x)[\vec{\lambda}]$:
\begin{align}
    Y_{\omega}(x)=(x-a_{\omega})\exp\left[\frac{\epsilon_1}{x}\nu_{\omega-1}+\sum_{I=1}^\infty \frac{\epsilon_1 D_{\omega-1}^{(I)}}{x^{I+1}}\right]
\end{align}
where
\begin{align}
    D_{\omega}^{(I)} = \sigma_{\omega}^{(I)} - \sigma_{\omega+1}^{(I)} + \sum_{i=1}^{I}
    \begin{pmatrix} I \\ i \end{pmatrix} \left(\frac{\epsilon_2}{N}\right)^i \sigma_{\omega}^{(I-i)}, \quad \sigma_{\omega}^{(I)} 
    & = \sum_{(\alpha,\Box)\in K_{\omega-1}} \sum_{j=0}^{I} \begin{pmatrix} I \\ j \end{pmatrix} \frac{\epsilon_1^{I-j}}{I+1-j}(c_{\Box})^j.
\end{align}
We derive using \eqref{eq:Y large x} 
\begin{align}
    \frac{1}{\epsilon_1}[x^{-1}]\mathcal{X}_\omega(x)= &
    D^{(1)}_\omega-\kq_\omega D^{(1)}_{\omega-1}+\frac{\epsilon_1}{2}(\nu^2_\omega+\kq_\omega\nu^2_{\omega-1}) \nonumber\\
    & -\kq_\omega(a_\omega-m_{\omega}^{+}-m_{\omega}^{-})\nu_{\omega-1}-a_{\omega+1}\nu_\omega+\kq_\omega\frac{P_\omega(a_\omega)}{\epsilon_1}. 
    \label{eq:x-1coeff}
\end{align}

The observable $\CalU(x)$ is defined as a linear combination of the fractional $qq$-characters $\CalX_{\omega}(x)$:
\begin{align}
    [x^{-1}]\, \mathcal{U}(x)= [x^{-1}]\sum_{\omega \in {\BZ}_{N}} u_\omega\mathcal{X}_\omega(x).
\end{align}
Using Eq.~\eqref{eq:x-1coeff} and Eq.~\eqref{eq:c_w} we obtain
\begin{align}\label{eq:largexcombine}
    \frac{1}{\epsilon_1}[x^{-1}]\,{\CalU} (x) 
    = & (1-\kq)\left[D^{(1)}+\sum_\omega\frac{\epsilon_1}{2}\left(\nu_\omega-\frac{a_{\omega+1}}{\epsilon_1}\right)^2-\frac{a_\omega^2}{2\epsilon_1}\right] \\
    & +\frac{1}{\epsilon_1}\sum_{\omega}\kq_{\omega+1} u_{\omega+1}\left[\epsilon_1\nu_{\omega}-a_{\omega+1}+m_{\omega+1}^+\right]\left[\epsilon_1\nu_{\omega}-a_{\omega+1}+m_{\omega+1}^-\right], \nonumber
\end{align}
with the bulk gauge invariant observable $D^{(1)}$ given by 
\begin{align}\label{eq:D-bulk}
    D^{(1)} = \sum_{\omega} D^{(1)}_{\omega} = \epsilon_2 \sum_{\omega}k_{\omega} = \epsilon_2 |\vec{\lambda}|.
\end{align}

As a linear combination of the fractional $qq$-characters $\CalX_{\omega}(x)$, the expectation value of the observable $\CalU(x)$ obeys the same Dyson-Schwinger equation \eqref{eq:X_w=0}, which translates into a second order differential equation obeyed by the surface defect instanton partition function $\CalZ$,
\begin{align}\label{Hamiltonian time}
    0 =  \left[x^{-1}\right] \langle{\CalU} (x) \rangle
    = & (1-\kq)\left[\epsilon_1\epsilon_2 D^{(1)} +\sum_\omega\frac{1}{2}\left(\epsilon_1\nabla^z_\omega-{a_{\omega+1}}\right)^2-\frac{a_{\omega+1}^2}{2}\right]\CalZ \nonumber\\ &+\left[\sum_{\omega}\kq_{\omega+1} u_{\omega+1}(\epsilon_1\nabla^z_{\omega}-a_{\omega+1}+m_{\omega+1}^+)(\epsilon_1\nabla^z_{\omega}-a_{\omega+1}+m_{\omega+1}^-)\right]\CalZ
\end{align}
with the variables $\{z_{\omega}\}_{\omega=0}^{N-1}$ defined in Eq.~\eqref{def:z_w}.

The Eq.~\eqref{Hamiltonian time} in the NS-limit becomes an eigenvalue problem of the normalized vev of the surface defect $\Psi_\ba$
\begin{align}
    & \left[ \sum_{\omega}-\frac{(1-\kq)}{2}\left[({\delta}_{\omega}-m_{\omega+1}^{+})^2 -a_{\omega+1}^2\right]-(\kq_{\omega+1} u_{\omega+1})\left(\delta_\omega\right)\left(\delta_\omega-m_{\omega+1}^{+}+m_{\omega+1}^{-} \right) \right] \Psi+{\ba,+}  \nonumber\\
    & = (1-\kq) E_2 \Psi+{\ba,+}
    \label{eq:Hamiltonian N}
\end{align}
with $\delta_{\omega} = \epsilon\nabla^z_{\omega} - a_{\omega+1} + m_{\omega+1}^+$. In particular $E_2 = \langle D^{(1)} \rangle$ is related to the twisted superpotential $\CalW$ by
\begin{align}
    E_2=\epsilon\kq\frac{\partial}{\partial\kq}\CalW({\ba},{\bm}^\pm,\tau;\epsilon_1).
\end{align}

\subsection{An example with one degree of freedom} \label{sec:gaudin}

Let us consider the $U(2)$ gauge theory with 4 fundamental flavors of masses $m_{0}^\pm, m_1^{\pm}$. We multiply normalized vev of the surface defect with perturbative factor
\begin{align}
    \psi(\vec{z},{\ba},{\bm}^\pm,\kq;\epsilon_1,\epsilon_2)=\kq^{-\sum_{\omega}\frac{a_{\omega}^2}{2\epsilon_1\epsilon_2}}\prod_{\omega=0}^{N-1} z_{\omega}^{-\frac{a_{\omega+1}}{\epsilon_1}} \CalZ, \nonumber
\end{align}
In the center of mass frame $\nabla_0^z + \nabla^z_1=0$, the Eq.~\eqref{eq:Hamiltonian N} becomes
\begin{align}\label{PDE Matter}
    0=&\left[-\frac{1}{z}(z-1)(z-\kq)\left(\epsilon z \frac{d}{dz}\right)^2+[\kq(z-1)(m_1^++m_1^-)-z(z-\kq)(m_0^++m_0^-)]\epsilon \frac{d}{dz}\right. \nonumber\\
    &\quad +\left.\kq\frac{z-1}{z}m_1^+m_1^--(z-\kq)m_0^+m_0^-+ (1-\kq) E_2\right]\psi \nonumber\\
    = & \hat{H}\psi
\end{align}
with $z=-\kq_{0}$. We identify Eq.~\eqref{PDE Matter}
as the well-known Gaudin Hamiltonian up to a canonical transformation
\begin{align}
    \left(\frac{d}{dz}+f\right)^2=\frac{d^2}{dz^2}+2f \frac{d}{dz}+\frac{df}{dz}+f^2.
\end{align}
By choosing
\begin{align}
    f(z)
    &=-\frac{m_1^++m_1^-+\epsilon}{2z}+\frac{m_0^++m_0^-}{2(z-1)}+\frac{m_1^++m_1^-}{2(z-\kq)}
\end{align}
the Hamiltonian in Eq.~\eqref{PDE Matter} becomes well recognized Gaudin Hamiltonian: 
\begin{align}
    \label{eq:liouville}
    \hat{H} 
    &=-\epsilon^2\frac{d^2}{dz^2}+\frac{\Delta_0}{z^2}+\frac{\Delta_1}{(z-1)^2}+\frac{\Delta_2}{(z-\kq)^2}+\frac{\Delta_3-\Delta_0-\Delta_1-\Delta_2}{z(z-1)}+\frac{(1-\kq)\mathfrak{u}}{z(z-1)(z-\kq)}
\end{align}
where the coefficients $\{\Delta_i\}$ and $\mathfrak{u}$ are
\begin{align}
    & \Delta_0=\frac{(m_1^+-m_1^-)^2}{4}-\frac{\epsilon^2}{4}, \qquad
    \Delta_1=\frac{(m_0^++m_0^--\epsilon)^2}{4}-\frac{\epsilon^2}{4}, \nonumber \\
    & \Delta_2=\frac{(m_1^++m_1^--\epsilon)^2}{4}-\frac{\epsilon^2}{4}, \qquad
    \Delta_3=\frac{(m_0^+-m_0^-)^2}{4}-\frac{\epsilon^2}{4}, \label{eq:Deltas}\\
    & \mathfrak{u}=-\frac{(m_1^++m_1^-+1)(m_1^++m_1^-)}{2\kq(1-\kq)}+\frac{(m_0^++m_0^-+1)(m_1^++m_1^-)}{2(1-\kq)^2}+\frac{m_1^+m_1^-}{\kq(1-\kq)}+E_2. \nonumber
\end{align}


\subsection{The Toda limit} \label{sec:Toda}

In the mass decoupling limit $m_f^\pm\to\infty$ of the $A_1$ quiver gauge theory, some or all fundamental hypermultiplets are integrated out with a simultaneous scaling of $\kq\to 0$ so that the product
$$
    \kq \prod_{f=1}^{N}m_f^+m_f^- = \Lambda^{2N}
$$
is kept finite.
The $qq$-character \eqref{eq:fundqq} in the mass decoupling limit becomes
\begin{align}
    \CalX(x) = Y(x+\epsilon_+) + (-1)^N\frac{\Lambda^{2N}}{Y(x)}.
    \label{eq:qqsym}
\end{align}
The $(-1)^N$ in the $qq$-character is a choice of convention. We explain the choice of such convention in appendix \ref{sec:basic}.
Let us demonstrate how Dyson-Schwinger equation of defect $qq$-character give rise to the Hamiltonian of $\hat{A}_{N-1}$ Toda lattice ($\hbar = \epsilon = \epsilon_1$) \eqref{H-Toda} in the NS-limit \cite{Nikita:V, Chen:2019vvt}. We introduce co-dimension two surface defect exactly as how we had done for SQCD. Orbifolded $qq$-character becomes
\begin{align}
    \CalX_\omega(x) = Y_{\omega+1}(x+\epsilon_+) - \frac{\kq_{\omega}}{Y_{\omega}(x)}, \quad \kq_{\omega} = \Lambda^{2}e^{\rx_{\omega} - \rx_{\omega-1}}
\end{align}
which satisfies Dyson-Schwinger equation
\begin{align}
    [x^{-1}]\langle \CalX_{\omega}(x) \rangle = 0.
\end{align}
We derive using Eq.~\eqref{eq:Y large x}
\begin{align}
    [x^{-1}]\CalX_{\omega}(x) = \epsilon_1 D_{\omega} + \frac{\epsilon_1^2}{2}\nu_{\omega}^2 - a_{\omega+1}\nu_{\omega} - \kq_{\omega}.
\end{align}
Taking expectation value and summation over $\omega$ gives
\begin{align}
    \left[\epsilon_1\epsilon_2\nabla^\kq + \epsilon_1\epsilon_2 \vec{\rho} \cdot \vec{\nabla}^z + \hat{H}_{\rm Toda} \right]{\varphi}(\vec{z},{\ba},\kq;\epsilon_1,\epsilon_2) = 0.
\end{align}
$\varphi$ is defined by multiplying surface defect partition function with perturbative factor:
\begin{align}
    \varphi(\vec{z},{\ba},\kq;\epsilon_1,\epsilon_2)=\kq^{-\sum_{\omega}\frac{a_{\omega}^2}{2\epsilon_2}}\prod_{\omega=0}^{N-1} z_{\omega}^{-\frac{a_{\omega+1}}{\epsilon_1}} \CalZ(\vec{z},{\ba},\kq;\epsilon_1,\epsilon_2). \nonumber
\end{align}
In the NS-limit, $\varphi$ has asymptotics
\begin{align}
    \varphi = e^{\frac{1}{\epsilon_2}\CalW({\ba},\kq;\epsilon_1)} \times \left(\Psi_{\ba,+}(\vec{z},{\ba},\kq;\epsilon_1,\epsilon_2) +\CalO(\epsilon_2) \right).
\end{align}

The normalized vev of the surface defect $\Psi_{\ba,+}$ is the eigenfuncton of the quantum Toda lattice
\begin{align}
    \hat{H}_{\rm Toda} \Psi_{\ba,+} = E_2 \Psi_{\ba,+}, \quad E_2 = \epsilon_1 \kq \frac{\partial }{\partial \kq}\CalW.
\end{align}

\section{The crossing formulas}\label{sec:cross}

In the familiar theory the FI-parameter ${\zeta}_{\BR}$ associated to real moment map is often set to be positive \cite{Nakajima:1994nid,Nekrasov:1998ss}. 
In the case of pure ${\CalN}=2$ SYM, ${\CalN}=2$ SQCD with $N_{f}<2N-1$ and ${\CalN}=2^{*}$ theories with gauge group $U(N)$, instanton partition does not care about the sign of the FI-parameter. 
Let us show here that the instanton partition function is independent of the sign of FI-parameter in the $U(N)$ super Yang-Mills theory. The instanton partition function has integral representation:
\begin{align}
    \CalZ_{\rm inst,+} = \sum_{k=0}^\infty  \frac{(\kq)^k}{k!} \oint_{\CalC} \prod_{i=1}^k\frac{d\phi_i}{2\pi {\ii}}\frac{\epsilon_+}{\epsilon_1\epsilon_2}\prod_{\alpha=1}^N \frac{1}{(\phi_i-\alpha_a)(\phi_i-a_\alpha+\epsilon_+)} \prod_{ j\neq i}^{k} \frac{(\phi_{i}-\phi_{j})(\phi_{i}-\phi_{j}+\epsilon_+)}{(\phi_{i}-\phi_{j}+\epsilon_1)(\phi_{i}-\phi_{j}+\epsilon_2)}.
\end{align}
How to pick up poles and evaluate residue in the contour integration is explained as follow: We first rewrite the integral with ordering 
\begin{align}\label{eq:Z-sym}
    \frac{1}{k!}\oint_{\CalC}\prod_{i=1}^k\frac{d\phi_i}{2\pi {\ii}} \to \oint \frac{d\phi_k}{2\pi{\ii}}\cdots \oint \frac{d\phi_2}{2\pi{\ii}} \oint \frac{d\phi_1}{2\pi{\ii}}. \nonumber
\end{align}
The first variable $\phi_1$ picks up the pole at $\phi_1=a_\alpha$ for some $\alpha=1,\dots,N$. The following ones are determined recursively as
$$
    \phi_{i} = a_{\beta(\neq\alpha)}, \quad \phi_{j(<i)}+\epsilon_1, \quad \phi_{j(<i)} + \epsilon_2.
$$
In the end, the poles are parametrized by $N$-tuples of Young diagrams $\vec{\lambda}$ where each box $(\ri,\rj)\in\lambda^{(\alpha)}$ corresponds to a pole at 
$$
    {\phi}_i = a_{\alpha} + ({\ri} - 1)\epsilon_1 + ({\rj} - 1)\epsilon_2.
$$
The contour integration can be evaluated by taking poles from the other side of the contour $\CalC$.
By doing so, the first variable $\phi_1$ picks up a pole at $\phi_{1} = a_\alpha - \epsilon_+$ for some $\alpha=1,\dots,N$. The following ones are determined recursively as
$$
    \phi_{i} = a_{\beta(\neq\alpha)}-\epsilon_+, \quad \phi_{j(<i)} - \epsilon_1, \quad \phi_{j(<i)} - \epsilon_2.
$$
Finally the poles form the $N$-tuples of Young diagrams $\vec{\lambda}_{\rm dual}$ in the negative FI-parameter chamber. Since both instanton partition functions in the two FI-parameter chambers are evaluated from the same integration formula, the two are equal 
\begin{align}
    \CalZ_{{\rm inst},-} 
    & = \sum_{k=0}^\infty  \frac{(\kq)^k}{k!} \oint_{\CalC} \prod_{i=1}^k\frac{d\phi_i}{2\pi {\ii}}\frac{\epsilon_+}{\epsilon_1\epsilon_2}\prod_{\alpha=1}^N \frac{1}{(\phi_i-a_\alpha)(\phi_i-a_\alpha+\epsilon_+)} \prod_{ j\neq i}^{k} \frac{(\phi_{i}-\phi_{j})(\phi_{i}-\phi_{j}+\epsilon_+)}{(\phi_{i}-\phi_{j}+\epsilon_1)(\phi_{i}-\phi_{j}+\epsilon_2)} \nonumber \\
    & = \sum_{k=0}^\infty  \frac{(\kq)^k}{k!} \oint_{\tilde{\CalC}} \prod_{i=1}^k\frac{d\phi_i}{2\pi {\ii}}\frac{\epsilon_+}{\epsilon_1\epsilon_2}\prod_{\alpha=1}^N \frac{1}{(\phi_i-a_\alpha)(\phi_i-a_\alpha+\epsilon_+)} \prod_{ j\neq i}^{k} \frac{(\phi_{i}-\phi_{j})(\phi_{i}-\phi_{j}+\epsilon_+)}{(\phi_{i}-\phi_{j}+\epsilon_1)(\phi_{i}-\phi_{j}+\epsilon_2)} \nonumber\\
    & = \CalZ_{{\rm inst},-}
\end{align}

A similar argument can be applied to $U(N)$ gauge theory with number of flavors $N_f<2N-1$. In the case of $\CalN=2^*$, a careful examination shows that there is no pole at infinity. Thus the instanton partition functions in the two chambers are equal.

\paragraph{}
However in the case of $N_f = 2N$ and $N_f = 2N-1$ in SQCD gauge theory, partition function evaluated with positive FI-parameter is different from partition function with negative FI-parameter.
To demonstrate the mismatch, we can simply take a $U(1)$ supersymmetric gauge theory in four dimension with moduli parameter $a$ and two fundamental flavors of masses $m^+$ and $m^-$ for demonstration. The instanton partition function in the positive FI-parameter $\zeta_\BR>0$ chamber has the following expansion in the instanton counting parameter $-\kq$
\begin{align}
    \CalZ_{{\rm inst},+} = 1 - \kq\frac{(a-m^+)(a-m^-)}{\epsilon_1\epsilon_2} + \CalO(\kq^2). \nonumber
\end{align}
In negative FI-parameter ${\zeta}_{\BR}<0$ chamber, the instanton partition function instead takes a different expansion
\begin{align}
    \CalZ_{{\rm inst},-} = 1 - \kq \frac{(a-\epsilon_+-m^+)(a-\epsilon_+-m^-)}{\epsilon_1\epsilon_2} + \CalO(\kq^2)\nonumber
\end{align}
with $\epsilon_+=\epsilon_1+\epsilon_2$. 
We will prove that the mismatch can be expressed by a crossing formula, which comes from the pole at infinity in the integration representation of the instanton partition function. 

\subsection{Crossing formula for a toy model}

Let us consider the following toy model, the vortex analogue of the ADHM instanton construction (cf. \cite{Shadchin:2006yz}): The vortex-ADHM data consists of the two vector spaces ${\bf K}=\mathbb{C}^k$ and ${\bf N}=\mathbb{C}^1$. In addition, we shall include the effect of one fundamental matter multiplet of mass $m$.  The only two matrices in the toy model are $I\in\operatorname{Hom}({\bf N}, {\bf K})$ and $B\in\operatorname{Hom}({\bf K},{\bf K})$. The real moment map is, naturally, cf. \cite{Hanany:2003hp}:
\beq
    \mu_{\mathbb{R}}=II^\dagger+[B,B^\dagger]={\zeta}_{\BR} 1_{\bf K}, 
    \label{eq:rvortex}
\eeq
We introduce the $U(1)_{a} \times U(1)_{\epsilon}$-symmetry which acts by: $(I,B)\mapsto (w^{-1}I,qB)$, which, together with the compensating transformations becomes, upon complexification:
\beq
    (I,B) \to \left( g^{-1} I w, q \, g^{-1} Bg \right)
\eeq
with $g\in GL(k,\BC)$ and $w\in GL(1,\BC)$. Accordingly, the $U(1)_{a} \times U(1)_{\epsilon}$-fixed points on the moduli space  are the solutions to
\beq
    I \, w \ =\  g \, I, \quad  q\, B =  g\, B\, g^{-1}\ .
\eeq
For each $k$ there is exactly one fixed point, in which the space ${\bf K}$ has the basis  
\beq
    B^{j-1}I, \quad j=1,2,\ldots, k \ .
\eeq
Each fixed point contributes the pseudo-measure
$$
    \bE \left[ N K^{*} - (1-q) KK^{*} - MK^{*} \right]
$$
where
\beq
    N = e^{a} = {\rm Tr}_{\bf N}(w)\, , \ K = {\rm Tr}_{\bf K} (g) = e^a \sum_{j=1}^k q^{j-1}, \quad M = e^{m}.
\eeq
The vortex partition function is the grand canonical ensemble (the Coulomb modulus is set to zero for simplicity, and we chose $(-{\kq})$ as a fugacity). 
\begin{align}
    \mathcal{Z}_{+}(m,\epsilon)=\sum_{k=0}^\infty \, (-{\kq})^k\ \mathbb{E}\left[ \left( q^{k} -e^{m} \right) K^* \right]=\sum_{k=0}^\infty \, (-{\kq})^{k}\ \prod_{j=1}^k\frac{m-(j-1)\epsilon}{j\epsilon}={(1-\kq)^\frac{m}{\epsilon}}.
    \label{eq:vortpf}
\end{align}
For negative ${\zeta}_{\BR}$ chamber the moduli space is obviously empty, since the Eq.~\eqref{eq:rvortex} implies $\Vert I \Vert^{2} = k {\zeta}_{\BR}$. 
The instanton partition function in the negative FI-parameter chamber is simply
$$
    \mathcal{Z}_{-}(m,\epsilon)=1.
$$
The instanton partition functions in the positive  and in the negative chambers differ by a simple crossing factor
\begin{align}
    \frac{\mathcal{Z}_{+}}{\mathcal{Z}_{-}}={(1-\kq)^{\frac{m}{\epsilon}}}.
\end{align}
It is instructive to obtain 
the crossing factor from the contour integral representation of the vortex partition function.
Denote by $K=\sum_{i=1}^ke^{\phi_i}$ the character of the ${\bf K}$ space (more precisely, it is the character of the compensator $g$). 
The integral representation of the vortex partition function \eqref{eq:vortpf} 
is known:
\begin{align}
    \CalZ_{+}
    &=\sum_{k=0}^\infty(-\kq)^k\frac{1}{k!}\oint_\CalC\prod_{i=1}^k\frac{d\phi_i}{2\pi {\ii}}\frac{1}{\epsilon} \frac{m-\phi_i}{\phi_i}\prod_{j\neq i}\frac{\phi_i-\phi_j}{\phi_i-\phi_j+\epsilon} \nonumber\\
    &=\sum_{k=0}^\infty(-\kq)^k\frac{1}{k!}\oint_{\tilde{\CalC}}\prod_{i=1}^k\frac{d\phi_i}{2\pi {\ii}}\frac{1}{\epsilon} \frac{m-\phi_i}{\phi_i}\prod_{j\neq i}\frac{\phi_i-\phi_j}{\phi_i-\phi_j+\epsilon}.
\end{align}
The deformed contour $\tilde{\CalC}$ enclosed pole at infinity clockwise. Residue for a single $\phi_i$ at infinity is  $r_\infty=\frac{m}{\epsilon}$. We identify
\begin{align}
    \frac{1}{k!}\oint_{\infty}\prod_{i=1}^k\frac{d\phi_i}{2\pi {\ii}}\frac{1}{\epsilon}\frac{m-\phi_i}{\phi_i}\prod_{j\neq i}\frac{\phi_i-\phi_j}{\phi_i-\phi_j+\epsilon}=
    \left(\!
    \begin{array}{c}
      r_\infty \\
      k
    \end{array}
  \!\right).
\end{align}
We then recover the crossing factor as the contribution of the residue at infinity
\begin{align}
    \sum_{k=0}^{\infty} (-\kq)^k \left(\!
    \begin{array}{c}
      r_\infty \\
      k
    \end{array}
  \!\right) = (1-\kq)^{r_\infty}.
  \label{eq:crosstoy1}
\end{align}


\subsection{Crossing in the bulk} \label{sec:cross-bulk}

The crossing factor applies to not only the surface contribution but also to the bulk partition function. 

In the case of a $U(1)$ gauge theory with two fundamental flavors. The theory is characterized by a moduli parameter $a$ and mass of fundamental flavors $m^\pm$. With the fugacity chosen to be $-\kq$,
the instanton partition function in the positive FI-parameter chamber has the following integration representation \cite{Nekrasov:2002qd,Nikita:I},
\begin{align}
    \CalZ_{{\rm inst},+} = \sum_{k=0}^\infty (-\kq)^k
    \frac{1}{k!}\oint_\CalC \prod_{i=1}^{k}\frac{d\phi_i}{2\pi {\ii}}\left(\frac{\epsilon_+}{\epsilon_1\epsilon_2}\right) \frac{(\phi_i-m^+)(\phi_i-m^-)}{(\phi_i-a)(\phi_i - a + \epsilon_+)} \prod_{j\neq i} \frac{\phi_{ij}(\phi_{ij} + \epsilon_+)}{(\phi_{ij} + \epsilon_1)(\phi_{ij} + \epsilon_2)}, \label{eq:U1part}
\end{align}
where $\phi_{ij} = \phi_i - \phi_j$. The same integral can be find in the Witten-index formula \cite{Hori:2014tda}. We explain the reason of choosing fugacity to be $-\kq$ in appendix  \ref{sec:basic}. 
The contour encloses the poles counterclockwise at $\phi = a+(\ri-1)\epsilon_1+(\rj-1)\epsilon_2$, $\ri,\rj=1,2,\dots$. \footnote{In general, there exists four sets of possible pole structure in \eqref{eq:U1part} characterized by sign of $\epsilon$-parameters: $a \pm (i-1) \epsilon_1 \pm (j-1) \epsilon_2$. The four sets are independent and have no mixing in between. It is enough to consider the one with $(+\epsilon_1,+\epsilon_2)$.}
The integration in Eq.~\eqref{eq:U1part} can also be evaluated by deformation of the contour
\begin{align}
    \CalZ_{{\rm inst},+}=
    & \sum_{k=0}^\infty (-\kq)^k\frac{1}{k!}
    \oint_{\tilde{\CalC}} \prod_{i=1}^{k}\frac{d\phi_i}{2\pi {\ii}} \left(\frac{\epsilon_+}{\epsilon_1\epsilon_2}\right) \frac{(\phi_i-m^+)(\phi_i-m^-)}{(\phi_i-a)(\phi_i - a + \epsilon_+)} \prod_{j\neq i} \frac{\phi_{ij}(\phi_{ij} + \epsilon_+)}{(\phi_{ij} + \epsilon_1)(\phi_{ij} + \epsilon_2)}.
\end{align}
The deformed contour encloses poles clockwise at $a  - {\rm i}\epsilon_1 - {\rm j} \epsilon_2$, ${\rm i},{\rm j}=1,2,\dots$ and at the infinity. 
Suppose there are $l$ integration variables $\phi_i$, $i=1,\dots,l$ picking up pole at the infinity, $l=0,1,\dots,k$. The residue  picked up by $\phi_i$ at infinity is equal to
\begin{align}
    \frac{\epsilon_+}{\epsilon_1\epsilon_2}\left[m^++m^- - 2a + \epsilon_+\right] = r_\infty.
\end{align}

The residue at infinity is evaluated following the procedure similar but not identical to \eqref{R_k,l}.
The first $\phi_1$ can pick up pole at the infinity directly, obtaining a factor of $-r_\infty$ and removed from integration completely. Or it may pick up either $\phi_1 = \phi_i + \epsilon_1$ or $\phi_1 = \phi_i + \epsilon_2$ for another $\phi_i$, $i-2,\dots,l$. The combination gives
\begin{align}
    - r_\infty + (l-1) \frac{\epsilon_+}{\epsilon_1\epsilon_2}\left(\frac{\epsilon_1^2(\epsilon_1^2 - \epsilon_+^2)}{\epsilon_1(\epsilon_1^2 - \epsilon_2^2)} + \frac{\epsilon_2^2(\epsilon_2^2 - \epsilon_+^2)}{\epsilon_2(\epsilon_2^2 - \epsilon_1^2)}\right) =  -r_\infty - (l-1).
\end{align}
The second $\phi_2$ is again given a choice of picking up pole at the infinity or taking residue at another $\phi_i$. 
The overall contribution from the pole at the infinity reads
\begin{align}
    \frac{1}{l!} \prod_{i=1}^{l}( - r_\infty - (i-1)) = 
    \left(\!
    \begin{array}{c}
      -r_\infty \\
      l
    \end{array}
  \!\right).  \nonumber
\end{align}
The instanton partition function in Eq.~\eqref{eq:U1part} for positive FI-parameter $\zeta_\BR>0$ and negative FI-parameter $\zeta_{\BR}<0$ are related by a crossing formula:
\begin{align}
    \CalZ_{{\rm inst},+} 
    & = \sum_{k=0}^\infty \sum_{l=0}^k (-\kq)^{l}\left(\!
    \begin{array}{c}
      -r_\infty \\
      l
    \end{array}
  \!\right) \times (-\kq)^{k-l} \CalZ_{{\rm inst},-,k-l} = (1-\kq)^{-r_\infty} \times \CalZ_{{\rm inst},-}.
  \label{eq:crossbulk1}
\end{align}

In particular, Eq.~\eqref{eq:crossbulk1} can be verified directly since the instanton partition function of the $U(1)$ gauge theory with two fundamental flavors is well known \cite{nekrasov2006seiberg,Nikita:I}, 
\begin{align}
    \CalZ_{{\rm inst},+} = (1-\kq)^{\frac{(a-m^+)(a-m^-)}{\epsilon_1\epsilon_2}}.
\end{align}
Flipping the sign of the FI-parameter ${{\zeta}_{\BR}}$ is equivalent to the sign flip of $\Omega$-parameters $\epsilon_{1,2}\mapsto-\epsilon_{1,2}$ and the shift of the moduli parameter $a\mapsto a - \epsilon_{+}$. The instanton partition function in the negative FI-parameter chamber is 
\begin{align}
    \CalZ_{{\rm inst},-} 
    & = (1-\kq)^{\frac{(a-\epsilon_+ - m^+)(a - \epsilon_+ - m^-)}{\epsilon_1\epsilon_2}} 
    = (1-\kq)^{r_\infty}\CalZ_{{\rm inst},+}
\end{align}
which substantiates Eq.~\eqref{eq:crossbulk1}. 


The crossing formula \eqref{eq:crossbulk1} can be easily extended to general $U(N)$ gauge group with $2N$ fundamental flavors. The fugacity is chosen to be $-\kq$. 
The instanton partition function in the positive FI-parameter chamber has the integration representation
\begin{align}
    &\CalZ_{{\rm inst},+} ({\ba},{\bm}^\pm,\epsilon_1,\epsilon_2;\kq) \nonumber\\
    & = \sum_{\vec{\lambda}} (-\kq)^{k} \oint_\CalC \prod_{i=1}^{k}\frac{d\phi_i}{2\pi{\ii}}\left(\frac{\epsilon_+}{\epsilon_1\epsilon_2}\right)^k \prod_{\alpha=1}^{N}\frac{(\phi_i - m_{\alpha}^+)(\phi_i - m_{\alpha}^-)}{(\phi_i-a_\alpha)(\phi_i - a_\alpha + \epsilon_+)}\prod_{j\neq i}\frac{\phi_{ij}(\phi_{ij}+\epsilon_+)}{(\phi_{ij}+\epsilon_1)(\phi_{ij}+\epsilon_2)}
\end{align}

The contour $\CalC$ is chosen to enclose poles counterclockwise located at $a_\alpha + ({\rm i}-1)\epsilon_1 + ({\rm j}-1)\epsilon_2$, $\alpha=1,\dots,N$, ${\rm i},{\rm j}=1,2,\dots$. The residue for a single integration variable $\phi_i$ at infinity is
\begin{align}
        r_\infty = \frac{\epsilon_+}{\epsilon_1\epsilon_2} \left(\sum_{\alpha=1}^N m_{\alpha-1}^+ + m_{\alpha-1}^- - 2a_\alpha + \epsilon_+\right).
\end{align}
When evaluating the contribution from pole at infinity after deformation of the contour, An integration variable $\phi_i$ has three options: picking up pole from infinity directly, picking up pole from another $\phi_n+\epsilon_1$, or picking up pole from another $\phi_n+\epsilon_2$. The structure of pole at infinity for general $U(N)$ gauge group has no difference from $U(1)$. 
Hence the crossing factor of $U(N)$ gauge theory is identical to the case of group $U(1)$:
\begin{align}
    \CalZ_{{\rm inst},+}({\ba},{\bm}^\pm,\epsilon_1,\epsilon_2;\kq) = (1-\kq)^{-r_\infty}\CalZ_{{\rm inst},-} ({\ba},{\bm}^\pm,\epsilon_1,\epsilon_2;\kq)
    \label{eq:crossbulkN}.
\end{align}

\subsection{More examples of crossing formulas} \label{sec: moreexamples}

Let us now look at a few explicit examples:

\subsubsection{$N=2$}
For $SU(2)$ gauge theory the pseudo-measure in the normalized vev of the surface defect $\Psi_{\ba,+}$ \eqref{chi} can be rewritten in terms of dual character $V_1$, 
\begin{align}
    &\kq_0^{|V_1|}\mathbb{E}\left[\frac{\Gamma_1(\Gamma_2-\Gamma_1)^*}{P_1^*}-\frac{M_0(\Gamma_2-\Gamma_1)^*}{P_1^*}\right] \nonumber \\
    &=\kq_0^{|V_1|}\mathbb{E}\left[\frac{(\Gamma_2-e^{A_1}-M_0)e^{-A_1}}{P_1^*}\right]\mathbb{E}\left[P_1(W-V_1)V_1^*+e^{A_0}V_1^*+q_1V_1e^{-A_1}-M_0V_1^*\right].
\end{align}
$W=\sum_{s=1}^{|W|}e^{b_s}$ is the character of jumps in the bulk. The dual character $V_1$ dependence can be written as the following contour integration by denoting $V_1=\sum_{i=1}^ve^{\phi_i}$.
\begin{align}\label{Contour V, N=2}
    \frac{\kq_0^v}{v!}\oint_\CalC\prod_{i=1}^v\frac{d\phi_i}{2\pi {\ii}}\frac{1}{\epsilon}\prod_{i=1}^v\frac{(\phi_i-m_0^+)(\phi_i-m_0^-)}{(\phi_i-A_0)(\phi_i-A_1+\epsilon)}\prod_{i>j}\frac{\phi_{ij}^2}{\phi_{ij}^2-\epsilon^2}\prod_{s=1}^{|W|}\frac{\phi_i-b_s-\epsilon}{\phi_i-b_s},
\end{align}
with $\phi_{ij}=\phi_{i}-\phi_{j}$. The contour $\CalC$ is chosen such that it only encloses poles counterclockwise from $A_0+({\rm i}-1)\epsilon$, ${\rm i}=1,2,\dots$, which generate the tail, or $b_s$, which generate the jumps. Eq.~\eqref{Contour V, N=2} can be calculated by picking up poles from the other side of the contour
\begin{align}\label{Countor outcome, N=2}
    &\frac{1}{v!}\oint_{\tilde{\CalC}}\prod_{i=1}^v\frac{d\phi_i}{2\pi {\ii}}\frac{1}{\epsilon}\prod_{i=1}^l\frac{(\phi_i-m_0^+)(\phi_i-m_0^-)}{(\phi_i-A_0)(\phi_i-A_1+\epsilon)}\prod_{i>j}\frac{\phi_{ij}^2}{\phi_{ij}^2-\epsilon^2}\prod_{s=1}^{|W|}\frac{\phi_i-b_s-\epsilon}{\phi_i-b_s}.
\end{align}
The dual contour $\tilde{\CalC}$ picks up poles clockwise at $A_1-{\rm i}\epsilon$, ${\rm i}=1,2,\dots$, and at the infinity. Residue at infinity for a single $\phi_i$ is
\begin{align}
    \frac{1}{\epsilon}(m_0^++m_0^--A_0-A_1+\epsilon+|W|\epsilon)=\frac{1}{\epsilon}(m_0^++m_0^--a_0-a_1+\epsilon):=r_\infty.
\end{align}
The dual contour integration is performed in the following steps: We choose $v-l$ integration variables $\phi_n$, $n=1,\dots,v-l$ to pick up poles at infinity. 
\begin{align}\label{eq:int-inft}
    & \frac{1}{(v-l)!}  \, \oint_{\infty}
    \prod_{n=1}^{v-l}\ \frac{d{\phi}_{n}}{2\pi{\ii}} \frac{1}{\epsilon} \frac{(\phi_n-m_0^+)(\phi_n-m_0^-)}{(\phi_n-A_0)(\phi_n-A_1+\epsilon)}\prod_{s=1}^{|W|}\frac{\phi_n-b_s-\epsilon}{\phi_n-b_s} \times \prod_{j=1,j>n}^{v} \frac{\phi_{nj}^2}{\phi_{nj}^2 - \epsilon^2} 
\end{align}
How to evaluate Eq.~\eqref{eq:int-inft} is explained as follows. First we rewrite integral with ordering 
\begin{align}
    \oint_\infty \frac{d\phi_{v-l}}{2\pi {\ii}} \cdots \oint_\infty \frac{d\phi_2}{2\pi {\ii}} \oint_\infty \frac{d\phi_1}{2\pi {\ii}}.
\end{align}
The first variables $\phi_1$ can pick up poles at infinity directly. Or it can take pole at another $\phi_n+\epsilon$ with $n=2,3,\dots v-l$. The contour circulates the poles clockwise for either of the choice. The former generate residue $-r_\infty$ while the second option gives 
\begin{align}
    \frac{1}{\epsilon} \frac{\epsilon^2}{2\epsilon} \frac{(\phi_n+\epsilon-m_0^+)(\phi_n+\epsilon-m_0^-)}{(\phi_n+\epsilon-A_0)(\phi_n+\epsilon-A_1+\epsilon)}\prod_{s=1}^{|W|}\frac{\phi_n+\epsilon-b_s-\epsilon}{\phi_n+\epsilon-b_s} \times \prod_{j=2,j\neq n}^{k} \frac{(\phi_{n}+\epsilon-\phi_j)^2}{(\phi_{n} - \phi_j)(\phi_n-\phi_j+2\epsilon)}. 
\end{align}
Residue of variable $\phi_n$ at infinity now becomes
\begin{align}
    \frac{1}{2} (-2r_\infty) = 1 \times (-r_\infty).
\end{align}
The factor $-r_\infty$ is credited to residue of $\phi_n$ at infinity. The residue for $\phi_1$ at another $\phi_n+\epsilon$ is effectively $1$. 
Combining with the multiplicity $v-l-1$, the total contribution from $\phi_1$ at the infinity is
\begin{align}
    -r_{\infty} + (v-l-1). \nonumber
\end{align}
The integration of the remaining variables $\phi_{n}$ are determined in the same trick as $\phi_1$. The only difference is that the multiplicity for variable $\phi_n$ is $v-l-n$. 
The combined residues at infinity give
\begin{align}\label{R_k,l}
    & \frac{1}{(v-l)!}  \, \oint_{\infty}
    \prod_{n=1}^{v-l}\ \frac{d{\phi}_{n}}{2\pi{\ii}}\frac{1}{\epsilon} \ \frac{(\phi_n-m_0^+)(\phi_n-m_0^-)}{(\phi_n-A_0)(\phi_n-A_1+\epsilon)}\prod_{s=1}^{|W|}\frac{\phi_n-b_s-\epsilon}{\phi_n-b_s} \times \prod_{j=1,j>n}^{v} \frac{\phi_{nj}^2}{\phi_{nj}^2 - \epsilon^2} \nonumber\\
    & = \frac{1}{(k-l)!}\prod_{j=0}^{v-l-1}(-r_\infty+j)
    =(-1)^{v-l}\left(\!
    \begin{array}{c}
      r_\infty \\
      v-l
    \end{array}
  \!\right)
\end{align}

The remaining contour integration variables $\phi_{n}$, $n=v-l+1,\dots,v$ now take poles at $A_1-\ri \epsilon$, $\ri = 1,\dots,l$, with the contour circulating these poles clockwise. The poles generate the dual Young diagram in the negative FI-parameter chamber
\begin{align}
    \frac{1}{l!}\oint_{\tilde{\CalC}\backslash\infty}\prod_{i=1}^l\frac{d\phi_i}{2\pi {\ii}}\frac{1}{\epsilon} \frac{(\phi_i-m_0^+)(\phi_i-m_0^-)}{(\phi_i-A_0)(\phi_i-A_1+\epsilon)}\prod_{j>i}\frac{\phi_{ij}^2}{\phi_{ij}^2-\epsilon^2}\prod_{s=1}^{|W|}\frac{\phi_i-b_s-\epsilon}{\phi_i-b_s}=\prod_{j=1}^l\frac{P_0(A_1-j\epsilon)}{Y(A_1-j\epsilon)}. \nonumber
\end{align}

Finally we find the normalized vev of the surface defect $\Psi_{\ba,+}$ of the form 
\begin{align}\label{chi_2 Matter Conjecture}
    \Psi_{\ba,+}
    &=\sum_{v=0}^\infty\sum_{l=0}^{v}(-\kq_0)^l\kq_0^{v-l}\left(\!
    \begin{array}{c}
      r_\infty \\
      v-l
    \end{array}
    \!\right)\frac{Q(A_1-\epsilon-l\epsilon)}{\mathfrak{M}_0(A_1-l\epsilon)}
    =(1-\kq_0)^{r_{\infty}}{\Psi_{\ba,-}}.
\end{align}
The fundamental matter contribution $M_0(x)$ is defined by
\begin{align}
    \mathfrak{M}_0(x) = \bE\left[\frac{M_0e^{-x}}{P_1^*}\right] = \prod_{\pm} \Gamma\left(\frac{x-m_0^\pm}{\epsilon}\right) \epsilon^{\frac{x-m_0^\pm}{\epsilon}}. \nonumber
\end{align}
We again see that the mismatch of the instanton partition functions can be factorized into a crossing factor.
We recover the toy model \eqref{eq:crosstoy1} by imposing the condition $m_0^- = A_1 - \epsilon$.

\subsubsection{$N=3$}

Next we consider the $SU(3)$ gauge theory, whose normalized vev of the surface defect is
\begin{align}\label{eq:N=3 ensemble}
    \Psi_{\ba,+} = \sum_{\vec{\lambda}}\sum_{\{U_1,U_2\}} \kq_0^{k_0-k_2}\kq_{1}^{k_1-k_2} \bE \left[\frac{\sum_{\omega=1}^2 \Gamma_{\omega}(\Gamma_{\omega+1} - \Gamma_{\omega})^* - M_{\omega-1}(\Gamma_3 - \Gamma_{\omega})^*}{P_1^*}\right].
\end{align}
The dual characters $V_{1}$ and $V_2$ dependence in the normalized vev of the surface defect are
\begin{align}
    &\sum_{v_1=0}^\infty\sum_{v_2=0}^\infty\kq_0^{k_1-k_3}\kq_1^{k_2-k_3}\sum_{|V_1|=v_1}\sum_{|V_2|=v_2}\mathbb{E}\left[L_0V_1^*+q_1L_1^*V_1+L_1V_2^*+q_1L_2^*V_2-M_0V_1^*-M_1V_2^*\right] \nonumber\\
    &\qquad\qquad\qquad\qquad\qquad\qquad\qquad\times\mathbb{E}\left[P_1WV_1^*+P_1V_1V_2^*-P_1V_1V_1^*-P_1V_2V_2^*\right] \\
    &=\sum_{v_1=0}^\infty\sum_{v_2=0}^\infty\kq_0^{v_1}\kq_1^{v_2}\frac{1}{v_1!}\oint_{\CalC_1}\prod_{i=1}^{v_1}\frac{d\phi^{(1)}_i}{2\pi{\ii}}\frac{1}{\epsilon} \frac{(\phi^{(1)}_i-m_0^+)(\phi^{(1)}_i-m_0^-)}{(\phi^{(1)}_i-A_0)(\phi^{(1)}_i-A_1+\epsilon)}\prod_{i'\neq i}\frac{\phi^{(1)}_{i}-\phi^{(1)}_{i'}}{\phi^{(1)}_{i}-\phi^{(1)}_{i'}+\epsilon}\prod_{b\in 
    W}\frac{b-\phi^{(1)}_{i}+\epsilon}{b-\phi^{(1)}_{i}} \nonumber\\
    &\qquad\qquad\times\frac{1}{v_2!}\oint_{\CalC_2} \prod_{j=1}^{v_2}\frac{d\phi^{(2)}_j}{2\pi{\ii}}\frac{1}{\epsilon}
    \frac{(\phi^{(2)}_{j}-m_1^+)(\phi^{(2)}_{j}-m_1^-)}{(\phi^{(2)}_{j}-A_1)(\phi^{(2)}_{j}-A_2+\epsilon)}\prod_{j'\neq j}\frac{\phi^{(2)}_{j}-\phi^{(2)}_{j'}}{\phi^{(2)}_{j}-\phi^{(2)}_{j'}+\epsilon}\times\frac{\phi^{(1)}_{i}-\phi^{(2)}_{j}+\epsilon}{\phi^{(1)}_{i}-\phi^{(2)}_{j}} \nonumber
\end{align}
We deform both contour $\CalC_1$ and $\CalC_2$ to evaluate the integral. The integrations after deforming the contour is performed according to the following steps. 
We start with the variables $\{\phi^{(2)}_{j}\}$. Suppose there exists $v_2-l_2$ variables $\phi^{(2)}_{j}$, $j=1,\dots,v_2-l_2$ picking up pole at infinity,
\begin{align}\label{v_2 infty, N=3}
    \frac{1}{(v_2-l_2)!}\oint_\infty 
    & \prod_{j=1}^{v_2-l_2}
    \frac{d\phi^{(2)}_j}{2\pi{\ii}}\frac{1}{\epsilon}
    \frac{(\phi^{(2)}_{j}-m_1^+)(\phi^{(2)}_{j}-m_1^-)}{(\phi^{(2)}_{j}-A_1)(\phi^{(2)}_{j}-A_2+\epsilon)} \prod_{j'=1,j'\neq j}^{v_2} \frac{\phi^{(2)}_{j}-\phi^{(2)}_{j'}}{\phi^{(2)}_{j}-\phi^{(2)}_{j'}+\epsilon}\prod_{i=1}^{v_1}\frac{\phi^{(1)}_{i}-\phi^{(2)}_{j}+\epsilon}{\phi^{(1)}_{i}-\phi^{(2)}_{j}}.
\end{align}
The residue picked up by the  $\phi^{(2)}_{j}$ at infinity is computed to be 
\begin{align}
    \frac{1}{\epsilon}(m_1^++m_1^--A_1-A_2+\epsilon)+v_1=r_\infty^{(2)}+v_1. \nonumber
\end{align}
The integration in the Eq.~\eqref{v_2 infty, N=3} has the same structure as in Eq.~\eqref{R_k,l}, which can be calculated by the same manner. The overall residue contribution at infinity of the quiver node ${\bf V}_2$ is given by
\begin{align}
    (-1)^{v_2-l_2}\left(\!
    \begin{array}{c}
      r_\infty^{(2)}+v_1 \\
      v_2-l_2
    \end{array}
    \!\right). \nonumber
\end{align}
Next the integration is performed over the remaining variables $\phi^{(2)}_j$, $j=v_2-l_2+1,\dots,v_2$, by taking the residues at the poles $A_2 - \ri \epsilon$, $\ri = 1,\dots,l_2$. Then summing over $(v_2,l_2)$ at the quiver node ${\bf V}_2$ yields
\begin{align}
    & \sum_{v_2=0}^\infty\sum_{l_2=0}^{v_2} (-1)^{v_2-l_2}
    \left(\!
    \begin{array}{c}
      r_\infty^{(2)}+v_1 \\
      v_2-l_2
    \end{array}
    \!\right)\kq_1^{v_2}
    \frac{1}{l_2!}\oint_{\tilde{\CalC}_2\backslash\infty} \prod_{j=1}^{l_2}\frac{d\phi^{(2)}_j}{2\pi{\ii}}\frac{1}{\epsilon}
    \frac{(\phi^{(2)}_{j}-m_1^+)(\phi^{(2)}_{j}-m_1^-)}{(\phi^{(2)}_{j}-A_1)(\phi^{(2)}_{j}-A_2+\epsilon)} \nonumber\\
    &\qquad \qquad\qquad \qquad \qquad\qquad\qquad \qquad \qquad\qquad\times \prod_{j'=1,j'\neq j}^{l_2}\frac{\phi^{(2)}_{j}-\phi^{(2)}_{j'}}{\phi^{(2)}_{j}-\phi^{(2)}_{j'}+\epsilon}\prod_{i=1}^{v_1}\frac{\phi^{(1)}_{i}-\phi^{(2)}_{j}+\epsilon}{\phi^{(1)}_{i}-\phi^{(2)}_{j}} \nonumber \\
    & = (1-\kq_1)^{r_\infty^{(2)}+v_1} \sum_{l_2=0}^\infty \kq_1^{l_2} \oint_{\tilde{\CalC}_2\backslash\infty} \prod_{j=1}^{l_2}\frac{d\phi^{(2)}_j}{2\pi{\ii}}\frac{1}{\epsilon}
    \frac{(\phi^{(2)}_{j}-m_1^+)(\phi^{(2)}_{j}-m_1^-)}{(\phi^{(2)}_{j}-A_1)(\phi^{(2)}_{j}-A_2+\epsilon)}\prod_{j'=1,j'\neq j}^{l_2}\frac{\phi^{(2)}_{j}-\phi^{(2)}_{j'}}{\phi^{(2)}_{j}-\phi^{(2)}_{j'}+\epsilon}\prod_{i=1}^{v_1}\frac{\phi^{(1)}_{i}-\phi^{(2)}_{j}+\epsilon}{\phi^{(1)}_{i}-\phi^{(2)}_{j}}. \nonumber
\end{align}
We then move on to the integration of the variables $\{\phi^{(1)}_{i}\}$ corresponding to the quiver node ${\bf V}_1$. We assume there exist $v_1-l_1$ variables $\phi^{(1)}_{i}$, $i=1,\dots,v_2-l_2$, that pick up pole at infinity. The residue picked up by the $\phi^{(1)}_{i}$ at infinity is equal to
\begin{align}
r_\infty^{(1)}+|W|-l_2.
\end{align}
The integration at infinity again follows the Eq. \eqref{R_k,l}. We denote
\begin{align}
    F_1(l_1) = \frac{1}{l_1!}\oint_{\tilde{\CalC}_1\backslash\infty} & 
    \prod_{i=1}^{l_1}\frac{d\phi^{(1)}_i}{2\pi{\ii}}\frac{-1}{\epsilon}\frac{(\phi^{(1)}_i-m_0^+)(\phi^{(1)}_i-m_0^-)}{(\phi^{(1)}_i-A_0)(\phi^{(1)}_i-A_1+\epsilon)} \nonumber\\
    & \times \prod_{i'=1,i'\neq i}^{l_1}\frac{\phi^{(1)}_{i}-\phi^{(1)}_{i'}}{\phi^{(1)}_{i}-\phi^{(1)}_{i'}+\epsilon}\prod_{b\in 
    W}\frac{b-\phi^{(1)}_{i}+\epsilon}{b-\phi^{(1)}_{i}} \frac{\phi^{(1)}_{i}-A_2+(l_2+1)\epsilon}{\phi^{(1)}_{i}-A_2+\epsilon} \nonumber
\end{align}
for integration over the remaining $l_1$ variables $\{\phi^{(1)}_{i}\}$, which take poles at $A_1 - \ri_1\epsilon$, $\ri_1=1, 2,\dots$, or $A_2-\ri_2\epsilon$, $\ri_2 = 1,\dots,l_2$.  
Summing over $(v_1,l_1)$ in the quiver node ${\bf V}_1$ results in
\begin{align}
    \sum_{v_1=0}^\infty\sum_{l_1=0}^{v_1}(-1)^{v_1-l_1}\left(\!
    \begin{array}{c}
      r_\infty^{(1)}+|W|-l_2 \\
      v_1-l_1
    \end{array}
    \!\right)
    \kq_0^{v_1}(1-\kq_1)^{v_1} F_1(l_1)
    = (1-\kq_0(1-\kq_1))^{r_\infty^{(1)}+|W|-
    l_2}\sum_{l_1=0}^\infty\kq_0^{l_1}(1-\kq_1)^{l_1}F_1(l_1). \nonumber
\end{align}

The normalized vev of the surface defect for $U(3)$ gauge theory now reads
\begin{align}\label{N=3 wave function}
    {\Psi}_{\ba,+}
    =&(1-\kq_0+\kq_0\kq_1)^{r_\infty^{(1)}+|W|}(1-\kq_1)^{r_\infty^{(2)}} \\
    &\times\sum_{l_1=0}^\infty\sum_{l_2=0}^\infty\kq_0^{l_1}\kq_1^{l_2}(1-\kq_1)^{l_1}(1-\kq_0+\kq_0\kq_1)^{-l_2}
    \left.\mathbb{E}\left[\frac{\sum_\omega\Gamma_\omega(\Gamma_{\omega+1}-\Gamma_\omega)^*}{P_1^*}-\frac{\sum_\omega M_{\omega-1}(\Gamma_N-\Gamma_\omega)^*}{P_1^*}\right]\right|_\text{Dual} \nonumber\\
    =&(1-\kq_0+\kq_0\kq_1)^{r_\infty^{(1)}+|W|}(1-\kq_1)^{r_\infty^{(2)}} \nonumber\\
    &\times\sum_{l_1=0}^\infty\sum_{l_2=0}^\infty\kq_{0,\text{eff}}^{l_1}\kq_{1,\text{eff}}^{l_2}
    \left.\mathbb{E}\left[\frac{\Gamma_NF_{\geq1}^*-\sum_{\omega=1}^2 F_{\geq\omega}(F_{\geq\omega}-F_{\geq\omega+1})^*}{P_1^*}\right]\mathbb{E}\left[-\frac{\sum_{\omega=1}^2M_{\omega-1}F_{\geq\omega}^*}{P_1^*}\right]\right|_\text{Dual}. \nonumber
\end{align}
The dual characters
\begin{align}
    \Gamma_N-\Gamma_\omega=\sum_{l\geq\omega}e^{A_l}q_1^{-d_{\omega}^{(l)}}=F_{\geq\omega}
\end{align}
label the dual Young diagram $\vec{\lambda}_{\rm dual}$ in the negative FI-parameter chamber with $l_\omega=\sum_{l\geq\omega}d_\omega^{(l)}$. 

The fractional couplings in the negative FI-parameter chamber, are relate to the original ones by
\begin{align}\label{eq: q-eff, N=3}
    \kq_{0,\text{eff}}=\kq_0(1-\kq_1),\quad  {\kq}_{1} = {\kq}_{1,\text{eff}} \left( 1 - {\kq}_{0, \text{eff}} \right) \ .
\end{align}


\subsubsection{Toy Model: the effective instanton counting}

Let us demonstrate that in the case of quiver gauge theories, it is not only natural but crucial to mutate the instanton counting parameters in crossing between the chambers to have the equality of the two instanton partition functions.

{}We consider the following simplified model of the $SU(3)$ gauge group with $6$ fundamental flavors by forcing the following conditions on the moduli parameters (sometimes this procedure is called simply  higgsing, since with such relations between the Coulomb moduli and masses some of the hypermultiplets become massless and can be given a vev, thereby higgsing the gauge group to a subgroup).  
\begin{align}
    &a_0+\epsilon_1=m_0^+,\quad  a_1=m_1^+, \quad a_2=m_2^+, \nonumber\\
    &m_0^-=a_1-\epsilon_+, \quad m_1^-=a_2-\epsilon_+-\epsilon_1, \quad m_2^-=a_0-\epsilon_+. \nonumber
\end{align}
This condition forces the only surviving Young diagrams in both positive and negative FI-parameter chambers to be the single rows in the $\pm\epsilon_2$ direction, 
\begin{align}
    & \vec{\lambda}_{\zeta_{\BR}>0} = \left(\footnotesize\yng(2)\cdots\footnotesize\yng(1), \ \emptyset, \ \emptyset\right), \quad \vec{\lambda}_{\zeta_{\BR}<0} = (\emptyset, \ \footnotesize\yng(2)\cdots\footnotesize\yng(1), \  \emptyset). \nonumber
\end{align}
A co-dimension two surface defect is introduced as a $\BZ_3$ orbifolding. 
The instanton paritition function in the positive FI-parameter ${{\zeta}_{\BR}}>0$ chamber is
\begin{align}
    \CalZ_{+}=\sum_{L=0}^\infty\kq^L\prod_{j=1}^L\frac{j\epsilon_2-\epsilon_1}{j\epsilon_2}\frac{a_2-2\epsilon_1-a_0-j\epsilon_2}{a_2-\epsilon_1-a_0-j\epsilon_2}\left[1-\kq_0+\kq_0\kq_1\frac{a_2-2\epsilon_1-a_0-L\epsilon_2}{a_2-\epsilon_1-a_0-L\epsilon_2}\right].
\end{align}
The instanton partition function in the negative FI-parameter ${{\zeta}_{\BR}}<0$ chamber is 
\begin{align}
    \CalZ_{-}=\sum_{L=0}^\infty\kq^L\prod_{j=1}^L\frac{j\epsilon_2-\epsilon_1}{j\epsilon_2}\frac{a_2-2\epsilon_1-a_0-j\epsilon_2}{a_2-\epsilon_1-a_0-j\epsilon_2}\left[1-\kq_{1,\text{eff}}+\kq_{0,\text{eff}}\kq_{1,\text{eff}}\frac{a_0+2\epsilon_1-a_2+L\epsilon_2}{a_0+\epsilon_1-a_2+L\epsilon_2}\right].
\end{align}
In the NS-limit we focus on the surface contributions
\begin{subequations}
\begin{align}
    {\Psi}_{\ba,+}
    &= \left[1-\kq_0+\kq_0\kq_1\frac{a_2-2\epsilon-a_0-L\epsilon_2}{a_2-\epsilon-a_0-L\epsilon_2}\right], \\
    {\Psi}_{\ba,-}
    &=\left[1-\kq_{1,\text{eff}}+\kq_{0,\text{eff}}\kq_{1,\text{eff}}\frac{a_0+2\epsilon-a_2+L\epsilon_2}{a_0+\epsilon-a_2+L\epsilon_2}\right].
\end{align}
\end{subequations}
By denoting $A_0=a_0+\epsilon$, $A_1=a_1$, $A_2=a_2$, and $W=e^{a_0+L\epsilon_2}$, the residues at infinity are
\begin{subequations}
\begin{align}
    r_\infty^{(1)}&=\frac{1}{\epsilon}(m_0^++m_0^--A_0-A_1+\epsilon)=0, \nonumber\\
    r_\infty^{(2)}&=\frac{1}{\epsilon}(m_1^++m_1^--A_1-A_2+\epsilon)=-1. \nonumber
\end{align}
\end{subequations}
We see that both the crossing factor and the transformation of the fractional couplings is needed  to recover the  partition function in the positive FI-parameter chamber from the negative chamber counterpart: 
\begin{align}
    &(1-\kq_0+\kq_0\kq_1)^{r_\infty^{(1)}+|W|}(1-\kq_1)^{r_\infty^{(2)}} \Psi_{\ba,-}(a_0,a_2,\epsilon,\kq_{0,\text{eff}},\kq_{1,\text{eff}}) \nonumber\\
    &=(1-\kq_0+\kq_0\kq_1)^{1}(1-\kq_1)^{-1}\left[1-\kq_{1,\text{eff}}+\kq_{0,\text{eff}}\kq_{1,\text{eff}}\frac{a_0+2\epsilon-a_2+L\epsilon_2}{a_0+\epsilon-a_2+L\epsilon_2}\right] \nonumber\\
    &=\frac{(1-\kq_0+\kq_0\kq_1)}{(1-\kq_1)}\left[1-\frac{\kq_1}{(1-\kq_0+\kq_0\kq_1)}+\frac{\kq_0\kq_1(1-\kq_1)}{(1-\kq_0+\kq_0\kq_1)}\frac{a_0+2\epsilon-a_2+L\epsilon_2}{a_0+\epsilon-a_2+L\epsilon_2}\right] \nonumber\\
    &=\frac{1-\kq_0+\kq_0\kq_1}{1-\kq_1}-\frac{\kq_1}{1-\kq_1}+\kq_0\kq_1\frac{a_0+2\epsilon-a_2+L\epsilon_2}{a_0+\epsilon-a_2+L\epsilon_2} \nonumber\\
    &=\Psi_{\ba,+}(a_0,a_2,\epsilon,\kq_0,\kq_1).
\end{align}

\section{Limit shape instanton configuration} \label{sec:limitshape}

In the presence of a full-type/regular surface defect, the bulk contribution is identified with the contribution in the representation $\CalR_{N-1}$ of $\BZ_N$ orbifolding. The projection ${\pi}_{N}$ of the moduli space of instantons in the presence of the surface defect to the moduli space of instantons in the bulk descends to the map ${\pi}_{N}: {\CalP}^{N} \to {\CalP}^{N}$ between the sets of fixed points. Thus, a new set of Young diagrams $\vec{\Lambda} = {\pi}_{N} ({\vec{\lambda}})$ can be constructed
\begin{equation}
\Lambda^{(\alpha)}_i=\left\lfloor\frac{\lambda^{(\alpha)}_{i}+c(\alpha)}{N}\right\rfloor,
\label{def:bulkyoung}
\end{equation}
where $\lfloor\cdot\rfloor$ is the floor operator. The bulk partition function is a grand canonical ensemble over the bulk intanton configuration
\begin{align}\label{def:bulkZ}
    \CalZ_{\rm bulk}({\ba},{\bm}^\pm.\kq) = \sum_{\vec{\Lambda}} \kq^{|\vec{\Lambda}|} \mu_{\rm bulk} ({\ba},{\bm}^\pm,\kq)[\vec{\Lambda}]
\end{align}
with the bulk pseudo-measure is of the form in Eq.~\eqref{def:bulk}. The full $Y(x)$ function is defined on the bulk Young daigram 
\begin{align}
    Y(x)[\vec{\Lambda}] = \prod_{\omega}\bE \left[-e^x\tilde{S}_{\omega}^*[\vec{\lambda}]\right] = \bE \left[ -e^{x}(\tilde{N} - P_{12}\tilde{K}_{N-1}[\vec{\lambda}])  \right] = \bE \left[ -e^x \tilde{S}^*[\vec{\Lambda}] \right].
\end{align}

In the NS-limit $\epsilon_2\to0$, with $\epsilon_1\equiv\epsilon$ fixed, the summation in the bulk partition function \eqref{def:bulkZ} is dominated by a limit-shape configuration $\vec{\Lambda}_*$
\begin{align}
    \CalZ_{\rm bulk} \approx \kq^{|\vec{\Lambda}_*|} \mu_{\rm bulk}[\vec{\Lambda}_*] + \cdots 
\end{align}

We denote $Y(x) \equiv Y(x)[\vec{\Lambda}_*]$ based on the limit shape configuration. $Y(x)$ satisfies 
\begin{align}
    (1+\kq)T(x) & = \langle \CalX(x) \rangle = \left\langle Y(x+\epsilon)  + \frac{\kq P(x)}{Y(x)} \right\rangle = Y(x+\epsilon)  + \frac{\kq P(x)}{Y(x)}.
\end{align}
Function $T(x)$ is a degree $N$ polynomial. The function $Y(x)$ can be expressed as ratio of two entire functions
\begin{align}
    Y(x) = \frac{Q(x)}{Q(x-\epsilon)}, \quad Q(x)=\mathbb{E}\left[-\frac{e^x\tilde{S}^*[\vec{\lambda}_*]}{P_1^*}\right].
\end{align}
In particular, function $Q(x)$ is the solution of T-Q equation \cite{Nikita-Shatashvili} 
\begin{align}\label{Baxter Q Hyper}
    Q(x+\epsilon)+\kq P(x)Q(x-\epsilon)=(1+\kq)T(x)Q(x),
\end{align}
which matches with the $T-Q$ Baxter equation of the $\spch$ spin chain \cite{Nikita-Shatashvili,Dorey:2011pa,HYC:2011}. See \cite{Nikita-Pestun-Shatashvili,Nikita:I,Nikita:II,Chen:2019vvt} for more details on the origin of the Baxter equation in gauge theory. 

The functions $Y(x)$ and $Q(x)$ are proven to be applicable tools to investigate the structure of the limit-shape instanton configuration $\vec{\Lambda}_*$. 
The zeros of $Y(x)$ and $Q(x)$ are separated by at least $N$ columns in the limit shape $\vec{\lambda}$ by the scaling of $\epsilon_2\to\frac{\epsilon_2}{N}$. 
The bulk $Q(x)$ function can be expressed in the context of new Young diagrams $\vec{\Lambda}_*$ :
\begin{align}\label{Q function}
	Q(x)=\prod_{\alpha=1}^{N}\frac{(-\epsilon)^\frac{x-\tilde{a}_\alpha}{\epsilon}}{\Gamma\left(-\frac{x-\tilde{a}_\alpha}{\epsilon}\right)}\prod_{\ri=1}^\infty\frac{x-\tilde{a}_\alpha-(\ri-1)\epsilon-\xi_{\alpha,\ri}}{x-\tilde{a}_\alpha-(\ri-1)\epsilon};\quad \xi_{\alpha,\ri}=\epsilon_2\Lambda_{*,\ri}^{(\alpha)}.
\end{align}
The asymptotic behavior of $\xi_{\alpha,\ri}=\epsilon_2\Lambda^{(\alpha)}_\ri$ is restricted by the Baxter equation \eqref{Baxter Q Hyper}. $\xi_{\alpha,\ri}$ can be determined via a order by order perturbation expansion of \eqref{Baxter Q Hyper} in $\kq$. In the zeroth order
\begin{align}
T_0(x)=\frac{Q_0(x+\epsilon)}{Q_0(x)}=Y_0(x+\epsilon)=Y(x+\epsilon)[\emptyset]=\prod_{\alpha=1}^N(x-\tilde{a}_\alpha+\epsilon).
\end{align}
The zeroth order $Q_0(x)$ is identified to the leading Gamma function in Eq.~\eqref{Q function}. And thus $\xi_{\alpha,\ri}$ are restrained to be at most of order $\kq$ for all $\alpha=1,\dots,N$ and $\ri=1,2,\dots$. In the $\kq^1$ order, we take $x=\tilde{a}_\alpha+\xi_{\alpha,1}$ which are the zeros of $Q(x)$ function : 
\begin{align}
0&=Q(\tilde{a}_\alpha+\xi_{\alpha,1}+2\epsilon)+\kq P(\tilde{a}_\alpha+\xi_{\alpha,1}) Q_0(\tilde{a}_\alpha-\epsilon+\xi_{\alpha,1}) \nonumber \\
&=Q_0(\tilde{a}_\alpha+\xi_{\alpha,1}-\epsilon)\left[\frac{1}{T_0(\tilde{a}_\alpha+\xi_{\alpha,1})T_0(\tilde{a}_\alpha+\xi_{\alpha,1}-\epsilon)}\frac{\epsilon}{\epsilon+\xi_{\alpha,1}}+\kq P(\tilde{a}_\alpha)\right] \nonumber
\end{align}
which we find 
\begin{align}\label{xi_1}
    \xi_{\alpha,1}
    &=-\frac{\kq P(\tilde{a}_\alpha)}{\epsilon\prod_{\beta\neq\alpha}(\tilde{a}_\alpha-\tilde{a}_\beta+\epsilon)(\tilde{a}_\alpha-\tilde{a}_\beta)}.
\end{align}
Similar procedure applied to the other zeros of $Q(x)$. It generates a order by order expansion that determines asymptotics of $\xi_{\alpha,\ri}$ by the recursive formula:
\begin{align}\label{xi-recursive}
\xi_{\alpha,\ri+1}=\frac{\kq P(\tilde{a}_\alpha+\ri\epsilon)\xi_{\alpha,\ri}}{\prod_{\beta=1}^N(\tilde{a}_\alpha+(\ri+1)\epsilon-\tilde{a}_\beta)(\tilde{a}_\alpha+\ri\epsilon-\tilde{a}_\beta)};\quad \ri=1,2,\dots.
\end{align}

Recursive relation \eqref{xi-recursive} shows that the partition $\vec{\Lambda}_*$ is very steep $\Lambda_{*,i+1}^{(\alpha)}\ll\Lambda^{(\alpha)}_{*,1}$, i.~e. its dual partition $\vec{\Lambda}^t$ has an almost flat plateau structure. The set of \emph{jumps in the bulk} $\{J'\}$ is defined by
\begin{align}
    \{J'\} = \left\{ J\in\BZ_{\geq0} | \quad \Lambda^{t,(\alpha)}_{*,J'}-\Lambda^{t,(\alpha)}_{*,J'+1}=1 \right\} \nonumber
\end{align} 
that locates where the elevation of dual partition $\vec{\Lambda}_*^t$ changes. Virtual character $\tilde{S} = \tilde{S}[\vec{\lambda}_*]$ now can be separated into two parts
\begin{align}\label{S-tilde}
	\tilde{S} = \tilde{S}[\vec{\lambda}] = \tilde{S}[\vec{\Lambda}_*]=\sum_{\alpha}e^{A_\alpha}
	+\sum_{\alpha}\sum_{\{J'\}}e^{\tilde{a}_\alpha}q_2^{J'}q_1^{\Lambda^{t,(\alpha)}_{J'+1}}(1-q_1)=F_N+P_1W,
\end{align}
with $A_\alpha = \tilde{a}_{\alpha} + \epsilon \Lambda_{*,1}^{t,(\alpha)}$. The $Y(x)$ function defined on the limit shape configuration $\vec{\Lambda}$ reads
\begin{align}
    Y(x) = \prod_{\alpha=1}^{N} (x-A_\alpha) \prod_{\{J'\}} \frac{x - \tilde{a}_{\alpha} - \epsilon_2 J' - \epsilon\Lambda_{*,J'+1}^{t,(\alpha)} }{x - \tilde{a}_\alpha - \epsilon_2 J' - \epsilon\Lambda_{*,J'+1}^{t,(\alpha)} - \epsilon}.
    \label{eq:Y-limit}
\end{align}

The bulk virtual character is of the form 
\begin{align}
	\Gamma_{N} =  \tilde{S}[\vec{\Lambda}_*]=\sum_{\alpha}e^{A_\alpha}
	+\sum_{\alpha}\sum_{\{J'\}}e^{\tilde{a}_\alpha}q_2^{J'}q_1^{\Lambda^{t,(\alpha)}_{*,J'+1}}(1-q_1)=F_N+P_1W.
\end{align}

Our main focus lies on the normalized vev of the surface defect, which considers all allowed surface configuration $\vec{\lambda}$ on top of the limit-shape bulk instanton $\vec{\Lambda}_*$: 
\begin{equation}\label{sandwich}
	\Lambda^{t,(\alpha)}_{*,J}=\lambda_{NJ-c(\alpha)}^{t,(\alpha)}\geq\lambda_{\omega+NJ-c(\alpha)}^{t,(\alpha)}\geq\lambda_{N+NJ-c(\alpha)}^{t,(\alpha)}=\Lambda_{*,J+1}^{t,(\alpha)}.
\end{equation}
The first $J=0$ is special since it lacks an upper bound. We find the virtual characters $\{\Gamma_\omega\}$ of the form
\begin{subequations}
\begin{align}
    \Gamma_N&=\tilde{S}=\sum_{\alpha}e^{\tilde{a}_\alpha}q_1^{\lambda^{t,(\beta)}_{N-c(\beta)}}+\sum_\beta\sum_{\{J'\}}e^{\tilde{a}_\beta}q_2^{J'}q_1^{\lambda^{t,(\beta)}_{N+NJ'-c(\beta)}}\left(1-q_1\right)=F_N+P_1W, \\
    \Gamma_\omega
	&=\tilde{S}_0+\cdots+\tilde{S}_\omega \nonumber\\
	&=\sum_{c(\alpha)<\omega} e^{\tilde{a}_\alpha}q_1^{\lambda^{t,(\alpha)}_{\omega-c(\alpha)}}+\sum_{\alpha}\sum_{\{J'_\omega\}}e^{\tilde{a}_\alpha}q_2^{J'_\omega}q_1^{\lambda^{t,(\alpha)}_{N+NJ'_\omega-c(\alpha)}}\left(1-q_1\right)=F_\omega+P_1U_\omega.
\end{align}
\end{subequations}
The $F_\omega$'s denote the $N-2$ Young diagrams $\vec{\lambda}_{\rm tail} = ({\lambda}^{(0)}_{\rm tail},{\lambda}^{(1)}_{\rm tail},\dots,{\lambda}^{(N-2)}_{\rm tail})$ attaching to first $J=0$ of limit shape $\vec{\Lambda}$, which we call a \emph{tail}. Each tail Young diagram $\lambda^{(\omega)}$, $\omega=0,1,\dots,N-2$, is the collection of row of boxes of non-negative length
$\lambda^{(\omega)}_{{\rm tail}} = (\lambda^{(\omega)}_{{\rm tail},1},\lambda^{(\omega)}_{{\rm tail},2},\dots)$ obeying
\begin{subequations}
\begin{align}
    & \lambda^{(\omega)}_{{\rm tail},1} \leq N-1-\omega; \nonumber \\ 
    & \lambda^{(\omega)}_{{\rm tail},i} \geq \lambda^{(\omega)}_{{\rm tail},i+1}, \quad i=1,2,\dots \nonumber
\end{align}
\end{subequations}
The jump sets $\{J'_\omega\}$ are defined by 
\begin{align}
    \{J'_{\omega}\} = \left\{ J\in\BZ_{\geq0}|\quad \lambda^{t,(\alpha)}_{NJ-c(\alpha)} - \lambda^{t,(\alpha)}_{\omega+NJ - c(\alpha)} = 1 \right\}. \nonumber
\end{align}
The jumps sets $\{J'_\omega\}$ are restricted by Eq.~\eqref{sandwich}:
\begin{equation}
	\{J'_1\}\subset\{J'_2\}\subset\cdots\subset\{J'_N\}=\{J'\},
\end{equation}

Once the intanton configuration in the bulk is locked to limit shape in the NS-limit, the surface defect partition function in Eq.~\eqref{eq:Z-defect} becomes
\begin{align}
    \CalZ  = e^{\frac{1}{\epsilon_2}\CalW({\ba},{\bm}^\pm,\kq)} \times \sum_{\vec{\lambda}_*}\prod_{\omega=0}^{N-2}\kq_{\omega}^{k_{\omega}-k_{N-1}} \mu_{\rm surface}({\ba},{\bm},\kq,\vec{z})[\vec{\lambda}_*]
\end{align}

The normalized vev of the surface defect $\Psi_{\ba}$ is identified as an ensemble over all allowed surface configurations, namely the arrangements of jumps $\{J_\omega'\}_{\omega=1}^{N-1}$ and tail Young diagrams $\lambda^{t,(\alpha)}_{\omega-c(\alpha)}$ connected to the very bottom of limit shape $\vec{\Lambda}_{*}$. See fig \ref{fig:jumps and tail} for illustration.
\begin{align}
	\Psi_{\ba}
	& = \sum_{\vec{\lambda}_*}\prod_{\omega=0}^{N-2} \kq_{\omega}^{k_{\omega} - k_{N-1}} \mu_\text{surface}[\vec{\lambda}] \nonumber\\
	&=\sum_{\rm \vec{\lambda}_{\rm tail}} \sum_{\{U_{\omega}\}}\prod_{\omega=0}^{N-2}\kq_\omega^{k_\omega-k_{N-1}}\mathbb{E}\left[\frac{\sum_{\omega=1}^{N-1}\Gamma_\omega(\Gamma_{\omega+1}-\Gamma_\omega)^*}{P_1^*}-\frac{\sum_{\omega=1}^{N-1}M_{\omega-1}(\Gamma_N-\Gamma_\omega)^*}{P_1^*}\right].
\end{align}

\section{Details of the second Hamiltonian} \label{sec:h2=H2}

We derived the second Hamiltonian from the gauge theory in \eqref{eq:Hamiltonian N}, which after resetting the zero point energy \eqref{eq:E_Izero} becomes
\begin{align}
    \hat{\rm H}_2 = \sum_{\omega}-\frac{(1-\kq)}{2}\left[({\delta}_{\omega})^2-2m_{\omega+1}^{+}\delta_{\omega}\right]-(\kq_{\omega+1} u_{\omega+1})\left(\delta_\omega\right)\left(\delta_\omega-m_{\omega+1}^{+}+m_{\omega+1}^{-} \right)
\end{align}
with $\delta_{\omega} = \epsilon \nabla^z_\omega - a_{\omega+1} + m_{\omega+1}^+$ such that the 
\begin{align}
    \hat{\rm H}_2 \Psi_{\ba} = (1-\kq)E_2 \Psi_{\ba}.
\end{align}
By multiplying the normalized vev of the surface defect partition function with a perturbative factor
$$
    \left[ \prod_{\omega=0}^{N-1} z_{\omega}^\frac{ - a_{\omega+1} + m_{\omega+1}^+}{\epsilon} \right] \Psi_\ba
$$
The operator $\delta_{\omega}$ becomes $\epsilon \nabla^z_{\omega}$.

The second Hamiltonian of the $\spch$ spin chain $\hat{h}_2$ is obtained from the trace of monodromy matrix \eqref{def:trans XXX}
\begin{align}
    \hat{h}_2
    =&\Tr \begin{pmatrix} \kq & 0 \\ 0 & 1 \end{pmatrix} \sum_{\omega>\omega'} (-\mu_{\omega}+\CalL_{\omega})(-\mu_{\omega'}+\CalL_{\omega'}) \\
    =&\sum_{\omega>\omega'}(\kq\gamma_{\omega'}-\gamma_{\omega})(\gamma_{\omega}-\gamma_{\omega'})\beta_{\omega}\beta_{\omega'} + \sum_{\omega>\omega'} -\kq (m_{\omega}^-\gamma_{\omega'}\beta_{\omega'}+m_{\omega'}^-\gamma_{\omega}\beta_{\omega})+ \kq (m_{\omega'}^- - m_{\omega'}^+ + \epsilon )\gamma_{\omega'}\beta_{\omega} \nonumber\\
    & + \sum_{\omega>\omega'}  (m_{\omega+1}^{+}-\epsilon)\gamma_{\omega'}\beta_{\omega'}+(m_{\omega'+1}^+-\epsilon)\gamma_{\omega}\beta_{\omega}+(m_{\omega}^--m_{\omega+1}^+ + \epsilon)\gamma_{\omega}\beta_{\omega'} + \kq m_{\omega}^- m_{\omega'}^- + (m_{\omega+1}^+-\epsilon)(m_{\omega'+1}^+-\epsilon) \nonumber
\end{align}

The coefficient of $\epsilon^2\partial^2_{z_{\omega}}$ is obtained by considering $\omega=\omega'+1$,
\begin{align}
    &-(\kq\gamma_{\omega'}-\gamma_{\omega'+1})(\gamma_{\omega'+1}-\gamma_{\omega'}) \frac{\partial^2}{\partial z_{\omega'}^2} \nonumber\\
    &= -\frac{1}{\kq-1}\left(\kq u_{\omega'} - \kq_{\omega'+1}u_{\omega'+1}\right) z_{\omega'}^2\frac{\partial^2}{\partial z_{\omega'}^2} \nonumber \\
    &= \kq (\nabla^z_{\omega'})^2 - \kq_{\omega'+1}u_{\omega'+1} (\nabla^z_{\omega'})^2 - \kq \nabla^z_{\omega'} + \kq_{\omega'+1}u_{\omega'+1}\nabla^z_{\omega'}.
\end{align}
The cross term between $\partial_{z_{\omega}}\partial_{z_{\omega'}}$ comes from 4 sources: $\beta_{\omega}\beta_{\omega'}$, $\beta_{\omega+1}\beta_{\omega'}$, $\beta_{\omega}\beta_{\omega'+1}$, and $\beta_{\omega+1}\beta_{\omega'+1}$. Coefficients are found by
\begin{align}
    &(\kq\gamma_{\omega'}-\gamma_{\omega})(\gamma_{\omega}-\gamma_{\omega'}) - (\kq\gamma_{\omega'}-\gamma_{\omega+1})(\gamma_{\omega+1}-\gamma_{\omega'}) \nonumber\\
    & - (\kq\gamma_{\omega'+1}-\gamma_{\omega})(\gamma_{\omega}-\gamma_{\omega'+1}) 
    + (\kq\gamma_{\omega'+1}-\gamma_{\omega+1})(\gamma_{\omega+1}-\gamma_{\omega'+1}) \nonumber\\
    &=(1+\kq)z_{\omega}z_{\omega'}.
\end{align}
We find the second order differential terms obeying 
\begin{align}
    h_2|_{\partial^2}=\frac{1+\kq}{2}(\epsilon\nabla^z_c)^2+\hat{\rm H}_2|_{\partial^2}, \quad \nabla_c^z :=  \sum_{\omega}\nabla^z_{\omega}.
\end{align}

For first order differentiation terms, $\partial_{z_{\omega}}$ in $\hat{\rm H}_2$ has coefficients:
\begin{align}
    &(1-\kq+\kq_{\omega+1}u_{\omega+1}) m_{\omega+1}^+ z_{\omega} - (\kq_{\omega+1}u_{\omega+1}) m_{\omega+1}^-  z_{\omega}  \nonumber\\
    &=(\kq-1)\gamma_{\omega}m_{\omega+1}^+- (\kq-1) \gamma_{\omega+1} m_{\omega+1}^-.
\end{align}
In $h_2$, a single $\beta_{\omega}$ has coefficient
\begin{align}
    &\left[\sum_{\omega'}-\kq m_{\omega'}^-\right] \gamma_{\omega} - \kq m_{\omega}^-\gamma_{\omega} + \left[\sum_{\omega'}(m_{\omega'+1}^+ - \epsilon)\right] \gamma_{\omega} + (m_{\omega+1}^+ - \epsilon)\gamma_{\omega} \nonumber \\
    & +  \sum_{\omega'<\omega} \kq(m_{\omega'}^--m_{\omega'+1}^+ + \epsilon)\gamma_{\omega'} + \sum_{\omega'>\omega} (m_{\omega'}^--m_{\omega'+1}^+ + \epsilon)\gamma_{\omega'} 
\end{align}
The $\epsilon\partial_{z_{\omega}}$ comes from 2 terms: $\beta_{\omega}$ and $\beta_{\omega+1}$, which we find
\begin{align}
    \left[\sum_{\omega'} - \kq m_{\omega'}^- + (m_{\omega'+1}^+ - \epsilon)\right]z_{\omega} + (1-\kq) m_{\omega+1}^- \gamma_{\omega+1} - (1-\kq) (m_{\omega+1}^+ + \epsilon)\gamma_{\omega}.
\end{align}
This leads to
\begin{align}
    \hat{h}_2 = & \hat{\rm H}_2 + \frac{1+\kq}{2}\left(\frac{\hat{\rm H}_1}{\kq-1}\right)^2 + (N-1)\epsilon\left(\frac{\hat{\rm H}_1}{1-\kq}\right) + (\kq m_c^- - m_c^+)\left(\frac{\hat{\rm H}_1}{1-\kq}\right) + \cdots
\end{align}
Last is the constant terms. $\hat{\rm H}_2$ has no constant term after resetting zero point energy.
The constant contribution in $\hat{h}_2$ is 
$$
    \sum_{\omega>\omega'} \kq m_{\omega}^-m_{\omega'}^- + (m_{\omega}^+-\epsilon) (m_{\omega'}^+-\epsilon).
$$
We have our final result as in Eq.~\eqref{eq:h2=H2}:
\begin{align}
    \hat{h}_2 = & \hat{\rm H}_2 + \frac{1+\kq}{2}\left(\frac{\hat{\rm H}_1}{\kq-1}\right)^2 + (N-1)\epsilon\left(\frac{\hat{\rm H}_1}{1-\kq}\right) \nonumber\\
    & +(\kq m_c^- - m_c^+)\left(\frac{\hat{\rm H}_1}{1-\kq}\right)+ \sum_{\omega>\omega'} \kq m_{\omega}^-m_{\omega'}^- + (m_{\omega}^+-\epsilon) (m_{\omega'}^+-\epsilon).
\end{align}

\section{Detail calculation of third Hamiltonian} \label{sec:h3=H3}

Third Hamiltonian of spin chain is defined in Eq.~\eqref{eq:h4}
\begin{align}
    \hat{h}_3 
    =& \Tr \begin{pmatrix} \kq & 0 \\ 0 & 1 \end{pmatrix} \sum_{\omega_1>\omega_2>\omega_3} (-\mu_{\omega_1}+\CalL_{\omega_1})(-\mu_{\omega_2}+\CalL_{\omega_2})(-\mu_{\omega_3}+\CalL_{\omega_3}) 
    \\
    =& \kq\left[(-\mu_{\omega_1}+\ell^{0}_{\omega_1})(-\mu_{\omega_2}+\ell^{0}_{\omega_2})(-\mu_{\omega_3}+\ell^{0}_{\omega_3}) + (-\mu_{\omega_1}+\ell^{0}_{\omega_1})\ell^{-}_{\omega_2}\ell^{+}_{\omega_3} + (-\mu_{\omega_2}-\ell^{0}_{\omega_2})\ell^{-}_{\omega_1}\ell^{+}_{\omega_3} + (-\mu_{\omega_3}+\ell^{0}_{\omega_3})\ell^{-}_{\omega_1}\ell^{+}_{\omega_2}\right] \nonumber\\
    & + \left[(-\mu_{\omega_1}-\ell^{0}_{\omega_1})(-\mu_{\omega_2}-\ell^{0}_{\omega_2})(-\mu_{\omega_3}-\ell^{0}_{\omega_3}) + (-\mu_{\omega_1}-\ell^{0}_{\omega_1})\ell^{+}_{\omega_2}\ell^{-}_{\omega_3} + (-\mu_{\omega_2}+\ell^{0}_{\omega_2})\ell^{+}_{\omega_1}\ell^{-}_{\omega_3} + (-\mu_{\omega_3}-\ell^{0}_{\omega_3})\ell^{+}_{\omega_1}\ell^{-}_{\omega_2}\right] \nonumber
\end{align}

On the gauge counter part, the third Hamiltonian is defined based on Eq.~\eqref{def:Hi} with $I=3$. The $[x^{-2}]$ coefficient of Laurent expansion of fractional fundamental $qq$-character $\CalX_{\omega}(x)$ reads: 
\begin{align}
    & [x^{-2}] \CalX_{\omega} (x) \nonumber\\
    & = \frac{1}{6}\epsilon_1^3\nu_{\omega}^3+\epsilon_1 D^{(2)}_{\omega} - (a_{\omega+1}+\epsilon_+)\left[\frac{1}{2}\epsilon_1^2\nu_{\omega}^2 + \epsilon_1 D^{(1)}_{\omega} \right] + \epsilon_+\epsilon_1 a_{\omega+1}\nu_{\omega} + \epsilon_1^2\nu_{\omega}D^{(1)}_{\omega} \\
    & + \kq_{\omega} \left[-\frac{1}{6}\epsilon_1^3\nu_{\omega-1}^3 - \epsilon_1 D_{\omega-1}^{(2)} + (a_{\omega}-m_{\omega}^{+} - m_{\omega}^{-}) \left(\frac{1}{2}\epsilon_1^2\nu_{\omega-1}^2 - \epsilon_1 D^{(1)}_{\omega-1}\right) - P_{\omega}(a_{\omega})(\epsilon_1\nu_{\omega-1}-a_{\omega})+\epsilon_1^2\nu_{\omega-1}D^{(1)}_{\omega-1}\right]. \nonumber
\end{align}
The Dyson-Schwinger equation restricts $\CalX_{\omega}$ to have vanishing expectation value
\begin{align}
    0 = \langle [x^{-2}] \CalX_{\omega}(x) \rangle, \quad \omega=0,\dots,N-1.
    \label{eq:[x-2]X}
\end{align}

We again consider the linear combination $\CalU(x) = \sum_{\omega}u_{\omega}\CalX_{\omega}(x)$ defined in \eqref{def:U}. By resetting the zero point energy \eqref{eq:E_Izero}, $\hat{\rm H}_3$ becomes:
\begin{align}
    \hat{\rm H}_3 = 
    & (\kq-1)\left[\sum_{\omega}\frac{1}{6}(\epsilon\nabla^z_{\omega}-m_{\omega+1}^+)^3 - \frac{1}{3}(m_{\omega+1}^+)^3 \right]  \\
    & + \sum_{\omega}  \epsilon u_{\omega} \left[ \frac{(\epsilon\nabla^z_{\omega}-m_{\omega+1}^+)^2}{2} - \frac{a_{\omega+1}^2}{2} + \epsilon \langle D^{(1)}_{\omega} \rangle \right] \nonumber\\
    & -(u_{\omega}+\kq_{\omega+1}u_{\omega+1})\left[\epsilon \langle (\epsilon\nabla^z_{\omega}-m_{\omega+1}^+)D^{(1)}_{\omega} \rangle - \frac{a_{\omega+1}^2}{2}(\epsilon\nabla^z_{\omega}-m_{\omega+1}^+)  \right]  \nonumber
    \\
    &+\kq_{\omega+1}u_{\omega+1}\left[ (m_{\omega+1}^+ + m_{\omega+1}^-)\left(\frac{(\epsilon\nabla^z_{\omega})^2}{2}+\frac{a_{\omega+1}^2}{2} - \epsilon_1 \langle D^{(1)}_{\omega} \rangle\right) -(m_{\omega+1}^+)^2\epsilon\nabla^z_{\omega} + \frac{(m_{\omega+1}^+)^2}{2}(m_{\omega+1}^+-m_{\omega+1}^-) \right] \nonumber
\end{align}

The $\langle D_{\omega}^{(1)} \rangle$ term can be solved by using expectation value of $[x^{-1}]\CalX_{\omega}$ in Eq.~\eqref{eq:x-1coeff}
\begin{align}
    &\epsilon_1\langle D_{\omega}^{(1)}-\kq_{\omega} D_{\omega-1}^{(1)} \rangle \nonumber\\
    &=-\frac{\epsilon_1^2}{2} (\nabla^z_{\omega})^2 - \frac{\epsilon_1^2}{2}\kq_{\omega}(\nabla^z_{\omega+1})^2+\kq_{\omega}(a_{\omega}-m_{\omega}^+-m_{\omega}^-)(\nabla^z_{\omega-1}) + a_{\omega+1}\nabla^z_{\omega} - \kq_{\omega}P_{\omega}(a_{\omega}) 
    & \label{def:E_w}
\end{align}
which solves out \eqref{eq:D_w}:
\begin{align}
    \epsilon_1\langle D_{\omega}^{(1)} \rangle
    = & \frac{1}{\kq-1}\left[\sum_{n=0}^{N-1}\kq_{\omega}\cdots\kq_{\omega-n+1}\epsilon\nabla_{\omega-n}^z(\epsilon\nabla^z_{\omega-n}-m_{\omega-n+1}^++m_{\omega-n+1}^-)\right] \nonumber\\
    & + \frac{(\epsilon\nabla^z_{\omega})^2}{2} + m_{\omega+1}^-\epsilon\nabla^z_{\omega} + \frac{a_{\omega+1}^2}{2} - \frac{(m_{\omega+1}^+)^2}{2}.
\end{align}

To construct relation between $\hat{h}_3$ and $\hat{\rm H}_3$, we will take order by order, term by term comparison in the derivatives. We start with highest derivative order, $(\nabla^z_{\omega})^3$ in $\hat{\rm H}_3$ has coefficients: 
\begin{align}
    \frac{1}{2}\frac{1+\kq}{1-\kq}(u_{\omega}+\kq_{\omega+1}u_{\omega+1}) - \frac{1-\kq}{6} =  -\frac{1+\kq}{2}\frac{\gamma_{\omega}+\gamma_{\omega+1}}{z_{\omega}} - \frac{1-\kq}{6}
\end{align}
and $(\nabla^z_{\omega})^2 \nabla^z_{\omega'}$: 
\begin{subequations}
    \begin{align}
    & -\frac{u_{\omega'}+\kq_{\omega'+1}u_{\omega'+1}}{\kq-1} \kq_{\omega'}\cdots \kq_{\omega+1} = - \frac{\gamma_{\omega'} + \gamma_{\omega'+1}}{z_{\omega}}, \quad \omega'>\omega \\
    & -\frac{u_{\omega'}+\kq_{\omega'+1}u_{\omega'+1}}{\kq-1} \kq_{\omega'}\cdots \kq_{\omega-N+1} = - \frac{\kq(\gamma_{\omega'} + \gamma_{\omega'+1})}{z_{\omega}}, \quad \omega'<\omega.
\end{align}
\end{subequations}
Last, $\hat{\rm H}_3$ has no $\nabla^z_{\omega_1}\nabla^z_{\omega_2}\nabla^z_{\omega_3}$ term when $\omega_1\neq\omega_2\neq\omega_3$.

In $\hat{h}_3$, degree 3 differentiation $\partial^3$ contribution comes from $\beta_{\omega_1}\beta_{\omega_2}\beta_{\omega_2}$, which has coefficient
\begin{align}
    &\kq\left[\gamma_{\omega_1}\gamma_{\omega_2}\gamma_{\omega_{3}} - \gamma_{\omega_1}\gamma_{\omega_3}^2 + \gamma_{\omega_2}\gamma_{\omega_3}^2 - \gamma_{\omega_3}\gamma_{\omega_2}^2 \right] + \left[-\gamma_{\omega_1}\gamma_{\omega_2}\gamma_{\omega_3} + \gamma_{\omega_1}\gamma_{\omega_2}^2 - \gamma_{\omega_2}\gamma_{\omega_1}^2 + \gamma_{\omega_3}\gamma_{\omega_1}^2\right] \nonumber\\
    &= (\gamma_{\omega_1}-\gamma_{\omega_2})(\gamma_{\omega_2}-\gamma_{\omega_3})\left(\kq \gamma_{\omega_3} - \gamma_{\omega_1} \right).
\end{align}

There are 8 possible combinations $\beta_{\omega_1}\beta_{\omega_2}\beta_{\omega_3}$, $\beta_{\omega_1+1}\beta_{\omega_2}\beta_{\omega_3}$, $\beta_{\omega_1}\beta_{\omega_2+1}\beta_{\omega_3}$, $\beta_{\omega_1}\beta_{\omega_2}\beta_{\omega_3+1}$, $\beta_{\omega_1+1}\beta_{\omega_2+1}\beta_{\omega_3}$, $\beta_{\omega_1+1}\beta_{\omega_2}\beta_{\omega_3+1}$, $\beta_{\omega_1}\beta_{\omega_2+1}\beta_{\omega_3+1}$, $\beta_{\omega_1+1}\beta_{\omega_2+1}\beta_{\omega_3+1}$ that give $\partial_{z_{\omega_1}}\partial_{z_{\omega_2}}\partial_{z_{\omega_3}}$. The coefficient is found by
\begin{align}
    &(\kq-1) (\gamma_{\omega_1+1}-\gamma_{\omega_{1}})(\gamma_{\omega_2+1}-\gamma_{\omega_{2}})(\gamma_{\omega_3+1}-\gamma_{\omega_{3}}) \frac{\partial}{\partial z_{\omega_{1}}}\frac{\partial}{\partial z_{\omega_{2}}}\frac{\partial}{\partial z_{\omega_{3}}} = (\kq-1) \nabla^z_{\omega_1}\nabla^{z}_{\omega_2} \nabla^{z}_{\omega_3}.
\end{align}
Next for $\partial_{z_{\omega}}^2\partial_{z_{\omega'}}$ terms, when $\omega>\omega'$ coefficient reads:
\begin{align}
    & -(\kq (\gamma_{\omega'} + \gamma_{\omega'+1} - \gamma_{\omega}) - \gamma_{\omega+1}) (\gamma_{\omega+1}  - \gamma_{\omega})(\gamma_{\omega'+1}-\gamma_{\omega'}) 
    = -\left[\frac{\kq(\gamma_{\omega'}+\gamma_{\omega'+1})}{z_{\omega}} - \frac{\kq\gamma_{\omega} + \gamma_{\omega+1}}{z_{\omega}} \right] z_{\omega}^2 z_{\omega'};
\end{align}
while for $\omega<\omega'$:
\begin{align}
    & (\kq\gamma_{\omega} - \gamma_{\omega'+1} - \gamma_{\omega'} + \gamma_{\omega+1}) (\gamma_{\omega+1} - \gamma_{\omega})(\gamma_{\omega'+1}-\gamma_{\omega'})
    = -\left[\frac{\gamma_{\omega'}+\gamma_{\omega'+1}}{z_{\omega}} - \frac{\kq \gamma_{\omega} + \gamma_{\omega+1}}{z_{\omega}}\right] z_{\omega}^2z_{\omega'}.
\end{align}

The $\hat{h}_3$ has no $\partial_{z_{\omega}}^3$. We find for the highest derivative terms: 
\begin{align}
    h_3|_{\partial^3} 
    & = \hat{\rm H}_3|_{\partial^3} - \left[\sum_{\omega}\frac{1}{2}\frac{\kq+1}{\kq-1}(u_{\omega}+\kq_{\omega+1}u_{\omega+1})(\epsilon\nabla^z_{\omega})^2\right] (\epsilon\nabla_c^z) + \frac{\kq-1}{6}(\epsilon\nabla_c^z)^3 \nonumber\\
    & = \hat{\rm H}_3|_{\partial^3} + (1+\kq) \frac{\hat{\rm H}_2}{1-\kq} \frac{\hat{\rm H}_1}{1-\kq} + \frac{1-\kq}{6} \left(\frac{\hat{\rm H}_1}{1-\kq}\right)^3 + \cdots
\end{align}
with $\hat{\rm H}_1 = (\kq-1)\epsilon\nabla_{c}^z$ and $\hat{\rm H}_2$ defined in \eqref{eq:Hamiltonian N}.

\paragraph{}

Let us move on to second order differentiation. The
second order differential, $(\nabla^z_{\omega})^2$ in $\hat{\rm H}$ has coefficient 
\begin{align}
    & -\frac{1}{z_{\omega}}\left[ \sum_{n>\omega+1} m_{n}^-\gamma_n - \sum_{n>\omega} m_{n+1}^+\gamma_n + \sum_{n<\omega+1}\kq  m_{n}^-\gamma_n - \sum_{n<\omega} \kq m_{n+1}^+  \gamma_n + (\kq m_{\omega+1}^-\gamma_{\omega} - m_{\omega+1}^+\gamma_{\omega+1})\right] \nonumber\\
    & -\frac{1+\kq}{z_{\omega}}\left[\gamma_{\omega+1}m_{\omega+1}^- - \gamma_{\omega} m_{\omega+1}^+\right]
    - \epsilon \frac{1}{z_{\omega}}\left[-\gamma_{\omega+1} +  \sum_{n>\omega}\gamma_n + \sum_{n<\omega}\kq \gamma_{n} \right]. \nonumber
\end{align}
The cross term $\nabla^z_{\omega'}\nabla^z_{\omega}$, $\omega'>\omega$, only comes from $\langle \nabla^z_{\omega} D_{\omega}\rangle$:
\begin{align}
    & -\frac{1}{z_{\omega}}(\gamma_{\omega'}+\gamma_{\omega'+1})(m_{\omega'+1}^--m_{\omega'+1}^+) -\frac{\kq}{z_{\omega'}}(\gamma_{\omega}+\gamma_{\omega+1})(m_{\omega+1}^- - m_{\omega+1}^+).
\end{align}

On spin chain side, the second order differentiations in $\hat{h}_3$ are: 
\begin{subequations}
\begin{align}
    \beta_{\omega_1}\beta_{\omega_2}: \quad & (\gamma_{\omega_1}-\gamma_{\omega_2})\left[\kq(m_{\omega_3}^- - m_{\omega_3+1}^+ + \epsilon)\gamma_{\omega_3} - \kq m_{\omega_3}^-\gamma_{\omega_2} + (m_{\omega_3+1}^+-\epsilon) \gamma_{\omega_1}\right], \\
    \beta_{\omega_1}\beta_{\omega_3}: \quad & (\kq\gamma_{\omega_3}-\gamma_{\omega_1})[(m_{\omega_2}^- - m_{\omega_2+1}^+ + \epsilon)\gamma_{\omega_2} - m_{\omega_2}^-\gamma_{\omega_1} + (m_{\omega_2+1}^+-\epsilon)\gamma_{\omega_3}], \\
    \beta_{\omega_2}\beta_{\omega_3}: \quad & (\gamma_{\omega_2}-\gamma_{\omega_3})[(m_{\omega_1}^- - m_{\omega_1+1}^+ + \epsilon)\gamma_{\omega_1} -  \kq m_{\omega_1}^-\gamma_{\omega_3} + (m_{\omega_1+1}^+ - \epsilon) \gamma_{\omega_2})].
\end{align}
\end{subequations}
For $\beta_{\omega}\beta_{\omega'}$, the coefficient is: 
\begin{align}
    & (\gamma_{\omega}-\gamma_{\omega'}) \sum_{n<\omega'} \left[\kq(m_{n}^--m_{n+1}^++\epsilon)\gamma_{n} - \kq m_{n}^-\gamma_{\omega'} + (m_{n+1}^+-\epsilon)\gamma_{\omega}\right] \nonumber\\
    & + (\gamma_{\omega}-\gamma_{\omega'}) \sum_{n>\omega} [(m_{n}^- - m_{n+1}^+ +\epsilon) \gamma_{n} - \kq m_{n}^-\gamma_{\omega'} + (m_{n+1}^+-\epsilon)\gamma_{\omega}] \nonumber\\
    & + (\kq\gamma_{\omega'}  -\gamma_{\omega}) \sum_{\omega'<n<\omega}[ (m_{n}^- - m_{n+1}^+ + \epsilon) \gamma_{n} - m_{n}^-\gamma_{\omega} + (m_{n+1}^+-\epsilon)\gamma_{\omega'}].
\end{align}
The $\partial_{\omega}^2$ has coefficients:
\begin{align}
    & - z_{\omega} \left[\sum_{n<\omega+1}\kq m_{n}^-\gamma_n  - \sum_{n<\omega}  \kq m_{n+1}^+\gamma_n + \sum_{n>\omega+1} m_{n}^- \gamma_n - \sum_{n>\omega} m_{n+1}^+\gamma_n + \kq m_{\omega+1}^-\gamma_{\omega} - m_{\omega+1}^+ \gamma_{\omega+1} \right]  \nonumber\\
    & - z_{\omega}\left[\left(\sum_{n}m_{n+1}^+ - \epsilon\right) \gamma_{\omega+1} - \left(\sum_{n} \kq m_n^- \right) \gamma_{\omega} \right] -\epsilon {z_{\omega}}\left[ \gamma_{\omega+1} +  \sum_{n<\omega}\kq\gamma_n + \sum_{n>\omega}\gamma_n\right]. 
\end{align}

The cross term
$\partial_{z_{\omega}}\partial_{z_{\omega'}}$ comes from 4 terms $\beta_{\omega}\beta_{\omega'}$, $\beta_{\omega}\beta_{\omega'+1}$, $\beta_{\omega+1}\beta_{\omega'}$, and $\beta_{\omega+1}\beta_{\omega'+1}$:
\begin{align}
    & (\gamma_{\omega+1}-\gamma_{\omega})(\gamma_{\omega+1}+\gamma_{\omega})( m_{\omega'+1}^+ - m_{\omega'+1}^- -\epsilon) + \kq (\gamma_{\omega'+1} - \gamma_{\omega'}) (\gamma_{\omega'+1} + \gamma_{\omega'})(m_{\omega+1}^+ - m_{\omega+1}^- -\epsilon ) \nonumber\\
    & + (1+\kq) \left[ (\gamma_{\omega+1}-\gamma_{\omega})(\gamma_{\omega'+1}m_{\omega'+1}^- - \gamma_{\omega'}(m_{\omega'+1}^+-\epsilon)) + (\gamma_{\omega'+1}-\gamma_{\omega'})(\gamma_{\omega+1}m_{\omega+1}^- - \gamma_{\omega}(m_{\omega+1}^+-\epsilon))\right] \nonumber\\
     & - (\gamma_{\omega+1}-\gamma_{\omega})(\gamma_{\omega'+1}-\gamma_{\omega'}) \sum_{n}(\kq m_n^- + m_{n+1}^+ - \epsilon).
\end{align}

Combining with $z_{\omega}^2z_{\omega'}\partial_{\omega}^2\partial_{\omega'} = (\nabla^z_{\omega})^2(\nabla^z_{\omega'}) - \nabla^z_{\omega}\nabla^z_{\omega'}$.
We conclude
\begin{align}
    \hat{h}_3 =
    & \, \hat{\rm H}_3 + (1+\kq)\frac{\hat{\rm H}_2}{1-\kq} \frac{\hat{\rm H}_1}{1-\kq} + \frac{1-\kq}{6} \left(\frac{\hat{\rm H}_1}{1-\kq}\right)^3 - 2 \epsilon \left(\frac{1}{2}\left(\frac{\hat{\rm H}_1}{1-\kq} \right)^2  +  \frac{\hat{\rm H}_2}{1-\kq}  \right) \nonumber\\
    & + (N\epsilon - m_c^+) \left(\frac{1}{2}\left(\frac{\hat{\rm H}_1}{1-\kq} \right)^2 +  \frac{\hat{\rm H}_2}{1-\kq}  \right) - (\kq m_c^-) \left(\frac{1}{2}\left(\frac{\hat{\rm H}_1}{1-\kq} \right)^2 -  \frac{\hat{\rm H}_2}{1-\kq}  \right) + \cdots 
\end{align}

\paragraph{}
The order one derivative $\nabla^z_{\omega}$ in $\hat{\rm H}_3$ has coefficient:
\begin{align}
    & -\frac{1}{z_{\omega}}( \kq m_{\omega+1}^- - m_{\omega+1}^+ ) ((m_{\omega+1}^-+\epsilon) \gamma_{\omega+1} - m_{\omega+1}^+ \gamma_{\omega}) \nonumber\\
    & - \frac{1}{z_{\omega}}(m_{\omega+1}^- - m_{\omega+1}^+ )\left[\epsilon(\kq\gamma_{\omega} - \gamma_{\omega+1}) + \sum_{n>\omega+1}m_{n}^-\gamma_n - \sum_{n>\omega} (m_{\omega+1}^+-\epsilon) \gamma_{n} + \kq\sum_{n<\omega+1} m_{n}^- \gamma_n + \kq \sum_{n<\omega} (m_{n+1}^+-\epsilon) \gamma_n \right]. \nonumber
\end{align}
The first order differential in $\hat{h}_3$ may come from:
\begin{subequations}
\begin{align}
    \beta_{\omega_1}: \quad & \kq \left[m_{\omega_2}^-m_{\omega_3}^-\gamma_{\omega_1} + (m_{\omega_3+1}^+ - m_{\omega_3}^- - \epsilon ) (m_{\omega_2+1}^+-\epsilon) \gamma_{\omega_3} + (m_{\omega_2+1}^+ - m_{\omega_2}^- -\epsilon)m_{\omega_3}^- \gamma_{\omega_2} \right] - (m_{\omega_2+1}^+ -\epsilon) (m_{\omega_3+1}^+ -\epsilon) \gamma_{\omega_1}, \\
    \beta_{\omega_2}: \quad & \kq \left[ m_{\omega_1}^-m_{\omega_3}^- \gamma_{\omega_2} + m_{\omega_1}^-(m_{\omega_3+1}^+ - m_{\omega_3}^- -\epsilon)\gamma_{\omega_3}\right] - (m_{\omega_1+1}^+-\epsilon) (m_{\omega_3+1}^+-\epsilon)\gamma_{\omega_2} + (m_{\omega_1+1}^+ - m_{\omega_1}^- -\epsilon)(m_{\omega_3+1}^+-\epsilon)\gamma_{\omega_1}, \\
    \beta_{\omega_3}: \quad & \kq m_{\omega_1}^- m_{\omega_2}^-\gamma_{\omega_3} - (m_{\omega_1+1}^+-\epsilon) (m_{\omega_2+1}^+-\epsilon) \gamma_{\omega_3} + (m_{\omega_2+1}^+ - m_{\omega_2}^--\epsilon)(m_{\omega_1+1}^+-\epsilon)\gamma_{\omega_2} + (m_{\omega_1+1}^+ - m_{\omega_1}^--\epsilon) m_{\omega_2}^-\gamma_{\omega_1},
\end{align}
\end{subequations}
which gives coefficients of $\beta_{\omega}$
\begin{align}
     & \left[\frac{\kq(m_c^-)^2 - (m_c^+-N\epsilon)^2}{2} - \sum_{n}\frac{\kq(m_n^-)^2 - (m_n^+-\epsilon)^2}{2} - \kq m_c^-m_{\omega}^- + (m_c^+-N\epsilon)(m_{\omega+1}^+-\epsilon) + \kq (m_{\omega}^-)^2 - (m_{\omega+1}^+-\epsilon)^2 \right] \gamma_{\omega} \nonumber\\
     & + \sum_{\omega>n>n'} \kq (m_{n+1}^+ - m_{n}^- - \epsilon)m_{n'}^- \gamma_{n} +  \kq (m_{n+1}^+ -\epsilon ) (m_{n'+1}^+ - m_{n'}^- - \epsilon) \gamma_{n'} \\
     & + \sum_{n>\omega>n'} (m_{n'+1}^+ - \epsilon) (m_{n+1}^+ - m_{n}^- - \epsilon )\gamma_{n} + \kq (m_{n+1}^+ - m_{n}^- - \epsilon) m_{n'}^- \gamma_{n'} \nonumber\\
     & + \sum_{n>n'>\omega} m_{n'}^- (m_{n+1}^+ - m_{n}^- - \epsilon) \gamma_{n} +  (m_{n}^+ - \epsilon)(m_{n'+1}^+ - m_{n'}^- - \epsilon) \gamma_{n'}. \nonumber
\end{align}
A single differential $\partial_{\omega}$ considers $\beta_{\omega+1}$ and $\beta_{\omega}$:
\begin{align}
    & z_{\omega} \left[\frac{\kq(m_c^-)^2 - (m_c^+-N\epsilon)^2}{2} - \sum_{n}\frac{\kq(m_n^-)^2 - (m_n^+-\epsilon)^2}{2} \right] +(m_{\omega+1}^+ - \kq m_{\omega+1}^- -\epsilon )(m_{\omega+1}^+\gamma_{\omega} - m_{\omega+1}^- \gamma_{\omega+1} - \epsilon\gamma_{\omega}) 
    \nonumber\\
    & \quad + (m_{\omega+1}^- - m_{\omega+1}^+ + \epsilon)\left[\kq \sum_{n<\omega+1} m_{n}^-\gamma_n - \kq \sum_{n<\omega} (m_{n+1}^+-\epsilon)\gamma_n + \sum_{n>\omega+1}m_{n}^-\gamma_n - \sum_{n>\omega} (m_{n+1}^+-\epsilon)\gamma_n \right] \nonumber\\
    & \quad + (N\epsilon - m_c^+ - \kq m_c^-) (m_{\omega+1}^+\gamma_{\omega} -  m_{\omega+1}^-\gamma_{\omega+1}) + \epsilon z_{\omega} (N\epsilon-m_c^+) - \epsilon^2 z_{\omega}.
\end{align}
Combining with contribution from higher derivative  $z_{\omega}^2\partial_{z_\omega}^2 = (\nabla^z_{\omega})^2 - \nabla^z_{\omega}$, we find
\begin{align}
    \hat{h}_3 =
    & \, \hat{\rm H}_3 + (1+\kq)\frac{\hat{\rm H}_2}{1-\kq} \frac{\hat{\rm H}_1}{1-\kq} + \frac{1-\kq}{6} \left(\frac{\hat{\rm H}_1}{1-\kq}\right)^3 - 2 \epsilon \left(\frac{1}{2}\left(\frac{\hat{\rm H}_1}{1-\kq} \right)^2  +  \frac{\hat{\rm H}_2}{1-\kq}  \right) \nonumber\\
    & + (N\epsilon - m_c^+) \left(\frac{1}{2}\left(\frac{\hat{\rm H}_1}{1-\kq} \right)^2 +  \frac{\hat{\rm H}_2}{1-\kq}  \right) - (\kq m_c^-) \left(\frac{1}{2}\left(\frac{\hat{\rm H}_1}{1-\kq} \right)^2 -  \frac{\hat{\rm H}_2}{1-\kq}  \right) \\
    & -\left[\kq\sum_{n>n'}m_{n}^-m_{n'}^-\right] \frac{\hat{\rm H}_1}{1-\kq} + \left[\sum_{n>n'} (\epsilon - m_{n}^+)(\epsilon - m_{n'}^+)\right]\frac{\hat{\rm H}_1}{1-\kq} - \epsilon(N\epsilon - m_c^+)\frac{\hat{\rm H}_1}{1-\kq} + \epsilon^2 \frac{\hat{\rm H}_1}{1-\kq} \cdots \nonumber
\end{align}

And last for the constant term in $\hat{\rm H}_3$: 
\begin{align}
    & \sum_{\omega}(1-\kq) \left[\frac{(m_{\omega+1}^+)^3}{3!} + \frac{(m_{\omega+1}^+)^3}{3}\right] -(u_{\omega} + \kq_{\omega+1}u_{\omega+1}) \frac{(m_{\omega+1}^+)^3}{2} + \kq_{\omega+1}u_{\omega+1} (m_{\omega+1}^+)^3 = 0.
\end{align}
And constant term in $\hat{h}_3$: 
\begin{align}
    \sum_{\omega_1>\omega_2>\omega_3} (\epsilon-m_{\omega_1}^+)(\epsilon-m_{\omega_2}^+)(\epsilon-m_{\omega_2}^+) - \kq m_{\omega_1}^-m_{\omega_2}^-m_{\omega_3}^-
\end{align}
We arrive at our conclusion of Eq.~\eqref{eq:h3=H3}:
\begin{align}
    \hat{h}_3 =
    & \, \hat{\rm H}_3 + (1+\kq)\frac{\hat{\rm H}_2}{1-\kq} \frac{\hat{\rm H}_1}{1-\kq} + \frac{1-\kq}{6} \left(\frac{\hat{\rm H}_1}{1-\kq}\right)^3 - 2 \epsilon \left(\frac{1}{2}\left(\frac{\hat{\rm H}_1}{1-\kq} \right)^2  +  \frac{\hat{\rm H}_2}{1-\kq}  \right) \nonumber\\
    & + (N\epsilon - m_c^+) \left(\frac{1}{2}\left(\frac{\hat{\rm H}_1}{1-\kq} \right)^2 +  \frac{\hat{\rm H}_2}{1-\kq}  \right) - (\kq m_c^-) \left(\frac{1}{2}\left(\frac{\hat{\rm H}_1}{1-\kq} \right)^2 -  \frac{\hat{\rm H}_2}{1-\kq}  \right) \nonumber \\
    & + \left[\sum_{n>n'} (\epsilon - m_{n}^+)(\epsilon - m_{n'}^+) - \kq m_{n}^-m_{n'}^-\right]\frac{\hat{\rm H}_1}{1-\kq}  - \epsilon(N\epsilon - m_c^+)\frac{\hat{\rm H}_1}{1-\kq} + \epsilon^2 \frac{\hat{\rm H}_1}{1-\kq} \nonumber\\
    & + \sum_{\omega_1>\omega_2>\omega_3} (\epsilon-m_{\omega_1}^+)(\epsilon-m_{\omega_2}^+)(\epsilon-m_{\omega_2}^+) - \kq m_{\omega_1}^-m_{\omega_2}^-m_{\omega_3}^-.
\end{align}




\section{Calculation of the fourth Hamiltonian} \label{sec:h4=H4}

We will demonstrate how it is required to consider both the fundamental and second order $qq$-characters for the proper definition of the fourth Hamilton $\hat{\rm H}_4$ as a differential operator. We only write out the highest order derivative terms for which is enough for the illustration. 

\subsection{The fundamental qq-characters}

To construct the fourth conserving Hamiltonian of normalized vev of the surface defect $\Psi_{\ba,+}$, we the consider $[x^{-3}]$ coefficient of fundamental $qq$-character $\CalX_{\omega}(x)$ under large $x$ expansion
\begin{align}
    \left[x^{-3}\right] \CalX_{\omega}(x)=
    &\frac{1}{4!}\delta_{\omega}^4 + \frac{\epsilon_1}{2!}\delta_{\omega}^2D^{(1)}_{\omega} + \frac{\epsilon_1^2}{2!} D^{(1)}_{\omega}D^{(1)}_{\omega} + \epsilon_1 D^{(3)}_{\omega} + \epsilon_1\delta_{\omega}D^{(2)}_{\omega} + \cdots  \nonumber\\
    &+\kq_{\omega}\left[\frac{1}{4!}\delta_{\omega-1}^4 - \frac{\epsilon_1}{2}\delta_{\omega-1}^2 D^{(1)}_{\omega-1} + \frac{\epsilon_1^2}{2!} D^{(1)}_{\omega-1}D^{(1)}_{\omega-1} - \epsilon_1 D^{(3)}_{\omega-1} + \epsilon_1\delta_{\omega-1}D^{(2)}_{\omega-1}+\cdots\right].
\end{align}
with $\delta_{\omega} = \epsilon\nabla^z_{\omega} - a_{\omega+1} + m_{\omega+1}^+$. By the non-perturbative Dyson-Schwinger equation
\begin{align}
    \langle \left[x^{-3}\right]  \CalX_{\omega}(x) \rangle = 0.
    \label{eq:[x-3]X}
\end{align}
We only list out the highest derivative terms with $\cdots$ denoting the lower derivatives.
Fourth Hamiltonian of gauge theory is defined in \eqref{eq:H4}
\begin{align}
    \hat{\rm H}_4 = (\kq-1)\left[\sum_{\omega}\frac{\epsilon_1}{2} \langle \delta_{\omega}^2 D_{\omega} \rangle \right] - \sum_{\omega} (u_{\omega}+\kq_{\omega+1}u_{\omega+1})\left[\frac{\delta_{\omega}^4}{4!} + \frac{\epsilon_1^2}{2} \langle D_{\omega}^2 \rangle + \epsilon_1 \langle \delta_{\omega} D_{\omega}^{(2)} \rangle \right] + \cdots 
\end{align}

\subsection{Second order $qq$-character}

To rewrite $\langle D_{\omega}^{(2)} \rangle$ into differential operators acting on $\Psi_{\ba,+}$, we use \eqref{eq:[x-2]X}
\begin{align}
    &\epsilon_1 \langle D_{\omega}^{(2)} - \kq_{\omega} D_{\omega-1}^{(2)} \rangle \nonumber\\
    &=-\frac{1}{6}\delta_{\omega}^3 - \langle \epsilon_1 \delta_{\omega}D^{(1)}_{\omega} \rangle + \frac{a_{\omega+1}^2}{2}\delta_{\omega} + \frac{a_{\omega}^3}{3} 
     \nonumber\\
    & \quad - \kq_{\omega}\left[-\frac{\delta_{\omega-1}^3}{6} + \langle \delta_{\omega-1}\epsilon_1D^{(1)}_{\omega-1} \rangle - \frac{a_{\omega}^2}{2}\delta_{\omega-1} + \frac{a_{\omega}^3}{3}
    - (m_{\omega}^++m_{\omega}^-)\left(\frac{\delta_{\omega-1}^2}{2} + \frac{a_{\omega}^2}{2} + \langle \epsilon_1 D^{(1)}_{\omega-1}\rangle\right) - m_{\omega}^+m_{\omega}^- \delta_{\omega-1}\right] \nonumber \\
    & := \epsilon_1\langle C_{\omega}^{(2)} \rangle .
    \label{def:C2_w}
\end{align}
We solve out:
\begin{align}
    \epsilon_1\langle D_{\omega}^{(2)} \rangle 
    & = \frac{\epsilon_1}{1-\kq}\left[\langle C_{\omega}^{(2)} \rangle+\kq_{\omega}\langle C_{\omega-1}^{(2)} \rangle + \cdots + \kq_{\omega}\kq_{\omega-1} \cdots \kq_{\omega-N+2}\langle C_{\omega-N+1}^{(2)} \rangle \right] \nonumber\\
    & = -\frac{1}{6} \delta_{\omega}^3 + \langle \delta_{\omega} D^{(1)}_{\omega} \rangle + \frac{2}{\kq-1}\sum_{n=0}^{N-1}\kq_{\omega}\cdots\kq_{\omega-n+1}\langle\delta_{\omega-n} D^{(1)}_{\omega-n} \rangle + \cdots
\end{align}

To rewrite $\langle (D^{(1)}_{\omega})^2 \rangle$ as proper differential operator w.r.t $z_{\omega}$, we will need to consider second order $qq$-character \eqref{eq:2ndqq}. 
We will consider $[x^{-2}]$ coefficient of $\CalX_{\omega_1,\omega_2}^{(2)}(x)$, whose highest derivative terms are
\begin{align}\label{eq:[x-2]X2}
    [x^{-2}]\CalX_{\omega_1,\omega_2}^{(2)}(x)  = 
    &\,  D_{\omega_1}^{(3)} + D_{\omega_2}^{(3)} + \frac{1}{2}(D_{\omega_1}+D_{\omega_2})^2 + (\delta_{\omega_1}+\delta_{\omega_2})(D_{\omega_1}^{(2)} + D_{\omega_2}^{(2)}) \\
    & \qquad \qquad+ \frac{1}{4!}(\delta_{\omega_1} + \delta_{\omega_2})^4 + \frac{1}{2}(\delta_{\omega_1}+\delta_{\omega_2})^2(D^{(1)}_{\omega_1}+D^{(1)}_{\omega_2}) \nonumber\\
    & + \kq_{\omega_1} R_{\omega_1,\omega_2}(\nu) \left[ -D_{\omega_1-1}^{(3)} + D_{\omega_2}^{(3)} + \frac{1}{2}(-D^{(1)}_{\omega_1-1}+D^{(1)}_{\omega_2})^2 + (-\delta_{\omega_1-1}+\delta_{\omega_2})(-D_{\omega_1-1}^{(2)} + D_{\omega_2}^{(2)}) \right. \nonumber\\
    & \left. \qquad \qquad  + \frac{1}{4!}(-\delta_{\omega_1-1} + \delta_{\omega_2})^4 + \frac{1}{2}(-\delta_{\omega_1-1}+\delta_{\omega_2})^2(-D^{(1)}_{\omega_1-1}+D^{(1)}_{\omega_2})\right] \nonumber\\
    & + \kq_{\omega_2} R_{\omega_2,\omega_1}(-\nu) \left[ D_{\omega_1}^{(3)} - D_{\omega_2-1}^{(3)} + \frac{1}{2}(D^{(1)}_{\omega_1}-D^{(1)}_{\omega_2-1})^2 + (\delta_{\omega_1}-\delta_{\omega_2-1})(D_{\omega_1}^{(2)} - D_{\omega_2-1}^{(2)}) \right. \nonumber\\
    & \qquad  \qquad \left.+ \frac{1}{4!}(\delta_{\omega_1} - \delta_{\omega_2-1})^4 + \frac{1}{2}(\delta_{\omega_1} - \delta_{\omega_2-1})^2 (D^{(1)}_{\omega_1} - D^{(1)}_{\omega_2-1}) \right] \nonumber\\
    & + \kq_{\omega_1}\kq_{\omega_2} \left[ - D_{\omega_1-1}^{(3)} - D_{\omega_2-1}^{(3)} + \frac{1}{2}(-D^{(1)}_{\omega_1-1}-D^{(1)}_{\omega_2-1})^2 + (-\delta_{\omega_1-1}-\delta_{\omega_2-1})(-D_{\omega_1-1}^{(2)} - D_{\omega_2-1}^{(2)}) \right. \nonumber\\
    & \qquad \qquad \left.+ \frac{1}{4!}(-\delta_{\omega_1-1} - \delta_{\omega_2-1})^4 + \frac{1}{2}(-\delta_{\omega_1-1} - \delta_{\omega_2-1})^2 (- D^{(1)}_{\omega_1-1} - D^{(1)}_{\omega_2-1}) \right] +\cdots  \nonumber
\end{align}

By the structure of $R_{\omega_1,\omega_2}(\nu)$ in \eqref{R}, we may take $\nu\to\infty$ such that $R_{\omega_1,\omega_2}(\nu)=1$ for any $\{\omega_1,\omega_2\}$. Using \eqref{eq:[x-3]X}:
\begin{align}
    0 & = \left\langle D_{\omega}^{(3)} - \kq_{\omega} D_{\omega}^{(3)} + \frac{1}{2} (D_{\omega}^2 + \kq_{\omega}D_{\omega-1}^2) + \frac{1}{2}(\delta_{\omega}^2 D_{\omega} - \delta_{\omega-1}^2D_{\omega-1}) + \frac{1}{4!}(\delta_{\omega}^4 + \kq_{\omega} \delta_{\omega-1}^2) \right\rangle +\cdots 
\end{align}
to simplify $\langle [x^{-2}] \CalX_{\omega_1,\omega_2}^{(2)} \rangle$. Define 
\begin{align}
    E_{\omega} = D^{(1)}_{\omega} - \kq_{\omega}D^{(1)}_{\omega-1}.
\end{align}
Now we have
\begin{align}
    0 
    = & \, \langle E_{\omega_1}E_{\omega_2} + (\delta_{\omega_1}-\kq_{\omega_1}\delta_{\omega_1-1}) C_{\omega_2}^{(2)} + (\delta_{\omega_2}-\kq_{\omega_2}\delta_{\omega_2-1})C_{\omega_1}^{(2)} + \frac{1}{2}(\delta_{\omega_1}^2 + \kq_{\omega_1}\delta_{\omega_1-1}^2) E_{\omega_2} + \frac{1}{2} (\delta_{\omega_2}^2 + \kq_{\omega_2}\delta_{\omega_2-1}^2) E_{\omega_1} \nonumber\\
    & + \frac{1}{6}(\delta_{\omega_2} - \kq_{\omega_2}\delta_{\omega_2-1}) (\delta_{\omega_1}^3 - \kq_{\omega_1}\delta_{\omega_1-1}^3) + \frac{1}{6}(\delta_{\omega_1} - \kq_{\omega_1}\delta_{\omega_1-1}) (\delta_{\omega_2}^3 - \kq_{\omega_2}\delta_{\omega_2-1}^3) + \frac{1}{4} (\delta_{\omega_1}^2 + \kq_{\omega_1}\delta_{\omega_1-1}^2)(\delta_{\omega_2}^2 + \kq_{\omega_2}\delta_{\omega_2-1}^2) \nonumber\\
    & + (\delta_{\omega_2}-\kq_{\omega_2}\delta_{\omega_2-1})(\delta_{\omega_1}D^{(1)}_{\omega_1} + \kq_{\omega_1}\delta_{\omega_1-1}D^{(1)}_{\omega_1-1}) + (\delta_{\omega_1}-\kq_{\omega_1}\delta_{\omega_1-1})(\delta_{\omega_2}D^{(1)}_{\omega_2} + \kq_{\omega_2}\delta_{\omega_2-1}D^{(1)}_{\omega_2-1}) +\cdots \rangle 
\end{align}
which can be decomposed into two parts
\begin{subequations}
\begin{align}
    & (\delta_{\omega_1} - \kq_{\omega_1}\delta_{\omega_1-1})\left[ \langle C_{\omega_2}^{(2)} \rangle + \frac{1}{6}(\delta_{\omega_2}^3 - \kq_{\omega_2}\delta_{\omega_2-1}^3) + \langle \delta_{\omega_2}D^{(1)}_{\omega_2} \rangle + \kq_{\omega_2} \langle \delta_{\omega_2-1}D^{(1)}_{\omega_2-1} \rangle \right] + (1\leftrightarrow 2) + \cdots \\
    & \langle E_{\omega_1}E_{\omega_2} \rangle + \frac{1}{4}(\delta_{\omega_1}^2+\kq_{\omega_1}\delta_{\omega_1-1})(\delta_{\omega_2}^2+\kq_{\omega_2}\delta_{\omega_2-1}) +\frac{1}{2}\langle(\delta_{\omega_1}^2 + \kq_{\omega_1}\delta_{\omega_1-1}^2) E_{\omega_2}\rangle + \frac{1}{2} \langle (\delta_{\omega_2}^2 + \kq_{\omega_2}\delta_{\omega_2-1}^2) E_{\omega_1} \rangle + \cdots
\end{align}
\end{subequations}
From calculation of $\hat{\rm H}_3$, we know on highest order derivatives $\langle \delta_{\omega_1} E_{\omega_2} \rangle = \langle E_{\omega_2} \rangle \delta_{\omega_1} + (\delta_{\omega_1}\langle E_{\omega_2} \rangle)$. Here we only considers highest order derivatives which gives
\begin{align}
    \langle E_{\omega_1}E_{\omega_2} \rangle & = - \frac{1}{4}(\delta_{\omega_1}^2+\kq_{\omega_1}\delta_{\omega_1-1})(\delta_{\omega_2}^2+\kq_{\omega_2}\delta_{\omega_2-1}) - \frac{1}{2} \langle E_{\omega_2} \rangle (\delta_{\omega_1}^2 + \kq_{\omega_1}\delta_{\omega_1-1}^2)  - \frac{1}{2} \langle E_{\omega_1} \rangle (\delta_{\omega_2}^2 + \kq_{\omega_2}\delta_{\omega_2-1}^2)  \nonumber + \cdots \\
    & = \langle E_{\omega_1} \rangle \langle E_{\omega_2} \rangle + \cdots
\end{align}
Multiplying matrix inverting $D^{(1)} \to E$ relations we get 
\begin{align}
    \langle D^{(1)}_{\omega_1} D^{(1)}_{\omega_2} \rangle = \langle D^{(1)}_{\omega_1} \rangle \langle D^{(1)}_{\omega_2} \rangle +\cdots
\end{align}
for all $\omega_1$, and $\omega_2$. In particular in our interests $\omega_1=\omega_2=\omega$. 

\subsection{Matching Hamiltonian at highest derivative}

In highest order derivative
$\langle (D^{(1)}_{\omega})^2 \rangle = \langle D^{(1)}_{\omega} \rangle^2 + \cdots$, which consists $\delta_{\omega}^4$ and $\delta_{\omega}^2\delta_{\omega'}^2$. Such derivative terms also come from $\delta_{\omega}D^{(2)}_{\omega}$. $\delta_{\omega}^4$, and $\delta_{\omega}^2 D^{(1)}_{\omega}$. We instead first deal with $\delta_{\omega}^3\delta_{\omega'}$, which only comes from $\delta_{\omega}D_{\omega}^{(2)}$:
\begin{align}
    -(u_{\omega}+\kq_{\omega+1}u_{\omega+1})\frac{2\kq}{(\kq-1)^2} - (u_{\omega'}+\kq_{\omega'+1}u_{\omega'+1})\frac{\kq+1}{(\kq-1)^2}\kq_{\omega'}\cdots\kq_{\omega+1}.
\end{align}
We notice that corresponding coefficients in $\frac{\kq+1}{\kq-1}\hat{\rm H}_3\delta_c^z$:
\begin{align}
    -\frac{1+\kq}{6} + (u_{\omega}+\kq_{\omega+1}u_{\omega+1})\frac{1}{2} \frac{(\kq+1)^2}{(\kq-1)^2} + (u_{\omega'}+\kq_{\omega'+1}u_{\omega'+1})\frac{\kq+1}{(\kq-1)^2}\kq_{\omega'}\cdots\kq_{\omega+1}.
\end{align}
$(D^{(1)}_{\omega})^2$ also lacks $\delta_{\omega}^2\delta_{\omega'}\delta_{\omega''}$, which also comes from $\delta_{\omega}D_{\omega}^{(2)}$: 
\begin{subequations}
\begin{align}
    \omega' > \omega'' > \omega : \quad & -\frac{2}{\kq-1}\frac{1}{z_{\omega}}(\gamma_{\omega'} + \gamma_{\omega'+1}), \\
    \omega'' > \omega' > \omega : \quad & -\frac{2\kq}{\kq-1}\frac{1}{z_{\omega}}(\gamma_{\omega'} + \gamma_{\omega'+1}), \\
    \omega' > \omega > \omega'' : \quad & -\frac{2\kq}{\kq-1}\frac{1}{z_{\omega}}(\gamma_{\omega'} + \gamma_{\omega'+1}), \\
    \omega'' > \omega > \omega' : \quad & -\frac{2\kq}{\kq-1}\frac{1}{z_{\omega}}(\gamma_{\omega'} + \gamma_{\omega'+1}), \\
    \omega > \omega' > \omega'' : \quad & -\frac{2\kq}{\kq-1}\frac{1}{z_{\omega}}(\gamma_{\omega'} + \gamma_{\omega'+1}), \\
    \omega > \omega'' > \omega' : \quad & -\frac{2\kq^2}{\kq-1}\frac{1}{z_{\omega}}(\gamma_{\omega'} + \gamma_{\omega'+1}),
\end{align}
\end{subequations}
In particular we notice from $\frac{\kq+1}{\kq-1}\hat{\rm H}_3\delta_c^z$: 
\begin{subequations}
\begin{align}
    \omega' > \omega'' > \omega : \quad & \frac{\kq+1}{\kq-1}\frac{1}{z_{\omega}}(\gamma_{\omega'} + \gamma_{\omega'+1}), \\
    \omega'' > \omega' > \omega : \quad & \frac{\kq+1}{\kq-1}\frac{1}{z_{\omega}}(\gamma_{\omega'} + \gamma_{\omega'+1}), \\
    \omega' > \omega > \omega'' : \quad & \frac{\kq+1}{\kq-1}\frac{1}{z_{\omega}}(\gamma_{\omega'} + \gamma_{\omega'+1}), \\
    \omega'' > \omega > \omega' : \quad & \frac{\kq^2+\kq}{\kq-1}\frac{1}{z_{\omega}}(\gamma_{\omega'} + \gamma_{\omega'+1}), \\
    \omega > \omega' > \omega'' : \quad & \frac{\kq^2+\kq}{\kq-1}\frac{1}{z_{\omega}}(\gamma_{\omega'} + \gamma_{\omega'+1}), \\
    \omega > \omega'' > \omega' : \quad & \frac{\kq^2+\kq}{\kq-1}\frac{1}{z_{\omega}}(\gamma_{\omega'} + \gamma_{\omega'+1}), 
\end{align}
\end{subequations}



Fourth Hamiltonian of spin chain $\hat{h}_4$:
\begin{align}
    \hat{h}_4 
    = & \Tr K \sum_{\omega_1>\omega_2>\omega_3>\omega_4} (-\mu_{\omega_1}+\CalL_{\omega_1}) (-\mu_{\omega_2}+\CalL_{\omega_2}) (-\mu_{\omega_3}+\CalL_{\omega_3}) (-\mu_{\omega_4}+\CalL_{\omega_4}) \\
    = & \kq\left[(-\mu_{\omega_1} + \ell^{0}_{\omega_1}) (-\mu_{\omega_2} + \ell_{\omega_2}^{0}) (-\mu_{\omega_3} + \ell_{\omega_3}^{0}) (-\mu_{\omega_4} + \ell_{\omega_4}^{0}) + \ell_{\omega_1}^{-}\ell_{\omega_2}^{+}(-\mu_{\omega_3} + \ell_{\omega_3}^{0})(-\mu_{\omega_4}^- + \ell_{\omega_4}^{0})\right.\nonumber\\
    & \quad + (-\mu_{\omega_1} + \ell_{\omega_1}^{0}) \ell_{\omega_2}^{-}\ell_{\omega_3}^{+}(-\mu_{\omega_4} + \ell_{\omega_4}^{0}) + \ell_{\omega_1}^{-}(-\mu_{\omega_2} - \ell_{\omega_2}^{0}) \ell_{\omega_3}^{+} (-\mu_{\omega_4} + \ell_{\omega_4}^{0}) \nonumber\\
    & \quad + (-\mu_{\omega_1} + \ell_{\omega_1}^{0}) (-\mu_{\omega_2} + \ell_{\omega_2}^{0}) \ell_{\omega_3}^{-}\ell_{\omega_4}^{+} + \ell_{\omega_1}^{-}\ell_{\omega_2}^{+}\ell_{\omega_3}^{-}\ell_{\omega_4}^{+} \nonumber\\
    & \quad \left. + (-\mu_{\omega_1} + \ell_{\omega_1}^{0})\ell_{\omega_2}^{-}(-\mu_{\omega_3}-\ell_{\omega_3}^{0})\ell_{\omega_4}^{+} + \ell_{\omega_1}^{-}(-\mu_{\omega_2} - \ell_{\omega_2}^{0})(-\mu_{\omega_3} - \ell_{\omega_3}^{0}) \ell_{\omega_4}^{+} \right] \nonumber\\
    & + \left[ \ell_{\omega_1}^{+} (-\mu_{\omega_2} + \ell_{\omega_2}^{0}) (-\mu_{\omega_3} + \ell_{\omega_3}^{0}) \ell_{\omega_4}^{-} + (-\mu_{\omega_1} - \ell_{\omega_1}^{0}) \ell_{\omega_2}^{+}(-\mu_{\omega_3} + \ell_{\omega_3}^{0}) \ell_{\omega_4}^{-}\right. \nonumber\\
    & \quad + \ell_{\omega_1}^{+} \ell_{\omega_2}^{-} \ell_{\omega_3}^{+} \ell_{\omega_4}^{-} + (-\mu_{\omega_1} - \ell_{\omega_1}^{0})(-\mu_{\omega_2} - \ell_{\omega_2}^{0}) \ell_{\omega_3}^{+} \ell_{\omega_4}^{-} \nonumber\\
    & \quad + \ell_{\omega_1}^{+}(\mu_{\omega_2} + \ell_{\omega_2}^{0}) \ell_{\omega_3}^{-} (-\mu_{\omega_4} - \ell_{\omega_4}^{0}) + (-\mu_{\omega_1} - \ell_{\omega_1}^{0}) \ell_{\omega_2}^{+} \ell_{\omega_3}^{-} (-\mu_{\omega_4} - \ell_{\omega_4}^{0}) \nonumber\\
    & \quad + \left. \ell_{\omega_1}^{+}\ell_{\omega_2}^{-} (-\mu_{\omega_3} - \ell_{\omega_3}^{0}) (-\mu_{\omega_4} - \ell_{\omega_4}^{0}) + (-\mu_{\omega_1} - \ell_{\omega_1}^{0}) (-\mu_{\omega_2} - \ell_{\omega_2}^{0}) (-\mu_{\omega_3} - \ell_{\omega_3}^{0}) (-\mu_{\omega_4} - \ell_{\omega_4}^{0}) \right]. \nonumber
\end{align}
The highest order derivative $\beta_{\omega_1}\beta_{\omega_2}\beta_{\omega_3}\beta_{\omega_4}$, $\omega_1>\omega_2>\omega_3>\omega_4$, has coefficient:
\begin{align}
    (\gamma_{\omega_1}-\gamma_{\omega_2}) ( \gamma_{\omega_2} - \gamma_{\omega_3}) (\gamma_{\omega_3}-\gamma_{\omega_4}) (\kq\gamma_{\omega_4}-\gamma_{\omega_1})
\end{align}
Coefficients of $\partial_{\omega}^2\partial_{\omega'}^2$ can be found by, $\omega>\omega'$:
\begin{align}
    z_{\omega}z_{\omega'}(\gamma_{\omega}-\gamma_{\omega'+1})(\kq\gamma_{\omega'}-\gamma_{\omega+1}).
\end{align}
Coefficient of $\partial_{\omega_1}\partial_{\omega_2}\partial_{\omega_3}\partial_{\omega_4}$: 
\begin{align} 
    (1+\kq)(\gamma_{\omega_1+1} - \gamma_{\omega_1}) (\gamma_{\omega_2+1} - \gamma_{\omega_2}) (\gamma_{\omega_3+1} - \gamma_{\omega_3}) (\gamma_{\omega_4+1} - \gamma_{\omega_4})
    = (1+\kq) z_{\omega_1}z_{\omega_2}z_{\omega_3}z_{\omega_4}.
\end{align}
Coefficients of $\partial_{\omega}^2\partial_{\omega'}\partial_{\omega''}$: 
\begin{subequations}
\begin{align}
    \omega>\omega'>\omega'': \quad & z_{\omega}z_{\omega'}z_{\omega''}[\kq(\gamma_{\omega'} + \gamma_{\omega'+1}) - \kq (\gamma_{\omega''} + \gamma_{\omega''+1}) - (\kq \gamma_{\omega} - \gamma_{\omega+1})] \\
    \omega'>\omega>\omega'': \quad & z_{\omega}z_{\omega'}z_{\omega''} [ (\gamma_{\omega'} + \gamma_{\omega'+1} ) - \kq (\gamma_{\omega''} + \gamma_{\omega''+1} ) - (\kq \gamma_{\omega} - \gamma_{\omega+1} )] \\
    \omega'>\omega''>\omega: \quad & z_{\omega}z_{\omega'}z_{\omega''}[(\gamma_{\omega'}+\gamma_{\omega'+1}) - (\gamma_{\omega''} + \gamma_{\omega''+1}) - (\kq \gamma_{\omega} - \gamma_{\omega+1})]
\end{align}
\end{subequations}
with 
\begin{align}
    -\kq \gamma_{\omega} + \gamma_{\omega+1} = \frac{1+\kq}{2} (\gamma_{\omega+1} - \gamma_{\omega}) + \frac{1-\kq}{2} (\gamma_{\omega+1} + \gamma_{\omega}) = z_{\omega}\left[\frac{1+\kq}{2} - \frac{1}{2}(u_{\omega} + \kq_{\omega+1}u_{\omega+1})\right].
\end{align}
From $D_{\omega}^2$ unrelated terms ($\delta_{\omega}^3\delta_{\omega'}$ and $\delta_{\omega}^2\delta_{\omega'}\delta_{\omega''}$), we find
\begin{align}
    \hat{h}_4 = \hat{\rm H}_4 + \frac{\kq+1}{\kq-1}\hat{\rm H}_3(\delta_c^z) + \frac{1}{2}\hat{\rm H}_2(\delta_c^z)^2 + \frac{1+
    \kq}{4!}(\delta_c^z)^4 + \cdots
\end{align}
For $D_{\omega}^2$ related terms, the first one is $\delta_{\omega}^4$, whose coefficient in $\hat{\rm H}_4$ is found to be
\begin{align}
    &\frac{1+\kq}{8} - \frac{1}{8} (u_{\omega}+\kq_{\omega+1}u_{\omega+1}) - \frac{1}{8}\left(\frac{\kq+1}{\kq-1}\right)^2 (u_{\omega}+\kq_{\omega+1}u_{\omega+1}) \nonumber\\
    & - \frac{1}{2}\frac{1}{(\kq-1)^2}\sum_{n<\omega} (u_{n} + \kq_{n+1} u_{n+1}) \frac{z_{n}^2}{z_{\omega}^2} - \frac{1}{2}\frac{1}{(\kq-1)^2}\sum_{n>\omega} (u_{n} + \kq_{n+1}u_{n+1}) \frac{\kq^2 z_{n}^2}{z_{\omega}^2} \nonumber\\
    =&\frac{1+\kq}{8} - \frac{1}{4}\left(1+\kq + \sum_{n>\omega}2\frac{z_{n}}{z_{\omega}} +  \sum_{n<\omega}2\kq\frac{z_{n}}{z_{\omega}} \right) - \frac{1}{2}\frac{\kq}{(\kq-1)^2} \left( 1+\kq +\sum_{n>\omega} 2\frac{z_{n}}{z_{\omega}} + \sum_{n<\omega} 2\kq\frac{z_{n}}{z_{\omega}} \right) \nonumber\\
    & -\frac{1}{2}\frac{1}{(\kq-1)^2}\sum_{n<\omega}\left(1+\kq + \sum_{n'>n} 2\frac{z_{n'}}{z_{n}} + \sum_{n'<n} 2\kq \frac{z_{n'}}{z_{n}} \right)\frac{z_{n}^2}{z_{\omega}^2} \nonumber\\
    & -\frac{1}{2}\frac{1}{(\kq-1)^2}\sum_{n>\omega}\left( 1+ \kq + \sum_{n'>n}2\frac{z_{n'}}{z_{n}} + \sum_{n'<n}2\kq \frac{z_{n'}}{z_{n}} \right) \kq^2\frac{z_{n}^2}{z_{\omega}^2} \nonumber\\
    =& -\frac{\kq+1}{(\kq-1)^2}\left[\frac{(\kq+1)^2}{8} + \frac{(1+\kq)}{4}\left(\sum_{n>\omega}2\frac{z_{n}}{z_{\omega}} + \sum_{n<\omega}2\kq\frac{z_{n}}{z_{\omega}}\right)+ \frac{1}{2}\left(\sum_{n>\omega}\frac{z_{n}^2}{z_{\omega}^2} + \sum_{n<\omega}\kq^2\frac{z_{n}^2}{z_{\omega}^2}\right)\right] \nonumber\\
    & -\frac{1}{(\kq-1)^2}\left[\sum_{n<\omega}\sum_{n'>n}\frac{z_{n}z_{n'}}{z_{\omega}^2} + \sum_{n<\omega}\sum_{n'<n}\kq\frac{z_{n}z_{n'}}{z_{\omega}^2} + \sum_{n>\omega}\sum_{n'>n}\kq^2\frac{z_{n}z_{n'}}{z_{\omega}^2} + \sum_{n>\omega}\sum_{n'<n}\kq^3\frac{z_{n}z_{n'}}{z_{\omega}^2} \right] \nonumber\\
    =& -\frac{1}{8}\frac{\kq+1}{(\kq-1)^2}\left[(1+\kq)^2 + 2 (1+\kq)\left(\sum_{n>\omega}2\frac{z_{n}}{z_{\omega}} + \sum_{n<\omega}2\kq\frac{z_{n}}{z_{\omega}}\right) + \left(\sum_{n>\omega}2\frac{z_{n}}{z_{\omega}} + \sum_{n<\omega}2\kq\frac{z_{n}}{z_{\omega}}\right)^2\right] \nonumber\\
    =& -\frac{1}{2}\frac{\kq+1}{(\kq-1)^2}\left[\frac{u_{\omega+1}+\kq_{\omega+1}u_{\omega+1}}{2}\right]^2.
\end{align}
The coefficient of $\delta_{\omega}^2\delta_{\omega'}^2$, with $\omega>\omega'$ reads:
\begin{align}
    &\frac{1+\kq}{4} + \frac{1}{2}\left[\frac{z_{\omega}}{z_{\omega'}} + \kq \frac{z_{\omega'}}{z_{\omega}} \right] - \frac{1}{2} \left[(u_{\omega} + \kq_{\omega+1}u_{\omega+1}) + (u_{\omega'} + \kq_{\omega'+1}u_{\omega'+1})\right] \nonumber \\
    & +\frac{1}{2}\frac{\kq+1}{(\kq-1)^2}(u_{\omega}+\kq_{\omega+1}u_{\omega+1})\frac{z_{\omega}}{z_{\omega'}} + \frac{1}{2}\frac{\kq+1}{(\kq-1)^2}(u_{\omega'}+\kq_{\omega'+1}u_{\omega'+1})\kq \frac{z_{\omega'}}{z_{\omega}} \nonumber\\
    & + \frac{1}{(\kq-1)^2} \sum_{n\neq\omega,\omega'}(u_{n} + \kq_{n+1} u_{n+1}) (\kq_{n}\cdots\kq_{\omega+1}) (\kq_{n}\cdots \kq_{\omega'+1}).
\end{align}
We identify
\begin{align}
    \hat{h}_4 = \hat{\rm H}_4 + (1+\kq)\frac{\hat{\rm H}_3}{1-\kq}\frac{\hat{\rm H}_1}{1-\kq} + \frac{1-\kq}{2}\frac{\hat{\rm H}_2}{1-\kq}\left(\frac{\hat{\rm H}_1}{1-\kq}\right)^2 + \frac{1+
    \kq}{4!}\left(\frac{\hat{\rm H}_1}{1-\kq}\right)^4 +\frac{1+\kq}{2}\left(\frac{\hat{\rm H}_2}{1-\kq}\right)^2 + \cdots
\end{align}
This agrees with highest derivative term in \eqref{eq:qq=SC2}.

\newpage
\bibliographystyle{utphys}
\bibliography{BPZ}

\end{document}